\def\fileversion{v2.25a} \def\filedate{2009/10/10}
\documentclass[edeposit,fullpage,prequest,10pt]{uiucthesis2009}
\usepackage{cite}
\usepackage{amsfonts}
\usepackage{amssymb}
\usepackage{amsmath}
\usepackage{hyperref}
\usepackage{graphicx}% Include figure files
\usepackage{epstopdf}
\usepackage{subfig}
\usepackage{makeidx}
\usepackage[page,toc,title,titletoc]{appendix}

\newcommand{\vk}{\ensuremath{\mathbf{k}}}

\providecommand{\vr}{\ensuremath{\mathbf{r}}}

\newcommand{\vp}{\ensuremath{\mathbf{p}}}

\newcommand{\vq}{\ensuremath{\mathbf{q}}}

\newcommand{\A}{\ensuremath{\mathbf{A}}}

\newcommand{\ket}[1]{\ensuremath{\left|#1\right>}}

\newcommand{\nth}[1]{\ensuremath{\frac{1}{#1}}}

\newcommand{\br}[1]{\ensuremath{\left(#1\right)}}
\newcommand{\mbr}[1]{\ensuremath{\left[#1\right]}}
\newcommand{\bbr}[1]{\ensuremath{\left\{#1\right\}}}

\newcommand{\av}[1]{\ensuremath{\bigl<{#1}\bigr>}}

\newcommand{\mtrx}[1]{\ensuremath{\begin{pmatrix}#1\end{pmatrix}}}
\newcommand{\cmtrx}[1]{\ensuremath{\br{\begin{smallmatrix}#1\end{smallmatrix}}}}

\providecommand{\abs}[1]{\ensuremath{\left\lvert{#1}\right\rvert}}

\newcommand{\dg}{\ensuremath{\dagger}}
\newcommand{\nG}{\ensuremath{\hat{\mathcal{G}}^{-1}}}

\providecommand{\pdiff}[2]{\ensuremath{\frac{\partial{#1}}{\partial{#2}}}}

\providecommand{\eef}[1]{Eq. \ref{#1}}

\providecommand{\sch}{{Schr\"{o}dinger }}

\providecommand{\sgn}{\ensuremath{\text{sgn}}}

\newcommand{\tr}{\ensuremath{\text{tr}}}
\newcommand{\Tr}{\ensuremath{\text{Tr}}}

\providecommand{\bigD}{\ensuremath{\mathcal{D}}}

\makeatletter

\newcommand{\Rmnum}[1]{\expandafter\@slowromancap\romannumeral #1@}
\makeatother

\newenvironment{unsure}{}{}

\begin{document}

\title{BEC-BCS Crossover with Feshbach Resonance for Three-Hyperfine-Species Model}
\author{Guojun Zhu}
\department{Physics}
\schools{B.S., University of Science and Technology of China, 2001\\
         M.S, University of Illinois at Urbana-Champaign, 2003}
\phdthesis
\advisor{Anthony Leggett}
\degreeyear{2012}
\committee{Professor Gordon Baym, Chair\\Professor Anthony Leggett, Director of Research\\Professor Brian DeMarco\\Professor Scott Willenbrock}
\maketitle

\frontmatter

\begin{abstract}
The BEC-BCS crossover problem has been intensively studied both theoretically and experimentally largely thanks to  Feshbach resonances which allow us to tune the effective interaction between alkali atoms.  In a Feshbach resonance, the effective s-wave scattering length grows when one moves toward the resonance point, and eventually diverges at this point.  There is one characteristic energy scale, $\delta_c$, defined as, in the negative side of the resonance point, the detuning energy at which the weight of the bound state shifts from predominatedly in the open-channel to predominated in the closed-channel.  When the many-body energy scale (e.g. the Fermi energy, $E_{F}$) is larger than $\delta_c$, the closed-channel weight is significant and has to be included in the many-body theory.  Furthermore, when two channels share a hyperfine species, the Pauli exclusion between fermions from two channels also needs to be taken into consideration in the many-body theory.  

The current  thesis addresses the above problem in detail. A set of gap equations and number equations  are derived at the mean-field level.  The fermionic and bosonic excitation spectra are then derived. Assuming that the uncoupled bound-state of the closed-channel in resonance is much smaller than the inter-particle distance, as well as the s-wave scattering length, $a_s$, we find that  the basic equations in the single-channel crossover model are still valid. The correction first comes from the existing of the finite chemical potential and additional counting complication due to the closed-channel.  These two corrections need to be included into the mean-field equations, i.e. the gap equations and the number equations, and be solved self-consistently.  Then the correction due to the inter-channel Pauli exclusion is in the  order of the ratio of the Fermi energy and the Zeeman energy difference between two channels, $E_F/\eta$, which can be analyzed perturbatively over the previous corrections.  

Fermionic and bosonic excitation modes are studied.  Similarly as the mean-field result, the basic structure follows that of the single-channel model, and the correction due to the inter-channel Pauli exclusion can be treated perturbatively with expansion parameter in the order of $E_F/\eta$.  In the bosonic excitation, a new out-of-sync phase mode emerges for the two-component order parameters.   It is nevertheless gapped at the the pair-breaking energy.  
\end{abstract}

%% Create a dedication in italics with no heading, centered vertically
%% on the page.
\begin{dedication}
To Jie, Ethan and Chloe
\end{dedication}

%% Create an Acknowledgements page, many departments require you to
%% include funding support in this.
\chapter*{Acknowledgments}
First I would like to deeply thank my adviser, Professor  Anthony J. Leggett.  This thesis would not have been possible without his guidance and patience.  I would like to thank Professor Monique Combescot from Institut des NanoSciences de Paris for her help, kindness and  invaluable advices.  My thanks also go to  Dr. Shizhong Zhang, Dr. Wei-Cheng Lee, Dr. Parag Ghosh and   Douglas
Packard  for their many discussions and suggestions.  Finally I wish to  thank my wife and my family for their constant support and affection.  

Part of this research presented in  this thesis  is supported  by NSF under grant No. DMR 09-06921. 

%% The thesis format requires the Table of Contents to come
%% before any other major sections, all of these sections after
%% the Table of Contents must be listed therein (i.e., use \chapter,
%% not \chapter*).  Common sections to have between the Table of
%% Contents and the main text are:
%%
%% List of Tables
%% List of Figures
%% List Symbols and/or Abbreviations
%% etc.

\tableofcontents
%\listoftables
\listoffigures

%% Create a List of Abbreviations. The left column
%% is 1 inch wide and left-justified
%\chapter{List of Abbreviations}
%
%\begin{symbollist*}
%\item[w.f.] wave function.
%\end{symbollist*}

%% Create a List of Symbols. The left column
%% is 0.7 inch wide and centered
\chapter{List of Symbols}

\begin{symbollist}[0.7in]
%\item[$\psi$] Wave function.
\item[$\alpha$]  Ratio of the closed-channel correlation $h_{2\vk}$ to the normalized two-body closed-channel bound state wave function, $\phi_{\vk}$, at high momentum.  $h_{2\vk} \xrightarrow{\text{high momentum}}\alpha\phi_{\vk}$. (Page \pageref{eq:pathInt2:hphi})
\item[$\gamma_{i\vk}$] Correction of the fermionic excitation spectrum over $\pm{}E_{\vk}$ and $\epsilon_{\vk}+\eta$. (Page \pageref{eq:pathInt2:xiExpand})
\item[$\delta_{c}$] Energy scale of the  detuning from resonance when the closed-channel takes substantial weight. (Page \pageref{eq:intro:deltaC})
\item[$\Delta_{1,2}$] Order parameters in the open-channel and closed-channel. (Page \pageref{eq:pathInt2:identity})
\item[$\zeta$] Small characteristic dimensionless  parameter for the inter-channel Pauli exclusion, $\zeta=\frac{\Delta_{2}^{2}}{\Delta_{1}\eta}$. (Page \pageref{eq:pathInt2:zetaDef})
\item[$\eta$] Absolute detuning between two channels. Note that $\eta=0$ is not where $a_{s}$ diverges. (Page \pageref{eq:intro:ham})
\item[$\kappa$] Momentum scale of the resonant bound state in the isolated close-channel, namely  $\kappa^{2}/2m=E_{b}$.
\item[$\lambda_{1},\,\lambda_{2}$] Two new parameters in the renormalized gap equation for two channels. They describe the inter-channel Pauli exclusion effects between the two channels. (Eqs. \ref{eq:pathInt2:lambda1} and \ref{eq:pathInt2:lambda2} in Page \pageref{eq:pathInt2:lambda1} and \pageref{eq:pathInt2:lambda2})
\item[$\mu$] Chemical potential.
\item[$\phi_{i}$] Bound-state wave functions of the isolated close-channel; especially, $\phi_{0}$ is the one at resonance. (Page \pageref{eq:pathInt2:phi})
\item[$a_{bg}$] The background s-wave scattering length in the open-channel when it does not coupled to the closed-channel. (Page \pageref{eq:intro:abg})

\item[$a_{c}$] Characteristic size of the bound state at resonance ($\phi_{0}$) if the close-channel is isolated. It is proportional  to the inverse of $\kappa$.
\item[$a_{s}$, $a_{s}^{(o)}$] The effective open-channel s-wave scattering length. Subscript ${}^{(o)}$ will be dropped when there is no ambiguity. Alternatively, it is  the single parameter in  the Bethe-Pierels boundary conditions. (Page \pageref{sec:intro:as}, \pageref{eq:intro:asK})
\item[$a_{0}$] Average inter-particle distance, $a_{0}k_{F}\sim1$.
\item[$E_{b}$, $E_{b}^{(i)}$] Binding energy of the $i^{th}$ two-body (bound) eigenstate in the isolated close-channel.  Superscript ${}^{(i)}$ is dropped when referring to the one in resonance. (Page \pageref{eq:intro:sch2})
\item[$E_{\vk}$] Defined as $E_{\vk}=\sqrt{\epsilon_{\vk}^{2}+\Delta^{2}}$, where $\epsilon_{\vk}=\hbar^{2}k^{2}/2m$ is the kinetic energy. $E_{\vk}$ corresponds the elementary fermionic excitation energy for $\vk$ in the single channel.   In the two-channel case, $E_{\vk}$ is defined in the similar way as, $E_{\vk}=\sqrt{\epsilon_{\vk}^{2}+\Delta_{1}^{2}}$, where $\Delta_1$ is the order parameters of the open-channel. However, it is then only the zeroth order   energy for the two elementary fermionic excitation modes. (Page \pageref{eq:pathInt:G0})

\item[$h_{1\vk}$, $ h_{2\vk}$] (Open and closed)-channel anomalous two-body correlations of the many-body system. (Page \pageref{eq:pathInt2:h2})
\item[$k_{F}$] Fermi momentum. $k_{F}=\hbar(3\pi^{2}n)^{1/3}$ in 3D.

\item[$n_{o(pen)}$] Density of the open-channel atoms.
\item[ $n_{c(lose)}$] Density of the  closed-channel atoms.
\item[ $n_{tot(al)}$] Density of all atoms.
\item[$r_{c}$] Potential extension.  All  potentials are taken as zero outside $r_{c}$. 
\item[$\mathcal{V}_{0}$]  Total volume of the system. 

\end{symbollist}

\mainmatter
% !TeX root =thesis.tex

\chapter{Introduction}
%Superfluidity in many-body fermionic system is one of the most dramatic and fascinating topic in physics. It calls the attention and effort from generation of physicists.  The study is dominated with one particular fermions, electrons.  However, this system suffers from one difficulty as the interaction  in a particular system is usually fixed and cannot be tuned experimentally.  
From  a methodological view, a physical system  would be very desirable for developing and  verifying a theory  if  it can be described with  as few parameters as possible,  and each parameter is as tunable as possible. One such system is the ultracold dilute fermionic alkali gas with the Feshbach resonance.  Dilute fermionic alkali gas was cooled into degenerate region in 1999 \cite{DeMarco1999}. Not long after that,  superfluidity was observed for such systems in 2003 \cite{Regal2003}.  In dilute ultracold fermionic alkali gas, it is sufficient in many phenomena to describe an atom-atom interaction with one single parameter, namely the s-wave scattering length, $a_{s}$, because the gas is very dilute and experiments are performed at very low temperatures.       

The other desirable property is the ability to  tune the effective interaction strength, i.e., in the present case, the s-wave scattering length, $a_{s}$ through  Feshbach resonances.  One atomic energy level  splits into several hyperfine levels under a magnetic field,  due to the hyperfine interaction between nuclear spins and electronic spins. Hyperfine spin indices provide a good set of quantum numbers for a single atom.  In the theory of  atom-atom interactions, a channel refers to a specific configuration of hyperfine spin indices of one atom pair. Different channels usually  have different magnetic moments and therefore have different Zeeman energies in the presence of a magnetic field.  The difference of the Zeeman energies can then be tuned by changing the magnetic field.  In addition, a channel is no longer an exact eigenstate when the atom-atom interaction is taken into consideration because the interaction is mostly due to the overlap between electronic parts of  the two-atoms' wave-functions. As a result, channels are hybridized.  The effective potential of each channel is also different.  The potential of one  channel  may be deep enough to sustain a bound state.  This channel is then called \emph{``closed-channel''}.  For a certain magnetic field,  the energy level of this bound state might be close to the zero-energy threshold of the other channel, usually called \emph{open-channel}, and two channels  thus strongly hybridized.  The low-energy scattering properties in the open-channel is then dramatically modified.   In such a situation, two atoms approaching each other in the open-channel may ``spend a certain amount of time in the closed-channel'' and then reemerge in the open-channel.  Atoms in the open-channel seem to feel an enhanced effective interaction.  This phenomenon is known as Feshbach resonances.    We will present a  quantitative analysis about alkali gas in Chpater \ref{sec:intro:one} and  Feshbach resonances in two-body context in Chapter \ref{sec:intro:twobody}.

A very desirable property of the Feshbach resonance is that the effective interaction is tunable experimentally through the Zeeman energy difference between channels which is in turn  tunable through  instruments such as a magnetic field.  
This unique ability gives physicists a rare opportunity to study  a many-body system under various interaction strengths,  and thus connect different physics originally developed separately.  Particularly for the fermionic gas, there are a series of  theoretical works about uniform treatment of  BEC and BCS since the 1960s \cite{Eagle,LeggettCrossover,Nozieres,RanderiaBEC}, for which dilute ultracold fermionic alkali gas with  Feshbach resonances provides the perfect testing grounds.  Indeed,  these theories works quite well  qualitatively.  

%One important characteristic quantity of Feshbach resonance is $\delta_{C}$N (see detail in Chapter \ref{sec:intro:twobody} for details): when detuning from resonance is smaller than it, open-channel atoms dominate and closed-channel can be neglected.  The effective interaction can still be characterized by $a_{s}$.  One seems to acquire a ``magic knob'' that can tune the interaction between atoms.  On the other hand, when negative tuning is much larger than $\delta_{C}$, atoms in closed-channel have comparable weight to that of open-channel or even dominate them.  Two channels need to be considered at the same time.  

%This thesis tries to look into the idiosyncrasy of the Feshbach resonance in contrast with a true ``Simple'' knob of the interaction strength.
\begin{figure}[htbp]
\begin{center}
\includegraphics[width=0.8\textwidth]{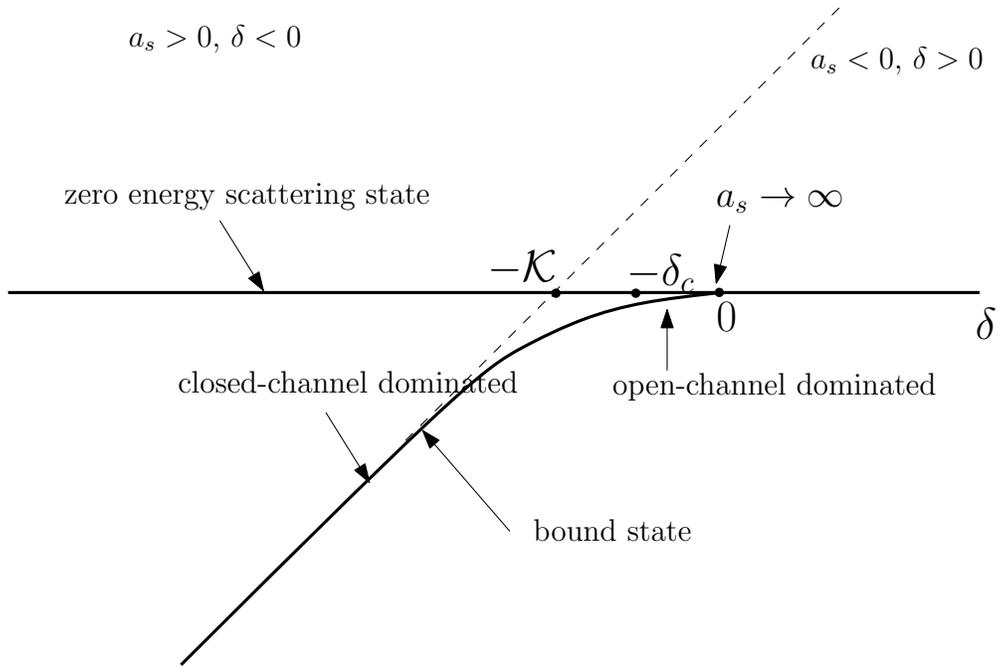}
\caption{Energy levels in a Feshbach resonance\label{fig:intro:levels}} 
\parbox{0.7\textwidth}{\small $\delta$ is the energy detuning from the resonance point, where the resonant point is defined as the point where the open-channel effective s-wave scattering length diverges, $a_s\to\pm\infty$.  The horizontal line stands for the zero energy s-wave scattering state, $\psi\sim\nth{r}-\nth{a_s}$, which exists for any detuning.  The lower line stands for the real bound state, which only exists for negative detuning ($\delta<0$, $a_s>0$). The dash line stands for the (uncoupled) closed-channel bound state.  An interesting point to notice is that the real bound state appears earlier than the cross point of the (uncoupled) closed-channel bound state level and zero energy. Another important point to notice is the negative detuning $-\delta_c$.  When the negative detuning is smaller than $\delta_c$, this real bound state is composed mostly with atoms in the open-channel and vice verse.  See Chapter \ref{sec:intro:twobody} for details about $\mathcal{K}$ and $\delta_c$.   }

\end{center}
\end{figure}

  The two-body theory of the Feshbach resonance has a characteristic  parameter, $\delta_c$, defined as  the detuning energy at which the weight of the bound state shifts from predominated in the open-channel to predominated in the closed-channel (see Fig. \ref{fig:intro:levels}).  Na\"{i}vely speaking, in the negative detuning side of any resonance (i.e. $\delta<0$), particles should mostly stay  in a (virtual) bound-state of the closed-channel (or ``virtual state'' in some other resonances).  However, at the resonance point  of the Feshbach resonance ($a_s\to\pm\infty$), atoms are mostly still in the open-channel, and they do so down to a negative detuning $\delta\sim-\delta_c$. Only when the detuning from resonance is much far away than $\delta_c$, do atoms have the majority weight in the closed-channel.    
  
  When considering a many-body problem, an important question is how this energy scale, $\delta_c$, compares to a typical many-body energy scale, namely, the Fermi energy of the free fermionic atoms. In the region not too far away from resonance ($\abs{\delta}\ll\delta_{c}$), the closed-channel weight is negligible if the Fermi energy is much smaller than $\delta_c$, (i.e., \emph{broad resonance}).  Crossover experiments are usually performed at detuning not too far from the resonance, and hence the closed-channel can be safely ignored at the many-body level. Eventually, when the detuning is too far away, $\abs{\delta}\gg\delta_{c}$, the bound state is almost like the    uncoupled closed-channel bound state with a little dressing from the open-channel.  We nevertheless do not concern such cases for the broad resonance because crossover phenomena have already be well covered in both BEC and BCS ends with $\abs{\delta}\ll\delta_{c}$. The problem can be well-described as a two-species fermion system with a tunable interaction.  The Feshbach resonance indeed serves as a simple ``magic'' knob to change the interaction strength.  The original  theories developed on  single-channel models  apply to this case directly.  This is also the situation for two  popular experiment cases (${}^{6}\text{Li}$ atoms at 834G, $^{40}\text{K}$ atoms at 224G).   Many theoretical works have been developed using the single-channel model as these original works or using the tow-channel model within the broad resonance assumption (e.g. \cite{Holland01,HoUniversal,Fuchs04}). On the contrary, when the Fermi energy of the free fermionic atoms is  comparable to or even larger than $\delta_c$, the closed-channel has to be included at the many-body level even for small detuning. Such a situation, previously considered in some works \cite{GurarieNarrow}, is the focus of the current thesis. 
  
  Nevertheless, one crucial simplification comes from  the fact  that relevant uncoupled closed-channel bound state is  tightly bound, with spatial extension much smaller than  many-body scales, e.g. the interparticle distance, (but often larger than the potential range).  This fact enable us to treat the Pauli exclusion between two channels perturbatively. It is not necessary to handle all the  fermion species simultaneously, which probably requires quite different techniques  other than those discussed in this thesis. 

To complicate the problem even further,  real experimental configurations often have one common hyperfine species between the two channels. There are three hyperfine species in the  two channels instead of four species (two for each channel).  Two most common setups (${}^{6}\text{Li}$ at 834G, $^{40}\text{K}$ at 224G) both contain three species of fermions although they are the broad resonance.  The Pauli exclusion principle prevents  atoms of both channels from occupying the same level simultaneously because of this common species.  The inter-channel Pauli exclusion has no counterpart in two-body physics. This peculiar effect  in many-body crossover problems has  received little theoretical attention up to now.    Nevertheless,   narrow resonances do exist \cite{ChinRMP} and it is not  inconceivable to perform many-body experiments using such resonances.  The central concern of this thesis is about these situations. 

Roughly speaking, turning from two-body systems to many-body systems brings three effects into the original two-body problem.  The first effect is closely associated with the Fermi energy:  For a many-body fermionic system at low temperature, most fermions are inactive; only the fermions close to the Fermi surface participate in the interaction processes. Therefore, energy often needs to be measured from the Fermi surface instead of from zero as in a two-body situation. This aspect has been extensively studied previously \cite{GurarieNarrow}.

The second effect is about counting. Unlike in the single-channel problem, there are two relevant densities in the two-channel problem: the density of atoms in the open-channel, $n_{o}$, and the density of atoms in the closed-channel, $n_{c}$. When the closed-channel weight is small (broad resonance), it is all right to treat the total density as the same as the open-channel density.  However, in the narrow resonance, where the closed-channel weight is not negligible, counting becomes complicated.  Extra care is required to specify which channel the  quantities, such as ``density'', belong to.  This aspect has been  also extensively studied previously \cite{GurarieNarrow}.

The last effect is unique for the three-species problem, where one common species is shared by both channels.  The phase spaces of two channels are overlapped because of the Pauli exclusion caused by the common species. This effect is controlled by the wave-function overlapping of states in the two channels. A rough estimate of this overlapping can be made: The uncoupled closed-channel bound-state which is in resonance with the open-channel zero energy threshold has  relatively small  spatial extension, $a_c$.  Its binding energy $E_b$ is close to the Zeeman energy difference between two channels, $\eta$.  On the other hand, fermions in the open-channel fill the lowest  momentum states up to typically the Fermi energy, $E_F$.  By a simple dimensional argument, the ratio $E_F/\eta$ must control the overlapping effect. This effect has  not been addressed in any theoretical work to this author's knowledge.  How it modifies the many-body picture is the central topic of this thesis. 

We can see this from a slightly alternative aspect using two-fermion molecule gas for the uncoupled closed-channel bound state. We assume that the molecule size is $a_{c}$ and the total number of molecules is $N$.  Assuming further that the bound-state is close to threshold,   the bound-state wave function can then be written as $A/(k^{2}+\kappa^{2})$, where $\hbar^{2}\kappa^{2}/2m=E_{b}$, (see Appendix \ref{sec:pathInt2:short-range}). The prefactor ``$A$'' can be determined  by normalization, $\sum_{k=0}^{1/a_{c}}\abs{\psi}^{2}\sim{}N$. Now  we consider all atoms in a typical many-body scale, e.g. the Fermi energy, $E_{F}$, which is going to overlap with levels occupied in the open-channel. Usually, the Fermi energy is much smaller than the energy scale of the closed-channel bound state, $E_{F}\ll{}E_b$.  The total number of atoms in $[0,E_F]$ is roughly $N\cdot(k_{F}a_{c})^{3}$, which is much smaller than $N$. This means that in the two-channel problem, the low momenta,  ($k\lesssim{}k_F$), are still dominated by the open-channel component even when the total number of atoms in the closed-channel is comparable or higher than the total number  of atoms in the open-channel because atoms in the closed-channel are mostly in  high-momentum states.     

The present thesis is divided as follows:
Chapter \ref{sec:intro:one} to  \ref{sec:intro:1channel} review several important concepts used in the thesis.  Chapters \ref{ch:path2} and \ref{ch:excitation} then present my main work  and Appendix \ref{ch:mean}  lists an earlier attempt using a roughly equivalent but less-flexible approach.  

More specifically, Chapter \ref{sec:intro:one} briefly reviews  dilute ultracold alkali gas.   Section \ref{sec:intro:as} in particular examines the idea of ``universality'', which is one of the central ideas in our treatment of the two-channel model.  Chapter \ref{sec:intro:twobody} goes over the Feshbach resonance in two-body physics and  the concept of  the narrow (broad) resonance is introduced. Chapter \ref{sec:intro:1channel} reviews the single-channel BEC-BCS crossover problem as well as the path-integral approach solving it. This chapter serves as the starting point for the solution of the two-channel model. After these reviews,   Chapter \ref{ch:path2} and Chapter \ref{ch:excitation} present my work on the three-species narrow Feshbach resonance within a many-body path-integral framework, in detail.   Chapter \ref{ch:path2} discusses the mean field result while Chapter \ref{ch:excitation} discusses fermionic and bosonic excitations. An earlier attempt   based on the BCS ansatz  approach in mean-field level is given in Appendix \ref{ch:mean}.  Chapter \ref{ch:conclusion} discusses our procedures and their conclusions.  

\chapter{Dilute ultracold alkali gas}\label{sec:intro:one}
  Since the 1990s, dilute  alkali gas has been cooled into quantum degenerate region where the thermal de Broglie wavelength ($\frac{h}{\sqrt{2\pi{m}k_{B}{T}}}$) is comparable or larger than the interparticle distance.  Not long after successfully cooling the bosonic atoms, fermionic alkali gas was also available in the degenerate region.  Because of the ultra-low temperature (in the order of nK), and the extreme diluteness ($10^{12}\sim10^{15}\text{cm}^{-3}$), the atoms are mostly \emph{free} except when they are  close.   This particular property simplifies theoretical analyses tremendously (see Sec. \ref{sec:intro:as} for details).  In this chapter, we review a few aspects of the dilute ultracold alkali gas that are closely related to the current thesis.     

\section{A single atom and its hyperfine levels}
\begin{figure}[htbp]
\begin{center}
\includegraphics[width=0.8\textwidth]{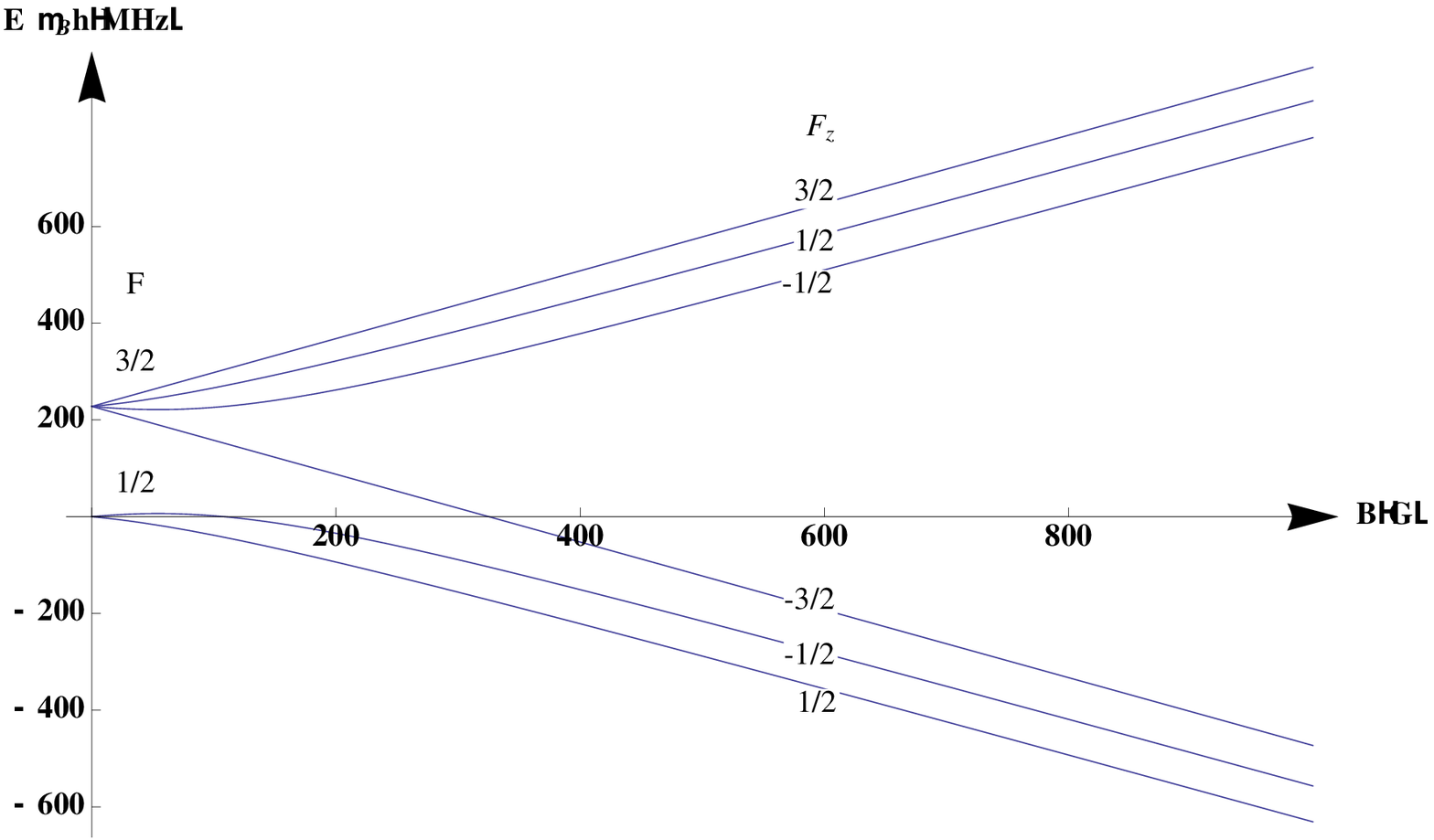}
\caption{Hyperfine structure of a single \textsuperscript{6}Li atom } 
Levels are marked with $F$ and $F_{z}$ {(see Footnote \ref{foot:intro:f} in page \pageref{foot:intro:f})} %Note that the energy of the $\ket{F=\nth{2},F_z=-\nth{2}}$ state first increases with the magnetic field, $B$, at low field then decrease at high field. In the same way, the energy of the $\ket{F=\frac{3}{2},F_z=-\nth{2}}$ state first decreases and then increases with the magnetic field.  
\label{fig:intro:li6}
\end{center}
\end{figure}

In experiments on ultracold alkali gas, a magnetic field ($\mathbf{B}$) is the most common physical quantity to manipulate.   Let us first study an   isolated atom in the presence of a magnetic field.  An alkali atom has only one electron in its outer shell.  All the rest electrons are in the filled inner shells which  has no total magnetic moment.  So we only need to consider the spin of the outermost electron,  $\mathbf{S}$, for interaction with the magnetic field. In addition, a magnetic field also interacts with the atom's nuclear spin, $\mathbf{I}$.  The full spin-part Hamiltonian is
\begin{equation}
\begin{split}\label{eq:intro:1atom}
H_{spin}&=A \mathbf{I}\cdot\mathbf{S}-\mu_{e}\mathbf{B}\cdot\mathbf{S}-{\mu}_{n}\mathbf{B}\cdot\mathbf{I}\\
&=A \mathbf{I}\cdot\mathbf{S}-\mu_{e}{B}{S_{z}}-{\mu}_{n}{B}{I_{z}}
\end{split}
\end{equation}
The first term with a characteristic energy $A$ describes the hyperfine interaction, while the next two terms describe Zeeman energies of the outer electron and nuclei respectively. $\mu_{e}$ is the electronic magnetic moment, while $\mu_n$ is the nuclear magnetic moment.  In the second line, we take the direction of the magnetic field as the z-direction. This Hamiltonian can be diagonalized by introducing the total spin 
\begin{equation}
\mathbf{F}=\mathbf{S}+\mathbf{I}
\end{equation}
When the magnetic field is zero, the two Zeeman energy terms in the above Hamiltonian vanish.  Thus, $(F,F_{z})$ are good quantum numbers and all states with the same $F$ are degenerate.   When the magnetic field is finite, $(F,F_{z})$ cannot diagonalize the Zeeman energy terms, and therefore are no longer good quantum numbers. Nevertheless, we can still label the atomic  states with these two numbers via the adiabatic connection to the levels in the zero magnetic field.  For a finite magnetic field, except for states with the highest and lowest $F_{z}$, namely, $\pm{}F$, each other state is a mixture of different $(S_{z}, I_{z})$ or $(F,F_z)$.  Fortunately, $S$ is just equal to $1/2$ for an alkali atom. So each state is mixed with at most  two sets of quantum numbers $(S_{z}, I_{z})$. At high magnetic fields, the first hyperfine coupling term in Eq. \eqref{eq:intro:1atom} is dominated by the last two terms of Zeeman energies and the eigenstates are approximately described by the quantum numbers $(S_{z},I_{z})$.  (See Fig. \ref{fig:intro:li6}.)

\section{Two-body interactions}
Things would be very boring if there was only the single-atom Hamiltonian.  Before discussing the interaction itself, let us introduce one important concept related to it.    The term ``channel'' is used to refer one  configuration of hyperfine spins for one atom pair in interaction, $\ket{F^{(1)},F_{z}^{(1)}}\otimes\ket{F^{(2)},F_{z}^{(2)}}$ . \footnote{Remember that $(F,F_{z})$ are only labels; they do not stand for the total angular momentum unless there is no magnetic field.\label{foot:intro:f}}    Channels are good basis for non-interaction pairs.  A pair of atoms in one channel would stay in this channel  forever in the absence of  interactions between atoms.  Considering the present case of alkali atoms, two alkali atoms mostly interact  through the overlapping of their electron wave-functions in the dilute limit.  Thus, besides the atom-atom distance, the interaction is mostly a function of electronic spins, with negligible dependence on nuclear spins.  Schematically the interaction can be written as 
\begin{equation}\label{eq:intro:two}
V=f(r)+g(r)\mathbf{S_{1}}\cdot\mathbf{S_{2}}
\end{equation}
Here the spatial part of the interaction is coupled with electronic spins. The hyperfine levels that diagonalize the single atom Hamiltonian are no longer eigenstates for this interaction.  In another word, the interaction has non-diagonal terms between channels and therefore hybridizes them. Instead, states with definite electronic spins form a good approximate basis for the atom-atom interaction.  Nonetheless, most experiments are performed in the so-called high-field region,  where the Zeeman energy terms dominate the hyperfine interaction term in the single-atom Hamiltonian (Eq. \ref{eq:intro:1atom}).   Recalling that one hyperfine state is at most mixture of the two $(S_{z}, I_{z})$ states.  In the high-field, one of them always dominate the other; therefore               the electronic spin $S_z$ is approximately a good quantum number for a hyperfine level, and the original hyperfine levels (channels) serve as a good starting point as the zeroth order approximation.  When the hybridization is taken into account, multichannel scatterings are possible.  The most interesting thing in these multi-channel scatterings is the possibility of a resonance.  One of them is the so-called Feshbach resonance:  The potential in one channel may be deep enough to sustain a bound-state. This channel is named ``closed-channel''. When this bound-state energy is close to the zero-energy threshold of another channel, named ``open-channel'', the low-energy scattering properties in the open-channel are dramatically modified.  This resonance turns out to be extremely useful in   cold atom experiments.  Chapter \ref{sec:intro:twobody} reviews its theory for  a two-body system. The more involving many-body problem is then the central theme of this thesis and is discussed in Chapter \ref{ch:path2} and \ref{ch:excitation} in detail. 

Let us illustrate above discussion in  one example.  In a common experimental setup for $^{6}$Li (Fig. \ref{fig:intro:li6}), experiments are usually  prepared with atoms in the two lowest hyperfine levels: described by the  direct product of two states, $\ket{F=\nth{2},F_{z}=-\nth{2}}\otimes\ket{F=\nth{2},F_{z}=+\nth{2}}$.  This is a good approximation until the two atoms are very close.  Recalling that the atom-atom interaction (Eq. \ref{eq:intro:two}) conserves the z-component of the total angular momentum, $F_{z}^{(1)}+F_{z}^{(2)}$, this channel mixes with four other possible channels of the same total z-direction angular momentum, i.e. $F_{z}^{(1)}+F_{z}^{(2)}=0$: $\ket{\nth{2},-\nth{2}}\otimes\ket{\frac{3}{2},+\nth{2}}$, $\ket{\frac{3}{2},-\nth{2}}\otimes\ket{\nth{2},+\nth{2}}$, $\ket{\frac{3}{2},+\frac{3}{2}}\otimes\ket{\frac{3}{2},-\frac{3}{2}}$, $\ket{\frac{3}{2},+\frac{1}{2}}\otimes\ket{\frac{3}{2},-\frac{1}{2}}$ (All states are labeled as $\ket{F,F_{z}}$).  Various resonances can take place.  Note that close to the resonance, it is normally sufficient to consider only the  channel that is in resonance, while neglect all other channels.  Another important aspect is whether the two channels share one single hyperfine species or not.  In the former case,  totally three hyperfine species are involved while in the later case, four hyperfine species are involved.   The closed channel in the most studied resonance with a magnetic filed close 834G, is approximately $\ket{\frac{3}{2},-\nth{2}}\otimes\ket{\nth{2},+\nth{2}}$ and the resonance is a three-species resonance\cite{ZhangThesis,ChinRMP}. 

\section{Universality,  Bethe-Peierls boundary conditions, the s-wave scattering length, and  two-body density matrices \label{sec:intro:as}}
One important aspect of the interaction in dilute ultracold alkali gas is that for many purposes, it is sufficient to characterize the interaction  by a single two-body parameter, namely, the s-wave scattering length, $a_s$,  because  both the density  and the temperature, are very  low. This is often interpreted as we can replace the real potential with a pseudo potential, $U(\vr)=\frac{4\pi{}a_{s}\hbar^{2}}{m}\delta(\vr)$\cite{pethick, LeggettBEC}.  Nevertheless, an alternative interpretation of $a_s$ \cite{LeggettBEC, Tan2008-1,Tan2008-2,CombescotTan} is more useful in this work.  For a short-range potential, where the potential range, $r_c$, is much smaller than the average interparticle distance, $a_0$,  particles are free-like  in the majority of the time.  They only interact  when two particles are close to each other.  We can thus schematically divide the space into two domains: $\mathcal{D}$, where any two particles are more than $r_c$ away from each other; and otherwise, $\mathcal{I}$. Most physical quantities would be very easy to calculate if only considering the free part, $\mathcal{D}$.  In a low-energy (ultracold) dilute system, we only need to consider  pair-wise interaction while neglect all the three-body or more-body interactions.  In addition,  $\mathcal{D}$ takes the majority of the space due to the same reason.     The effect of the potential on wave-function in the short-range region, $\mathcal{I}$, can be taken simply as   a boundary condition on the wave-function in the free part $\mathcal{D}$, $\psi(r)\xrightarrow{r\to0}\psi_{0}(r)$.  For an isometric $\psi_{0}(r)$, the lowest order in the radial coordinate $r$ is $\nth{r}$. \footnote{The extra $\nth{r}$ factor is there for  radial wave function in 3D.}  Including the next order, a constant,  we have $\psi_{0}(r)\propto\nth{r}-\nth{a_s}$ barring the normalization.  All these consideration gives us the simplest non-trivial boundary condition on  a wave function
\begin{equation}\label{eq:intro:Bethe}
\psi(r)\xrightarrow{r\to0}A\br{\nth{r}-\nth{a_s}}
\end{equation}
which is also known as the Bethe-Peierls boundary condition\cite{BethePeierls}.    $a_{s}$ is fully determined by two-body physics.  This simple boundary condition applies to two-body, few-body, as well as many-body systems, and has been proved to be a very powerful tool in solving various problems.  

Eq. \ref{eq:intro:Bethe} coincides the zero-energy s-wave scattering wave function, which explains the name of parameter ``$a_s$'', the s-wave scattering length. Nevertheless,    we did not mention anything about zero energy so far, although $a_{s}$ is defined for zero energy in the scattering theory context.  In fact,  this boundary condition (Eq. \ref{eq:intro:Bethe}) applies generally to  any low (positive or negative) energy solutions as long as the energy involved is much lower than the energy scale in the interaction domain $\mathcal{I}$.  Hence, this boundary condition applys  to close-to-threshold bound states as well.  The s-wave wave function of a weak bound-state reads $\psi(r)=\nth{r}e^{-r/a_s}$ in $\mathcal{D}$, which matches the Bethe-Peierls boundary condition with a positive $a_{s}$ for  $r\ll{}a_{s}$. The exponential decay factor of the wave function, $a_{s}$, is directly related to the binding energy, $E_{b}$, with the often cited relation.
\begin{equation}
 E_{b}=\frac{\hbar^{2}}{2m_{r}a_{s}^{2}}
\end{equation}
Here $m_{r}$ is the reduced mass for center of mass, which is equal to half of the atom mass for a pair of the same atoms.  This immediately clears one often confusing and counter-intuitive fact, that a positive  $a_s$ is associated to a bound state.  In the standard scattering theory, a positive $a_s$  is usually associated with a repulsive interaction, which obviously does not support a bound state.\footnote{This seeming paradox can be resolved carefully within scattering theory as follows. In the scattering theory, only when the interaction is weak, and phase shift as well as $a_s$ are small, we have the fact that a repulsive interaction leads to a positive phase shift and a positive $a_s$; while an attractive interaction leads to a negative phase shift and a negative $a_s$.  No simple relationship of signs holds for a strong interaction, where a bound state might form.  In fact, the phase shift changes as much as $2\pi$ when a bound state starts to form; therefore, $a_s$ is large and  changes sign over the threshold. }

  Eq. \ref{eq:intro:Bethe}, does not fix the normalization on the wave function. This normalization factor, encapsulating effects from particles outside the immediate interacting pair, appears in many physical quantities.  In a dilute and low-energy system,  its square is proportional to the so-called \emph{integrated contact intensity}, $C$, introduced by Tan \cite{ Tan2008-1,Tan2008-2,CombescotTan}.  For the limit $r_c\to0$, the integrated contact intensity, $C$, and the s-wave scattering length, $a_s$, are sufficient to describe several important physical quantities, such as  internal energy.  A particular useful one for this thesis is the limit at high-momentum of the particle number distribution of particles, 
 \begin{equation}
 \lim_{k\to\infty}n_k=\frac{C}{k^4}
 \end{equation}
 Note that here \emph{high-momentum} does not mean the absolutely high-momentum, it means lower than the characteristic momentum of potential $1/r_c$, but higher than any other scale, $1/a_0$,...  
 Indeed, when the short-range approximation and low-energy approximation apply, we expect that the two-body correlation at  high-momentum ($\gg{k_{F}}$) does not change much from a two-body system to a many-body system.  In this high-momentum region, we can always use   the two-body wave function as good approximation. 
 
 In many-body physics, various physical observable quantities are related to one set of  quantities, namely, density matrices,  $\av{\Psi^\dg(x_{1})\cdots\Psi^\dg(x_{N})\Psi(x'_{N})\cdots\Psi(x'_{1})}$. In fermionic system,  the one-body density matrix ($N=1$) for an interacting system is often very close to the one for the free particles, although the difference can be important for some theories, such as Landau Fermi liquid theory.  A two-body density matrix ($N=2$) is often  used because it is often different from the two-body density matrices of the free fermions qualitatively, such as in the case of BCS pairing. Formally, we can decompose a two-body density matrix into an orthogonal basis
 \begin{equation}\label{eq:intro:2bodyDM}
 \av{\Psi^\dg(x_1)\Psi^\dg(x_2)\Psi(y_2)\Psi(y_1)}=\sum_nC_n\phi_n^\dg(x_1,x_2)\phi_n(y_1,y_2)
 \end{equation}     
 When one or a few $C_n$ are macroscopic,  some special quantum phenomena often emerges.  Especially when only one parameter, $C_{0}$, is macroscopic, the system can often be interpreted as one macroscopic wave function, $\phi_{0}(x_{1},x_{2})$, (which is often called  order parameters).\cite{Leggett}  This can serve as a starting point for several phenomena, such as  BCS superconducting,...
 
Zhang and Leggett developed independently another universality theory based on two-body density matrices \linebreak[2] \cite{shizhongUniv}, which actually take a more general form of boundary conditions than the Bethe-Peierls boundary condition with $a_{s}$.   They asserted that for a short-range potential and a low temperature, as in the case of  dilute ultracold alkali gas, the basis wave functions $\phi_n$  in Eq. \eqref{eq:intro:2bodyDM} follows the two-body wave function at short-range. This is actually similar to Eq. \eqref{eq:intro:Bethe}.  Instead of requiring the simplest form of $\psi_0$ given in Eq. \eqref{eq:intro:Bethe}, they require a more general wave-function that solves the  Hamiltonian at two-body level.  Not surprisingly, many physical properties are determined by the normalization factors  as using the Bethe-Peierls boundary condition.  

In this thesis, similar idea is used.  We assume that the closed-channel correlations follow the its two-body bound-state wave-function at high energy (i.e. short-range).   An open-channel correlation follows its two-body  wave-function at short-range, but its intermediate range does not do so because of the sensitive nature of resonance.   The current thesis focus on how this wave-function are modified.

 \chapter{The  Feshbach resonance in two-body physics\label{sec:intro:twobody}}
 As discussed in Sec. \ref{sec:intro:as}, a two-particle interaction in a dilute system is often approximated by a pseudo-potential characterized with a s-wave scattering length $a_{s}$.   The drastic  change  of $a_{s}$  obtained by tuning the energy difference  between two channels  through a magnetic field in the Feshbach resonance gives  experimentalists a rare ability to tune the interaction strength between two atoms.  And this possibility is extremely useful to study BEC-BCS crossover where the interaction varies from weak to strong.  
 
Here we briefly review the Feshbach resonance in a two-body system.   As discussed in Chapter \ref{sec:intro:one}, a hyperfine level is an eigenstate for an isolated single atom.  However, when two atoms interact, most of the interaction comes from  electrons with only negligible effects from nucleons .  As a result, hyperfine levels are no longer true eigenstates of the two-body system.  Nevertheless, hyperfine levels can serve as  good approximated quantum numbers and we are going to call a pair of hyperfine indices a ``channel''.  Different channels in general have different  interactions.  They are decoupled at the lowest order.  In a magnetic field, different channels differ in energy at threshold, where two atoms are infinitely away from each other, due to the Zeeman energy which is mostly determined by the electronic magnetic moment because  the electronic magnetic moment is much larger than the nuclear magnetic moment.  This energy difference is easy to tune through a magnetic field.  

When the mixing between channels are taken into consideration, the simple single-channel scattering becomes the multi-channel scattering.  Especially, when the one channel's threshold is close to a bound-state in the other channel, the low-energy scattering property of that channel is dramatically altered.  Its phase shift  changes $2\pi$;  its s-wave scattering length $a_{s}$ blows to infinity and then jumps to the infinity of the opposite sign.  This is essentially what happens in a Feshbach resonance, studied by Fano\cite{Fano} and Feshbach \cite{nuclear} for nuclear and atomic physics in 1960s.  Here we  mostly follow the treatment by Leggett \cite{Leggett} (with some different symbols to comply with the rest of this thesis). 
\footnote{In order to be consistent  with  other parts of the thesis, we here use some different symbols comparing to the original works by Leggett \cite{Leggett}.  Here we list them, with the symbols from \cite{Leggett} in parenthesis.  $U$ ($=-V$): open-channel interaction; $V$ ($=-V_{c}$): closed-channel interaction; $Y$ (=-$g\cdot{}f$): inter-channel coupling; $E_{b}$ ($=\epsilon_{0}$): binding energy of the closed-channel bound state; $r_{c}$ ($=r_{0}$): potential range; $\eta$ ($=\epsilon_0+\tilde\delta$): the Zeeman energy difference between two channels; $\mathcal{K}$ ($=\kappa$) see its definition in Eq.\ref{eq:intro:kappa}.}

The  Hamiltonian for  the coupled open- and  closed- channel can be written as a $2\times2$ matrix 
\begin{equation}\label{eq:intro:ham}
\hat{H}(r)=
\begin{pmatrix}
-\frac{\hbar^{2}}{2m_{r}}\nabla^{2}-U(r)&\;&-Y(r)\\
-Y(r)&\;&-\frac{\hbar^{2}}{2m_{r}}\nabla^{2}+\eta-V(r)
\end{pmatrix}
\end{equation}
where the  zero energy is taken as the energy of two atoms in the open-channel with infinite separation. $\eta$ is the  difference in the Zeeman energies of the two channels.  The first column (row) stands for the open-channel and the second column (row) stands for the closed-channel.  All the interactions are short-range.  For a s-wave solution, we have the radial part as 
\begin{equation}
\psi(r)=\nth{r}\mtrx{\chi(r)\\\chi_{c}(r)}
\end{equation}
The coupled time-independent \sch equations in the radial direction read as:
\begin{align}
-\frac{\hbar^{2}}{2m_{r}}\chi''-U\chi-Y\chi_c&=E\chi\label{eq:intro:open}\\
-\frac{\hbar^{2}}{2m_{r}}\chi_c''+\eta\chi_c-V\chi_c-Y\chi&=E\chi_c\label{eq:intro:close}
\end{align}

We now expand the closed-channel component $\chi_{c}$ over the eigenstates of the isolated closed-channel Hamiltonian, $\chi_{c}=\sum_{i}c_{i}\phi_{i}$, where $\phi_{i}$ satisfies \sch equation of the isolated closed-channel
\begin{equation}\label{eq:intro:sch2}
-\frac{\hbar^{2}}{2m_{r}}\phi_{i}''-V \phi_{i}=-E_{b}^{(i)}\phi_{i}
\end{equation}
We denote $\phi_{0}$  the wave function  in resonance and $c_{0}$ its coefficient.  Here we assume that the energy differences between eigenstates, $\phi_{i}$'s, are larger than any other energy scales in the problem.  This guarantees $c_{0}$ dominates any other $c_{i}$'s.  So, $\chi_{c}\approx{}c_{0}\phi_{0}$.  Na\"{i}vely speaking, the resonance is expected to happen at the point where the closed-channel bound state has its energy exactly at the threshold of the open-channel.  This leads us to introduce the relative detuning, $\tilde\delta=\eta-E_{b}$. By comparing Eq. \ref{eq:intro:open} and Eq. \ref{eq:intro:sch2}, it is easy to show 
\begin{equation}\label{eq:intro:closeCoeff}
\chi_{c}=\frac{\phi_{0}}{E-\tilde\delta}\int{dr'}\,\phi_{0}^{*}(r')Y(r')\chi\br{r'}
\end{equation}
provided that  $\phi_{0}$ is normalized (for radial component).  Inserting the expression of $\chi_{c}$ into the \sch equation of the open-channel component $\chi$, Eq. \ref{eq:intro:open}, we get 
\begin{equation}\label{eq:intro:chi}
-\frac{\hbar^{2}}{2m_{r}}\chi''-(U+E)\chi+\frac{1}{E-\tilde\delta}\int_{0}^{\infty}K(r\,r')\chi(r')dr'=0
\end{equation}
where the kernel $K(r\,r')$ is
\begin{equation}\label{eq:intro:Krr}
K(r\,r')\equiv\phi_{0}(r)\phi_{0}(r')Y(r')Y(r)
\end{equation}
The  \sch equation of the  decoupled open-channel is
\begin{equation}\label{eq:intro:chi0}
-\frac{\hbar^{2}}{2m_{r}}\chi''-(U+E)\chi_{0}=0
\end{equation}
Note that $\chi_{0}(r)\overset{r\to0}\rightarrow0$ and $\chi_{0}(r)\overset{r\to\infty}\rightarrow{A}(1-r/a_{bg})\label{eq:intro:abg}$ , where $a_{bg}$ is the background s-wave scattering length of the isolated open-channel.\footnote{Note that we are dealing with the internal wave function $\chi_{0}$ here, i.e. the wave-function in the region $\mathcal{I}$, instead of the external wave function as in Sec. \ref{sec:intro:as}; therefore, the boundary condition for $r\rightarrow0$ in Sec. \ref{sec:intro:as} actually corresponds the boundary condition $r\rightarrow\infty$ here.}  Multiply Eq. \ref{eq:intro:chi} by $\chi_{0}$ and Eq. \ref{eq:intro:chi0} by $\chi$, integrate from $r=0$ to a  distance $r_{0}$ much larger than the potential range, $r_{c}$, subtract them, we find,
\begin{equation*}
\int_{0}^{r_{0}}dr\mbr{\chi_{0}(r)\chi''(r)-\chi_{0}''(r)\chi(r)}+\frac{2m_{r}E}{\hbar^{2}}\int_{0}^{r_{0}}dr\chi_{0}(r)\chi(r)
=\frac{2m_{r}/\hbar^{2}}{E-\tilde\delta}\int_{0}^{r_{0}}dr\int_{0}^{r_{0}}dr'\chi_{0}(r)K(rr')\chi(r')
\end{equation*}
Using the Green's theorem on the first integral, we find 
\begin{equation}
\chi_{0}(r_{0})\chi'(r_{0})-\chi_{0}'(r_{0})\chi(r_{0})+\frac{2m_{r}E}{\hbar^{2}}\int_{0}^{r_{c}}dr\chi_{0}(r)\chi(r)
=\frac{2m_{r}/\hbar^{2}}{E-\tilde\delta}\int_{0}^{r_{0}}dr\int_{0}^{r_{0}}dr'\chi_{0}(r)K(rr')\chi(r')
\end{equation}
Here we can use the boundary condition by $\chi_{0}(r)\overset{r\to\infty}\rightarrow{A}(1-r/a_{bg})$, $\chi(r)\overset{r\to\infty}\rightarrow\tilde{A}(1-r/a_{s})$.  $Y(r)$ is a short-range interaction and thus $K(r,r')$ only picks the short-range parts of $\chi(r)$ and $\chi_{0}(r)$, which varies little for different detuning. Consequently,  the R.H.S approaches a constant.  For the scattering solution at  $E=0$, the last terml on the L.H.S. is zero, and we have 
\begin{equation}
\nth{a_{s}}-\nth{a_{bg}}=\frac{2m_{r}/\hbar^{2}}{\tilde\delta}\int_{0}^{\infty}dr\int_{0}^{\infty}dr'\chi_{0}(r)K(rr')\chi(r')
\end{equation}
Contrary to the previous guess, $a_{s}$ does not diverge at the point $\tilde\delta=0$ because of the original interaction in the open-channel, $a_{bg}$.  We can introduce a quantity $\mathcal{K}$, the detuning where $a_{s}$ diverges, through the implicit equation ($\mathcal{K}$ shows up in R.H.S as well)
\begin{equation}\label{eq:intro:kappa}
\mathcal{K}\equiv-\frac{2m_{r}a_{bg}}{\hbar^{2}}\int_{0}^{\infty}{dr}\int_{0}^{\infty}dr'\chi_{0}(r)K(rr')\chi_{\tilde\delta=\mathcal{K}}(r')
\end{equation}
And if we define the ``real detuning'' $\delta\equiv\tilde\delta-\mathcal{K}$, which is the detuning from the real resonant point, where $a_{s}\to\pm\infty$, we have 
\begin{equation}\label{eq:intro:asK}
a_{s}(\delta)=a_{bg}\br{1+\frac{\mathcal{K}}{\delta}}
\end{equation}
This is consistent with the empirical formula of the Feshbach resonance
\begin{equation}
a_{s}(B)=a_{bg}\br{1+\frac{\Delta{B}}{B-B_{0}}}
\end{equation}
where $B_{0}$ is the magnetic field at which the $a_{s}$ diverges, i.e., the  resonant point.  Comparing the above two equations, we see that $\Delta{B}=\mathcal{K}(\partial\delta/\partial{B})^{-1}$.  %Here $\partial\delta/\partial{B}$ is the magnetic momentum difference between two channels.  
\begin{figure}[htbp]
\begin{center}
\includegraphics[width=0.8\textwidth]{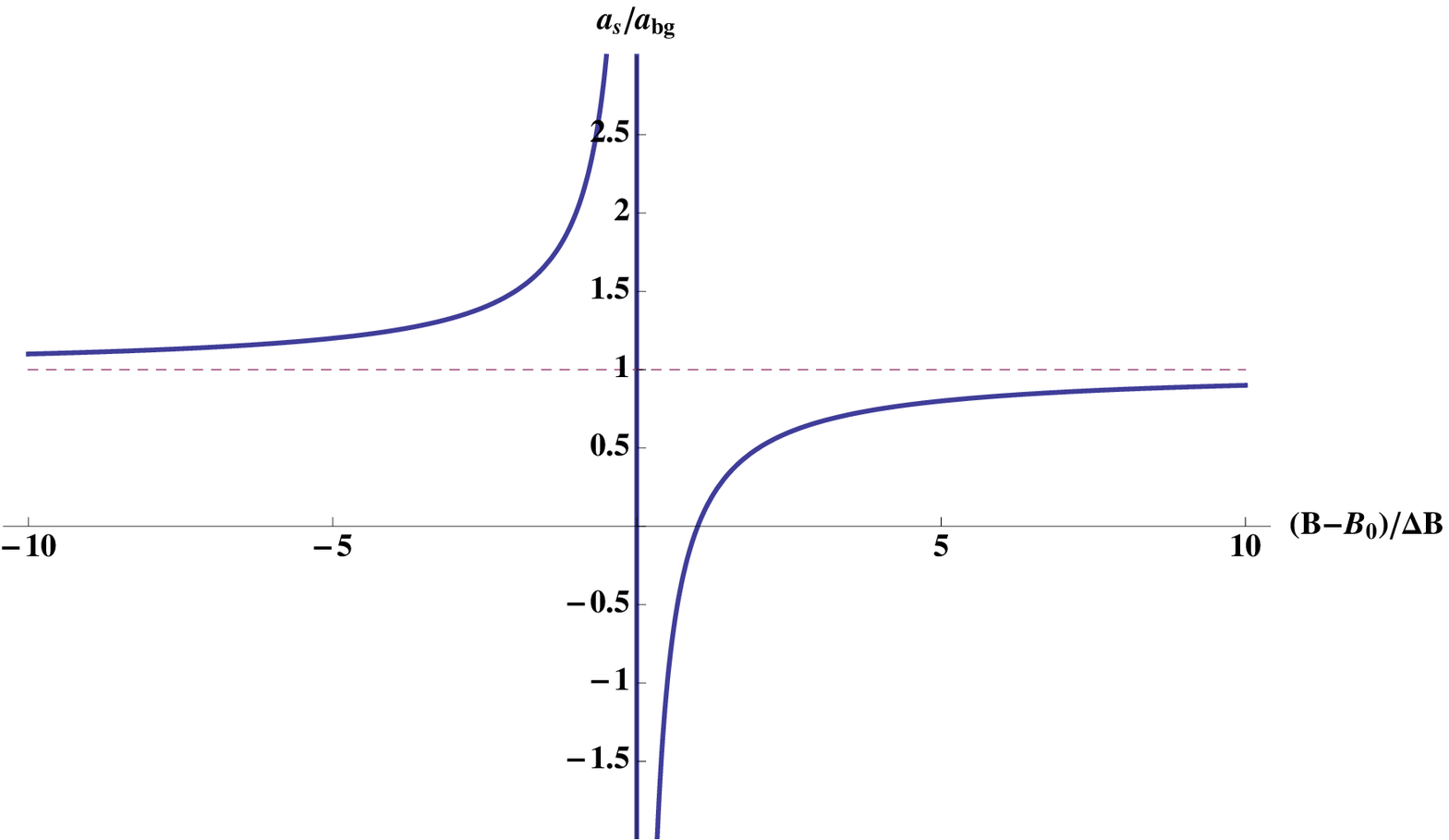}
\caption{S-wave scattering length in Feshbach resonance} 
\label{fig:intro:Feshbach}
{\small The dashed line is $y=a_{bg}$.}
\end{center}
\end{figure}

Let us now consider the bound state,\footnote{We wish to stress again the difference between a uncoupled closed-channel bound-state $\phi_{i}$ and a real bound state  formed by atoms from both open- and closed-channel.   The closed-channel bound state, $\phi_{i}$, is the eigenstate of the  isolated closed-channel Hamiltonian.  It is not a real eigenstate for the full two-channel Hamiltonian.  On the other hand, the full  Hamiltonian has bound eigenstates (i.e. $E<0$) at large negative detuning (in the two-body context).  The bound state solution for a two-channel eigenstate has components in both the open-channel and the closed-channel.  The open-channel weight is often larger when close to the resonance; consequently, such bound-states are often called open-channel bound-states.  Only at large negative detuning, the real two-channel bound state, mostly composed of close-channel component, coincides with the close-channel bound-state, $\phi_{i}$ to a large degree.}
where $E<0$.  We define
\begin{equation}\label{eq:intro:ab}
a_{b}(E)\equiv\frac{\hbar}{(2m_{r}\abs{E})^{1/2}}
\end{equation}
 Here, we only study the bound state close to threshold with a binding energy much smaller than the binding energy of the uncoupled closed-channel bound state, $\phi_{0}$, namely, $\abs{E}\ll{}E_{b}$.  Therefore, we have  $a_{b}\gg{a_{c}}$.  Outside the range of potential $r_{c}$,  the wave function is proportional to $e^{-r/a_{b}}$. For $r_{c}\ll{}r\ll{}a_{b}$, this wave-function can be expanded as  $1-\frac{r}{a_{b}}$, just as the Bethe-Peierls boundary condition (or the s-wave scattering wave function) we discussed in Chapter \ref{sec:intro:as}.  Note that $a_{b}(E)$ is not identified as $a_{s}$ a priori.  Through a  procedure similar as previous,  it is not hard to find
\begin{equation}
\frac{a_{bg}}{a_{b}}-1=\frac{-\mathcal{K}}{\delta+\mathcal{K}-E}
\end{equation}
We have assumed the short-range part of the bound-state wave function $\chi$ does not change much from the short-range part of the scattering state \footnote{This is guaranteed by the boundary condition of $\chi(r)\to1-r/a_{s}$, which fixes the normalization of the short-range part of the wave function ($\chi{r\to0}$) to be the same, namely $1$. As a result, all the integral term over the short-range kernel $\iint{drdr'}K(r,r')\chi_{0}(r)\chi(r')$  give the similar value as  $\mathcal{K}$ at the resonant point, Eq. \ref{eq:intro:kappa}.};  therefore $\mathcal{K}$ stays like a constant.  Provided both $\delta$ and $\abs{E}$ are much smaller than $\mathcal{K}$, this yields\begin{equation}\label{eq:intro:abKE}
a_{b}=a_{bg}\frac{\mathcal{K}}{\delta-E}
\end{equation}
It is not hard to see that $a_{b}$ indeed  coincides with the s-wave scattering length $a_{s}$ in Eq. \ref{eq:intro:asK} when $\abs{E}\ll\abs{\delta}$. Thus we will use $a_{b}$ and $a_{s}$ interchangeably hereafter. This is actually an example of our discussion about the Bethe-Peierls boundary condition in Chapter \ref{sec:intro:as}. Using the concept of the Bethe-Peierls boundary condition, we do not need to distinguish the zero energy scattering state and the negative energy bound state, and we can arrive the same conclusion easily.  

  Using Eqs. \ref{eq:intro:ab} and \ref{eq:intro:abKE}, it is easy to obtain an equation for $E$
\begin{equation}
(\abs{E}+\delta)^{2}-2\delta_{c}\abs{E}=0 
\end{equation}
where $\delta_{c}$ is defined as 
\begin{equation}\label{eq:intro:deltaC}
\delta_{c}\equiv\frac{\mathcal{K}^{2}}{\hbar^{2}/m_{r}a_{bg}^{2}}
\end{equation}
And the solution for the negative detuning, i.e. $\delta<0$, is
\begin{equation}\label{eq:intro:bindE}
\abs{E}=\delta_{c}-\delta-\sqrt{\delta_{c}^{2}-2\delta\delta_{c}}
\end{equation}
We can also calculate the ``relative weight'' on probability of the closed-channel for a normalized open-channel component, $\chi_{n}$, of the wave function ($\int{}\chi_{n}(r)^{2}dr=1). $\footnote{Note here the normalization on $\chi(r)$ is different from the most cases of this section, where $\chi(r)$ is normalized by  requiring $\chi{}\overset{r\rightarrow\infty}\rightarrow{}(1-r/a_{b})\text{ or }e^{-r/a_{b}} $ here. The difference is an extra $(a_{b})^{-1/2}$ for the wave function and that contributes the extra factor $a_{b}^{-1}$ in Eq. \ref{eq:intro:lambda}.} 
\begin{equation}
\label{eq:intro:closeWeight}
\lambda=\br{\frac{1}{E-\tilde\delta}}^{2}\abs{\int{}dr'\phi_{0}(r')Y(r')\chi_{n}(r')}^{2}
\end{equation}
Comparing this with Eq. \ref{eq:intro:kappa}, and assuming the short-range part of $\chi_{n}(r)$ does not differ from $\chi_{o}(r)$ much except the normalization, we  find 
\begin{equation}\label{eq:intro:lambda}
\lambda=\nth{(E-\tilde\delta)^{2}}\frac{\hbar^{2}}{2m_{r}a_{bg}}\frac{\mathcal{K}}{a_{b}}
\end{equation}
For $\abs{\delta}\lesssim\delta_{c}\ll\mathcal{K}$,  we have 
\begin{equation}\label{eq:intro:shiftK}
E-\tilde\delta\approx\mathcal{K}
\end{equation}
We can simplify $\lambda$ further
\begin{equation}\label{eq:intro:lambda}
\lambda\approx\frac{\hbar^{2}}{2m_{r}a_{bg}}\nth{(\mathcal{K}a_{b})}=\br{\frac{\abs{E}}{2\delta_{c}}}^{1/2}
\end{equation}
This shows that  $\delta_{c}$ is a ``characteristic'' energy scale. When $\abs{\delta}\gg\delta_{c}$, Eq. \ref{eq:intro:bindE} gives $E\approx\delta$ , and the above equation gives $\lambda=\sqrt{|\delta|/(2\delta_{c})}\gg1$: this simply means that when the negative detuning is large, weight is predominently in the closed-channel; the real mixed bound state is a closed-channel bound state ($\phi_{0}$) slightly dressed with the open-channel component; therefore the binding energy is very close to the binding energy  of the uncoupled closed-channel bound state, $\phi_{0}$.  The often quoted relation between $a_{s}$ and the binding energy, Eq. \ref{eq:intro:ab},  is not valid here.  On the contrary, when $\abs{\delta}\ll\delta_{c}$, Eq. \ref{eq:intro:bindE}  gives $\abs{E}\approx\delta^{2}/\delta_{c}\ll\abs{\delta}$,  and Eq. \ref{eq:intro:lambda} then gives $\lambda\approx{\delta/(\sqrt{2}\delta_{c})}\ll1$.  Here, the closed-channel weight is much smaller than that of the open-channel; the real bound-state is essentially an open-channel affair with little dress-up from the closed-channel.  Eq. \ref{eq:intro:ab} is valid in this case. % It is not hard to  estimate $\delta_{c}$. 

When dealing with  many-body problems, another important energy scale comes into play, namely, the Fermi energy, $E_{F}$.  When $E_{F}$ is much smaller than $\delta_{c}$, i.e. a \emph{broad resonance},  the closed-channel has only negligible weight close to resonance, which is the situation of interest. In such a case, it is a good approximation to take the Feshbach resonance as only a knob to tweak the interaction in the open-channel.  On the contrary, when $E_{F}$ is close or even larger than $\delta_{c}$, i.e. a \emph{narrow resonance}, the closed-channel weight can be  significant clsoe to  resonance; as a result, it is necessary to explicitly take the closed-channel into account in the many-body framework.  

One important remark about the above discussion is that whether a resonance is ``broad'' or ``narrow'' is a many-body concept and is relative.  For one particular resonance with a fixed $\delta_{c}$, in a infinite dilute system (i.e. a two-body system) with a zero Fermi energy,  $E_{F}=0\ll\delta_{c}$, it is a broad resonance.  As the system becomes denser and denser, $E_{F}$ increases and the resonance becomes less and less broad; finally, when $E_{F}$ is in the order or even larger than $\delta_{c}$, the resonance becomes a narrow one.  In the real experimental system of the ultracold alkali gases, the achievable density has a certain range, so does the Fermi energy range, $E_{F}$; therefore the ``broadness'' or ``narrowness'' of a resonance is not as flexible.

\chapter{The single-channel BEC-BCS crossover\label{sec:intro:1channel}}
In this chapter, we briefly review the BEC-BCS crossover in a single channel.   The idea to describe the BEC and BCS on the same footing stems back several decades \cite{Eagle, LeggettCrossover, Nozieres, RanderiaBEC}.  %One way to interpret BCS is through the existing macroscopic eigenvalue of two-body density matrix (see Sec. \ref{sec:intro:as})\cite{Leggett}.   This eigenvalue corresponds the more familiar anomalous expectation, $\av{a_{\uparrow-\vk}a_{\downarrow\vk}}$.  If simply taking it as wave function of a two-fermion molecule, this eigenfucntion, $\ket{\psi}=(\sum_{\vk}c_{\vk}a^{\dg}_{\uparrow\vk}a^{\dg}_{\downarrow-\vk})\ket{0}$ is very large in size (coherent length), much larger than the average particle distance.  Therefore overlap between different ``molecules'' is significant.   
The BCS theory can be understood as the following: fermions form  ``giant molecules'' and those molecules then condense simultaneously.    It is not hard to show that the  BCS ansatz is equivalent to  the  coherent state of a two-body pair $\psi^{\dg}=\sum_{\vk}c_{\vk}a^{\dg}_{\uparrow\vk}a^{\dg}_{\downarrow-\vk}$
\begin{equation}\label{eq:intro:BCScoherent}
\prod_{\vk}(u_{\vk}+v_{\vk}a^{\dg}_{\uparrow\vk}a^{\dg}_{\downarrow-\vk})\ket{0}=A\exp{}(\sum_{\vk}c_{\vk}a^{\dg}_{\uparrow\vk}a^{\dg}_{\downarrow-\vk})\ket{0}
\end{equation}
where $c_{\vk}=( v_{\vk}/u_{\vk})$. The size of these ``giant molecules'' is  much larger than the interparticle distance.
Moving away from the BCS end toward the BEC end, pairs shrink as the interparticle attraction becomes stronger.  On the BEC side, two-fermion molecules are smaller than the interparticle distance and therefore a well-defined object. However, the binding energy now is much larger than the typical many-body energy scale and condensation does not happen at the same time when molecules form.  Nevertheless,  at low enough temperature, we can still consider the formation of molecules and their condensation  at the same time, compatible with the  the same framework used for BCS.  

In the single-channel BEC-BCS crossover model, one imagines a ``magic'' knob that can tune the interaction strength along the crossover.  The many-body fermion  system sweeps from  BCS of a fermionic atom system to BEC of diatomic molecules  in response to the increase of attraction.   This directly applies to the broad-resonance of the two-channel case as well, where the closed-channel weight is negligible and only serves  to modify the effective interaction strength in the open-channel.  We mostly follow the path-integral treatment by Randeria and the company \cite{RanderiaBEC, Randeria1997, Randeria2008}.
%\subsection{Path integral for one channels}
% !TeX root =thesis.tex
%\subsection{Path integral approach for single channel\label{sec:pathInt}}
\label{sec:pathInt}
They have studied this problem with a path integral approach which is proved to be a  nice tool for the problem due to its flexibility and readiness to be   extended for the higher order fluctuations.  In the next two chapters, this method will be adapted  further for the two-channel model. 

We start with an attractive $\delta$-potential in the coordinate space.  This potential is not equivalent to the  reduced pairing potential used in the original BCS work.  The reduced pairing potential only couples  particles of the opposite momentum and does not support simple form of Hubbard-Stratonovich transformation, which is essential to solve the problem in the path integral formulation. 

  The Hamiltonian with the chemical potential of the system can be written as   
\begin{equation}
\hat{H}-\mu\hat{N}=\sum_{\sigma}\int{d^{d}\vr}c^{\dagger}_{\sigma}(\vr)\br{-\nth{2m}\nabla^{2}-\mu}c^{}_{\sigma}(\vr)-g\int{d^{d}\vr}c^{\dagger}_{\uparrow}(\vr)c^{\dagger}_{\downarrow}(\vr)c^{}_{\downarrow}(\vr)c^{}_{\uparrow}(\vr)
\end{equation}
Introducing the quantum partition function $\mathcal{Z}=\int{\bigD(\bar\psi,\psi)\exp\br{-S[\bar\psi,\psi]}}$, where $\bigD(\bar\psi,\psi)$ denotes the functional integral over all possible wave function $\psi$ and $\bar\psi$, and the action $S[\bar\psi,\psi]$ can be written down from the Hamiltonian
\begin{equation}\label{eq:pathInt2:actionPsi}
S[\bar\psi,\psi]=\int^{\beta}_{0}d\tau\int{d^{d}\vr}\mbr{\sum_{\sigma}\bar\psi_{\sigma}(\vr,\tau)\br{\partial_{\tau}-\nth{2m}\nabla^{2}-\mu}\psi_{\sigma}(\vr,\tau)-g\bar\psi_{\uparrow}(\vr,\tau)\bar\psi_{\downarrow}(\vr,\tau)\psi^{}_{\downarrow}(\vr,\tau)\psi^{}_{\uparrow}(\vr,\tau)}
\end{equation}
The fermion fields $\psi_{\sigma}$ and $\bar\psi_{\sigma}$ are two independent Grassmann variables. Notice that  they are not complex conjugate to each other as in the usual operator language because  complex conjugate is not a well-defined concept for Grassmann variables. 

This system can be solved with Hubbard-Stratonovich transformation.   Introduce an auxiliary field (functional variable) $\Delta(\vr,\tau)$ coupled with a pair $\psi_{\uparrow}(\vr,\tau)\psi_{\downarrow}(\vr,\tau)$. %Here we follow the normal notation from path integral, $r$ is four tempo-space coordinator.  
We write down first the Gaussian integral of $\Delta$
\begin{equation}
1=\int{\bigD(\bar\Delta,\Delta)}\exp\br{-\nth{g}\int{d\tau{d}^{d}r}\bar\Delta\Delta}
\end{equation}
Note that we absorb the extra constant of integration into the measure of $\bigD(\bar\Delta,\Delta)$.
And with a shift of $\Delta(\vr,\tau)\rightarrow\Delta(\vr,\tau)-g\psi_{\uparrow}(\vr,\tau)\psi_{\downarrow}(\vr,\tau))$, we have 
\footnote{$\int{\bigD(\bar\Delta,\Delta)}\cdot1$ is only a constant factor on partition function $\mathcal{Z}$ and has no effect on real physical quantity; therefore, we can take it as 1. (This is equivalent to  divide the $\mathcal{Z}$ by a constant)}
\begin{equation}\label{eq:pathInt:expHS}
\exp\br{g\int{d\tau{}d^{d}\vr}\bar{\psi}_{\uparrow}\bar\psi_{\downarrow}\psi_{\downarrow}\psi_{\uparrow}}=
\int{\bigD(\bar\Delta,\Delta)}\exp\bbr{-\int{d\tau{d^{d}\vr}}\mbr{\nth{g}{\bar\Delta}{\Delta}-\br{\bar\Delta\psi_{\downarrow}\psi_{\uparrow}+\Delta\bar\psi_{\uparrow}\bar\psi_{\downarrow}}}}
\end{equation}
Note that  $\Delta(\vr,\tau)$ (or $\bar\Delta(\vr,\tau)$) comes from  Grassmann fields $\psi(\vr,\tau)$ (or $\bar\psi(\vr,\tau)$). Therefore, they are not related to each other as complex conjugate either.  Nevertheless, at the mean field level or only at the phase fluctuation around the mean field values, $\Delta$  and $\bar\Delta$ are indeed complex conjugate.  Consequently, we will just take $\Delta$  as normal bosonic field in the following and often simply treat $\bar\Delta$ as $\Delta$'s complex conjugate. Now the interaction term can be replaced.
\begin{align*}
\mathcal{Z}=&\int{}\bigD(\bar\psi,\psi)\int{\bigD(\bar\Delta,\Delta)}\\
&\;\exp\bbr{-\int{d\tau{d^{d}\vr}}\mbr{\sum_{\sigma}\bar\psi_{\sigma}\br{\partial_{\tau}-\nth{2m}\nabla^{2}-\mu}\psi_{\sigma}+\nth{g}{\bar\Delta}{\Delta}-\br{\bar\Delta\psi_{\downarrow}\psi_{\uparrow}+\Delta\bar\psi_{\uparrow}\bar\psi_{\downarrow}}}}
\end{align*}
At the expense of introducing an auxiliary field ($\Delta$) which has contact-type coupling to the original field $\psi$, we eliminate the four-field interaction term formally.  $\Delta$ field is like a \emph{local potential} for $\psi$, although this \emph{local potential} has to be calculated from the original field self-consistently.  Nevertheless, $\Delta$ couples to a pair of fermionic field $\psi$, and thus it extracts a special degree of freedom from the $\psi$ field.  When properly selected, this degree of freedom is highly non-trivial and has macroscopic importance, which serves as the  ``order parameter'' for the system.  The above formula for partition function is bilinear to $\psi$, and we can rewrite it into a nicer form in Nambu spinor representation
\begin{equation}
\bar\Psi=\begin{pmatrix}\bar{\psi}_{\uparrow}&\psi_{\downarrow}\end{pmatrix}\text{,  }\qquad
\Psi=\begin{pmatrix}{\psi}_{\uparrow}\\\bar\psi_{\downarrow}\end{pmatrix}
\end{equation}
\begin{equation}\label{eq:pathInt:ZDeltaPhi}
\mathcal{Z}=\int{\bigD(\bar\Psi,\Psi)}\int{\bigD(\bar\Delta,\Delta)}\exp
	\bbr{-\int{d\tau{d^{d}\vr}}\mbr{\nth{g}{\bar\Delta}{\Delta}-\bar\Psi \nG\Psi}}
\end{equation}
where 
\begin{equation}\label{eq:pathInt:nG}
\nG=\begin{pmatrix}
[\hat{G}_{0}^{(p)}]^{-1}&\Delta\\\bar\Delta&[\hat{G}_{0}^{(h)}]^{-1}
\end{pmatrix}
\end{equation}
is known as the Gor'kov Green function. $[\hat{G}_{0}^{(p)}]^{-1}=-\partial_{\tau}+\nth{2m}\nabla^{2}+\mu$, and $[\hat{G}_{0}^{(h)}]^{-1}=-\partial_{\tau}-\nth{2m}\nabla^{2}-\mu$ represent the non-interacting Green's functions of the particle and the hole respectively. 

Before going further, we would like to discuss one confusing point about the possible one-or-two indices for quantities such as ${G}$ or $\Delta$ in Eq. \ref{eq:pathInt:nG}.  As a matrix, such a quantity has two indices $(x,x')$ or $(p,p')$, which have no ambiguity in usage. On the other hand, there are often ambiguity when only one index $x$ or $p$ is used. In some cases, the one index means the relative value of the two indices. For example, an  interaction, $U(\vr_{1},\vr_{2})$, normally only depends on the relative coordinate, $\vr=\vr_{1}-\vr_{2}$. So $U(\vr)$ means $U(\vr_{0}+\vr,\vr_{0})$.  In other cases, especially common in the current thesis, the one index stands for its repetition.  In this case, the difference is always zero. For example, the free Green's function, $G_{0}(p)$ stands for $G_{0}(p,p')\delta({p-p'})$.  Similarly,  the order parameter, $\Delta(x)$, only couples to $\bar\psi(x)\bar\psi(x)$ (Eq. \ref{eq:pathInt:expHS}).  When used in the matrix context (Eq. \ref{eq:pathInt:nG}), it means $\Delta(x)\delta(x-x')$. Interestingly, its Fourier transformation in momentum space does not have the same properties.  In fact, it means $\Delta(p,p')=\Delta(p'-p)$.

Now action in Eq. \ref{eq:pathInt:ZDeltaPhi} is bilinear to $\Psi$; so it can be integrated out formally and the partition function then only depends on the  field $\Delta$.  
\begin{equation}\label{eq:pathInt:DeltaPF}
\mathcal{Z}=\int{\bigD(\bar\Delta,\Delta)}\exp
	\bbr{-\mbr{\br{\int{d\tau{d^{d}r}}\nth{g}{\bar\Delta\Delta}}-\ln\det\nG}}
\end{equation}
And the action becomes
\begin{equation}\label{eq:pathInt:DeltaAction}
S[\bar\Delta,\Delta]=
	{\mbr{\br{\int{d\tau{d^{d}r}}\nth{g}{\bar\Delta\Delta}}-\ln\det\nG}}
\end{equation}
Note that the determinant in $\ln\det\nG$ runs through both the normal coordinate space and $2\times2$ Nambu spinor space.  The above formulas are exactly equivalent to the original partition function (action) in the fermion field $\psi$ (Eq. \ref{eq:pathInt2:actionPsi}). It looks nice and compact. Nevertheless, $\ln\det\nG$ term is highly non-trivial and contains all the many-body physics.

\section{Mean field results\label{sec:pathInt:meanfield}}
The saddle point equation of Eq. (\ref{eq:pathInt:DeltaPF}) gives the mean-field result of the system.  First we need to find the derivative of $\ln\det\nG$.  We notice the identity
\begin{equation}
\ln\det\hat{A}=\tr\ln\hat{A}
\end{equation}
and differential rule of the function ``$\tr\ln$''
\begin{equation}\label{eq:pathInt:diffTr}
\frac{\delta}{\delta\phi_q}\tr\ln(\nG)=\tr(\hat{\mathcal{G}}\frac{\delta}{\delta\phi_q}\nG)
\end{equation}
Using the above relations, we can write the saddle equation of Eq. (\ref{eq:pathInt:DeltaPF}) (differential with respect to $\Delta$) as
\begin{equation}
\nth{g}\bar{\Delta}(\vr,\tau)-\tr\mbr{\hat{\mathcal{G}}(\vr,\tau,\vr,\tau)\begin{pmatrix}0&1\\0&0\end{pmatrix}}=0
\end{equation}
Here this matrix is in the Nambu spinor space.  At the  mean field level, we seek a tempo-spacial homogeneous solution of $\Delta(x)=\Delta_{0}$.  At this level,  $\Delta(p)$ becomes a $\delta$-function in the frequency-momentum space, and has non-zero elements only for two fermions with the same momentum.  (Please See the discussion in the previous section about one vs. two indices. This is not generally true in other situations, as we show it when discussing collective modes in sec. \ref{sec:collective1})
We can find the Gor'kov Green function from Eq. (\ref{eq:pathInt:nG}) in momentum space at the mean-field level
\begin{equation}\label{eq:pathInt:G0}
G_{0\;p,p'}=\nth{(i\omega_n)^2-E_\vp^2}
\begin{pmatrix}
	i\omega_n+\xi_\vp&\;&-\Delta_0\\
	-\bar{\Delta}_0&\;&i\omega_n-\xi_\vp
\end{pmatrix}
\delta_{p=p'}
\equiv{}G_{0}(p)\delta_{p=p'}
\end{equation}
Here $p$ is the frequency-momentum, $p=(\omega_{n},\vp)$, and $\omega_n$ is the Matsubara frequency of Fermions.  $\xi_{\vk}=\epsilon_{\vk}-\mu$, $\epsilon_{\vk}=\vk^{2}/2m$,  $E_\vp=\sqrt{\xi_\vp^2+\abs{\Delta_0}^2}$.  And the saddle point equation can be rewritten as 
\begin{equation}
\nth{g}\bar{\Delta}_0=\frac{T}{\mathcal{V}_{0}}\sum_{\vp,n}\frac{\bar\Delta_0}{\omega_n^2+E_\vp^2}
\end{equation}
Here $T$ is the temperature, and $\mathcal{V}_{0}$ is the volume in $d$-dimension.  The summation of the Matsubara frequency  can be evaluated\footnote{\label{foot:intro:sum}The summation of the Matsubara frequency of a function $h(i\omega_{n})$ is carried out by the normal trick.  We  multiply $h(z)$ with the Fermi distribution function $n_{F}(z)$,  the summation is the sum of residuals at the imaginary axis of $n_{F}(z)$.  The contour can be deform into a contour over the rest of singular points of $h(z)$. We just need to find the residuals of the total function $h(z)n_{F}(z)$ over those singular points to find the Matsubara summation.   However, due to zero temperature, the  $n_{F}(z)$ is only nonzero at the negative singular points of $h(z)$, $-E_{\vk}$ in this case.  (The other singular point, $E_{\vk}$, gives $n_{F}(E_{\vk})=0$ for zero temperature.)}  and we find 
\begin{equation}
\nth{g}=\nth{\mathcal{V}_{0}}\sum_{\vp}\frac{1-2n_f(E_p)}{2E_p}=\nth{\mathcal{V}_{0}}\sum_{\vp}\frac{\tanh{(E_p/2T)}}{2E_p}
\label{eq:pathInt:gap}
\end{equation}
where $n_f(\epsilon)$ is the fermi distribution function.  This is exactly the famous gap equation obtained from other methods as well.  On the other hand, $\nG$ in Eq. (\ref{eq:pathInt:nG})  is the inverse of the  fermion-fermion correlation of $\Psi$.  In the mean field, $G_{0}$ as Eq. (\ref{eq:pathInt:G0}) can be diagonalized in the momentum space with a canonical (Bogoliubov) transformation.  We can make an analytic continuation of $i\omega_{n}\rightarrow\omega+0^{+}$.  Eq. (\ref{eq:pathInt:G0})  then has poles ($\pm{}E_{p}$) where  $\omega^2-E_\vp^2=0$,  which determine  the spectrum of fermionic excitations.  Indeed, in the BCS-like states ($\mu>0$), the spectrum is gapped at $\Delta$; while in the BEC-like states ($\mu<0$), the fermionic excitation starts from the molecule binding energy $\sqrt{\mu^{2}+\Delta^{2}}\approx\abs{\mu}$.

\begin{figure}[htbp]
\begin{center}
\includegraphics[width=0.8\textwidth]{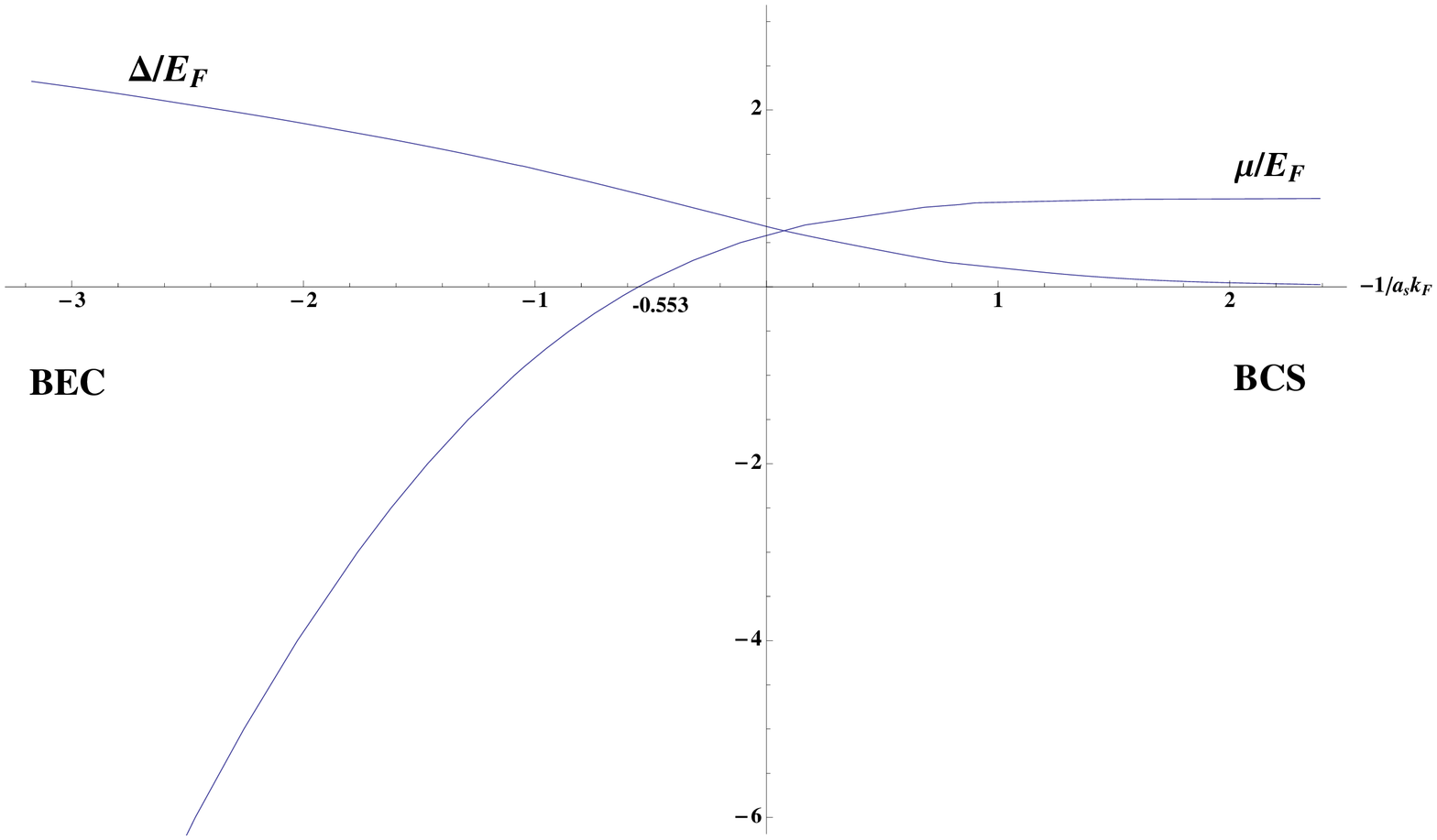}
\caption{The chemical potential $\mu$ and gap $\Delta$ in the mean field level over crossover} 
\label{fig:pathInt:meanField}
{\small All quantities in the unit of energy ($\mu$, $\Delta$) are rescaled with the Fermi energy $E_{F}$ and the s-wave scattering length $a_{s}$ is rescaled with $1/k_{F}$.  }
\end{center}
\end{figure}

 The summation in Eq. (\ref{eq:pathInt:gap})  does not converges  in 3D because the summand does not decreases fast enough.  This is because our assumption of contact interaction breaks down for the scale smaller than real potential range $r_{c}$, i.e., the summation of momentum is capped at some high momentum $\Lambda$ related to $1/r_{c}$.  Notice that in 3D, we have a similar relation that connect the bare potential $g$ to a more physically observable quantity, the s-wave scattering length $a_{s}$
\begin{equation}\label{eq:pathInt:as}
\frac{m\mathcal{V}_{0}}{4\pi{}a_{s}}=-\nth{g}+\sum_{k<\Lambda}\nth{2\epsilon_{\vk}}
\end{equation}
Here $\mathcal{V}_{0}$ is the total volume.  We can renormalize Eq. \ref{eq:pathInt:gap} with this relation
\begin{equation}\label{eq:pathInt:gapRenormalized}
-\frac{m\mathcal{V}_{0}}{4\pi{}a_{s}}=\sum_{\vk}\mbr{\frac{\tanh{(E_k/2T)}}{2E_k}-\nth{2\epsilon_{\vk}}}
\end{equation}
Now the gap equation has proper decay in high momentum and no artificial cutoff is necessary.  There are two unknown parameters, $\mu$ and $\Delta$,  in the equation.  We need another equation in order to pin them down. To complement the gap equation, we can introduce the number equation, $N=-\partial\Omega/\partial\mu$. At the saddle point, the thermodynamic potential is $\Omega_{0}=S[\Delta_{0}]/\beta$, and we have the number equation
\begin{equation*}
N=-\nth{\beta}\tr\br{{G_{0}\pdiff{G_{0}^{-1}}{\mu}}}
\end{equation*}
Similarly the summation (due to the trace) over the Mastubara frequency can be evaluated and we have the number equation
\begin{equation}
N=\nth{L^{d}}\sum_{\vk}\mbr{1-\frac{\epsilon_{\vk}}{E_{\vk}}\tanh{(\frac{E_{\vk}}{2T})}}
\end{equation}
This equation has no divergence at high momentum.  The number equation  and the renormalized gap equation Eq. (\ref{eq:pathInt:gapRenormalized}) compose the implicit equations for two unknown parameters, gap $\Delta$ and chemical potential $\mu$.  It is not hard to find the zero temperature analytic result at both ends.  At the BCS end ($1/k_{F}a_{s}\rightarrow-\infty$), we obtain $\mu\approx{}E_{F}$ and $\Delta\propto\exp(-\pi/2k_{F}\abs{a_{s}})$; at the BEC end ($1/k_{F}a_{s}\rightarrow+\infty$),  $\mu=-\hbar^{2}/2ma_{s}^{2}$, i.e. half of the binding energy of a molecule, while $\Delta\propto{}n^{1/2}a_{s}^{-1/2}$ no longer has  much physical significance.  In the more general crossover region, these two equations can only  be solved numerically.  They have no singularity in the whole region, which indicates it is a crossover instead of any simple phase transition.  Please see Fig. \ref{fig:pathInt:meanField} for detail. 

\section{Gaussian fluctuation and collective modes}\label{sec:collective1}
We can expand the partition function Eq. (\ref{eq:pathInt:DeltaPF}) around the mean-field value, $\Delta(\vr,\tau)=\Delta_{0}+\theta(\vr,\tau)$. The linear order of the  expansion is zero because $\Delta_{0}$ is the saddle point.  The next order gives us the bilinear terms on $\theta$, i.e., correlation of bosonic fields $\Delta$.  Note that here the Hamiltonian only has a contact-type potential, therefore it cannot cover the situation of a charged system where long-range Columnb interaction cannot be neglected.  We limit ourselves to the neutual case.  Nevertheless, it is conceivable that a more realistic short-range potential only renormalizes some parameters in the following calculation while leaves the qualitative result unmodified.  

Notice that we can expand the second term in Eq. (\ref{eq:pathInt:DeltaPF}) for $\hat{G}{}^{-1}=\hat{G}_{0}^{-1}+\hat{K}$
\begin{equation}\label{eq:pathInt:expand}
\tr\ln \hat{G}^{-1}=\tr\ln\hat{G_{0}}^{-1}+\tr(\hat{G_{0}}\hat{K})-\nth{2}\tr(\hat{G_{0}}\hat{K}\hat{G_{0}}\hat{K})+\cdots
\end{equation}
In our case,
\begin{equation}
\hat{K}=\begin{pmatrix}
0&\theta\\
\theta^{*}&0
\end{pmatrix}
\end{equation}
Here the linear terms of $\hat{K}$ or $\theta$ ($\theta^{*}$) are zero as the saddle point condition.  To the second order, the action is 
\begin{equation}\label{eq:pathInt:DeltaActionGaussian}
S[\Delta_{0},\theta,\theta^{*}]=S[\Delta_{0}]+
	\nth{2g}\tr(\hat{K}\hat{K})+\nth{2}\tr(\hat{G_{0}}\hat{K}\hat{G_{0}}\hat{K})
\end{equation}
Write the last term into the momentum representation
\begin{equation}
\tr(\hat{G_{0}}\hat{K}\hat{G_{0}}\hat{K})=\sum_{q,p}\Tr\br{G_{0}({p})K_{q}G_{0}{}({p-q})K_{-q}}
\end{equation}
Notice that the second ``$\Tr$'' and following ``$\Tr$'' in this section only runs in Nambu spinor space and $q={(\vq,q_{l})}$, $p=(\vp,p_{n})$ are all four momentum, where $q_{l}$ is the bosonic Matsubara frequency while $p_{n}$ is the fermionic Matsubara frequency.
\begin{equation}
K_{p_{0},p_{0}+q}=K_{q}=\begin{pmatrix}
0&\theta_{q}\\
\theta^{*}_{-q}&0
\end{pmatrix}
\end{equation}
And we remember that $G_{0}(p)=G_{0}{}_{p,p}$
If we introduce  a new vector 
\begin{equation}
\theta{(q)}=\begin{pmatrix}\theta_{q}\\\theta^{*}_{-q}\end{pmatrix}\qquad
\theta^{\dg}{(q)}=\begin{pmatrix}\theta^{*}_{q}&\theta_{-q}\end{pmatrix}
\end{equation}
the action can be rewritten into a more compact form
\begin{equation}
S[\Delta_{0},\theta,\theta^{*}]=S[\Delta_{0}]+\nth{2}\sum_{q}\mbr{\theta^{\dg}(q)\mathbf{M}(q)\theta(q)}
\end{equation}
Notice that we can always choose a real $\Delta_{0}$ and therefore $G_{0}{\ _{12}}(p)=G_{0}{\ _{21}}(p)$, we have 
\begin{equation}
\mathbf{M}_{q,q}=\mathbf{M}(q)=
\begin{pmatrix}
\nth{g}+\sum_{p}G_{0}{\ }_{11}(p)G_{0}{\ }_{22}(p-q)&\sum_{p}G_{0}{\ }_{12}(p)G_{0}{\ }_{12}(p-q)\\
\sum_{p}G_{0}{\ }_{12}(p)G_{0}{\ }_{12}(p-q)&\nth{g}+\sum_{p}G_{0}{\ }_{11}(p-q)G_{0}{\ }_{22}(p)
\end{pmatrix}
\end{equation}
The summation over  the (fermionic) Matsubara frequency of $p_{n}$ can be carried out at zero temperature

\begin{equation}
\begin{split}
M_{11}(q)&=M_{22}(-q)\\
	&=\nth{g}+\sum_{\vp{,}p_{n}}G_{0}{\ }_{11}(p)G_{0}{\ }_{22}(p-q)\\
	&=\nth{g}+\sum_{\vp}\br{\frac{u^{2}u'^{2}}{iq_{l}-E-E'}-\frac{v^{2}v'^{2}}{iq_{l}+E+E'}}
\end{split}
\end{equation}
\begin{equation}
\begin{split}
M_{12}(q)&=M_{21}(q)\\
	&=\sum_{\vp{,}p_{n}}G_{0}{\ }_{12}(p)G_{0}{\ }_{12}(p-q)\\
	&=\sum_{\vp}uvu'v'\br{\nth{iq_{l}+E+E'}-\nth{iq_{l}-E-E'}}
\end{split}
\end{equation}
where $u=u_{\vp}$, $v=v_{\vp}$, $E=E_{\vp}$ and $u'=u_{\vp-\vq}$, $v'=v_{\vk-\vq}$, $E'=E_{\vk-\vq}$.  $u_{\vk}$, $v_{\vk}$, $E_{\vk}$ are  defined as usual BCS literature. 
\begin{equation}
v_{\vk}^{2}=1-u_{\vk}^{2}=\nth{2}\br{1-\frac{\xi_{\vk}}{E_{\vk}}}
\end{equation}
 The $G^{(M)}=\mathbf{M}^{-1}$ is the correlation function of $\theta$ (or $\Delta$) and its poles give the spectrum of collective modes as every  $\theta_{q}$ (or $\Delta_{q}$) involves many fermions moving in a coherent manner.  So the spectrum of collective modes can be determined by finding poles of $G^{(M)}$, $\det{M(\omega,\vq)}=0$, after we analytically continue for the frequency $iq_{l}\rightarrow\omega+i0^{+}$.  
 
For low energy modes, where $\omega,\,\abs{\vq}^{2}$ both are much smaller than $\min\bbr{E_{\vk}}=\Delta_{0}$ (or $\sqrt{\mu^{2}+\Delta^{2}}$ for $\mu<0$), we can expand $M$ with $\omega$ and $\vq$.  The lowest order has the form $\omega\approx{}c\,q$, which suggests a sound wave as expected for any Goldstone mode.  At BCS side, $c=v_{F}/\sqrt{3}$, where $v_{F}$ is the Fermi velocity.  This coincides with the famous Anderson-Bogoliubov mode.  At the BEC side, we get $c^{2}=\Delta^{2}/8m\abs{\mu}=v_{F}^{2}(k_{F}a_{s})/3\pi=4\pi{}n_{B}a_{B}/m_{B}$, which fits the low momentum part of Bogoliubov spectrum of bosons gas.  Here $m_{B}=2m$ is the molecule mass, $n_{B}=n/2$ is the molecule density and $a_{B}=2a_{s}$ is the inferred  interaction between molecules.  This value differs from the result of more accurate calculation from the few-body theory, $a_{B}=0.6a_{s}$ \cite{Petrov}, which indicates the possible deficiency of the current theory.

\section{An alternative method to invert the Green's function\label{sec:diagonalizeGreen1}}
In the above section, we inverted the  Gor'kov green function matrix Eqs. (\ref{eq:pathInt:nG}, \ref{eq:pathInt:G0}) directly and it is not hard to do as a $2\times2$ matrix in the  momentum space.   Alternatively, we can use a different approach which proves to be more convenient in the two-channel problem.  First, we diagonalize $\nG$ with a unitary transformation $T$, in the momentum space
\begin{equation}
\nG=\mtrx{i\omega_{n}-\xi_{k}&\Delta\\\bar\Delta&i\omega_{n}+\xi_{k}}=T^{\dg}BT
\end{equation}
It is easy to show that such $T$ and $B$ satisfying above equation are
\begin{equation}
T=\mtrx{u_{k}&v_{k}\\-v_{k}^{*}&u_{k}}\qquad{}B=\mtrx{i\omega_{n}+E_{k}&0\\0&i\omega_{n}-E_{k}}
\end{equation}
where $u_{k}^{2}(v_{k}^{2})=\nth{2}(1\pm\xi_{k}/E_{k})$ and $E_{k}=\sqrt{\xi^{2}_{\vk}+\Delta^{2}}$ are conventionally defined quantities in the BCS theory.   Actually, this transformation is nothing but the Bogoliubov canonical transformation, and the $B$ matrix simply describes the spectrum of the fermionic quasi-particles.  Now it is easy to invert $\nG$
\begin{equation}
\mathcal{G}=T^{\dg}B^{-1}T
\end{equation}
Green's function $\mathcal{G}$ takes a more conventional form $A/(i\omega_{n}\pm{}E_{k})$ here without any dependency on frequency in nominator as Eq. (\ref{eq:pathInt:G0}). Matsubara frequency summation over $G_{0}(k)$ in the mean-field and $G_{0}(k)G_{0}(k+q)$ in the Gaussian order are then easier to perform  as in text-book.

\begin{subappendices}
\end{subappendices}

%\chapter{Path-Integral approach}

% !TeX root =thesis.tex

\newcommand{\htd}{\tilde{h}}
\chapter{The two-channel three-species many-body model, the mean field\label{ch:path2}}

For the narrow Feshbach resonance, atoms have considerable weight in the closed-channel and the Pauli exclusion between two channels cannot be neglected.  A many-body framework needs to include both channels.  Before diving into the detailed calculation, let us make some rough estimates about scales in order to  build some intuition.  The same problem at the two-body level is well understood as briefed in Chapter \ref{sec:intro:twobody}.  %Extended to many-body, three different types of Pauli exclusion requires to be considered: Pauli exclusion within open-channel, Pauli exclusion within closed-channel and Pauli exclusion between two channels. Let us estimate the scale of each with an artificial question.  We  calculate how much is reduced in two-body wave function by one single type of Pauli exclusion when a new pair is added.  This quality is quite unphysical without any real measurable counterpart, but it serves as a starting point to build intuition. 
We estimate how much each channel would change  in a many-body system.  Compared to a two-body system, the two-body correlation of a (cold) many-body fermion system in the momentum representation is modified mostly in  low momentum, i.e.  around or below the Fermi momentum, while staying  just like  the two-body wave-function intact in high momentum.  The open-channel component  of the two-body scattering wave-function is like a zero-energy free wave ($k=0$) plus a small short-range kernel of size $r_c$.  In the momentum space, this wave function is like a $\delta$-function at $k=0$ with a small  tail in high momentum.  The occupation number in the first available level, $k=0$, is close to one (for two hyperfine spins) and this level cannot accommodate one more pair.  The next pair has to occupy the next available level in $k$ instead of the lowest energy two-body level, $k=0$.  This suggests that the open-channel requires a full many-body treatment.  The situation is quite different in the closed-channel.   One crucial assumption is that  the closed-channel bound state is much smaller than the interparticle distance in the real representation. A closed-channel bound state,  $\phi_{0}$, in  the  momentum,  spreads its most weight  in the range $[0,1/a_{c}]$.  The weight in each momentum level is so tiny that it is even smaller than $1$ when it is multiplied by total atoms number $N$, $N\abs{\phi_{k}}^{2}\ll1$.             It only has a very small fraction of its weight within the Fermi energy range ($<k_{F}$).  Hence, the closed-channel only has a small  overlap with atoms in  the open-channel, as well as with other atoms in the closed-channel.  The smallness of this overlap ensures that many-body effects in the closed-channel can be  treated perturbatively.  As discussed in Sec. \ref{sec:intro:as} and Appendix \ref{sec:pathInt2:short-range}, the high momentum part of the two-body correlation just follows the two-body wave function;  therefore, we can write two-body correlation as 
\begin{equation}\label{eq:pathInt2:hphif}
h_{\vk}\sim\phi_{0\vk}f(\vk)
\end{equation}
 with $f(\vk)=1$ for $k\gg{k_{F}}$.  $f(\vk)$ deviates from $1$ in low energy and represents all the many-body correction. 

In addition to the two channel description, the model needs to be explicitly expressed with the three hyperfine species at least in some part in order to address the inter-channel  Pauli exclusion between two channels due to the common species.  This effect has received little  attention in theoretical research and a model as such is a useful addition to our knowledge of the BEC-BCS crossover and the Feshbach resonance.  Nevertheless, as we just discussed, and will illustrate more quantitatively later, the effects of the inter-channel Pauli exclusion between the two channels is relatively minor and can be treated perturbatively.  Consequently, it is not necessarily to carry the three species description all through the calculation.  The description of two channels and the description of three species are used in different phases of the calculation to solve the problem.  

\section{The extremely narrow resonance\label{sec:pathInt2:extremelyNR}}
Before the quantitative model, we discuss qualitatively a simple yet revealing case,   the extremely narrow resonance, where the inter-channel coupling approaches zero, $Y\rightarrow0$.  In this case, two channels are almost independent to each other except sharing the same chemical potentials.  From the two-body discussion in Chapter \ref{sec:intro:twobody} (Eqs. \ref{eq:intro:Krr}, \ref{eq:intro:kappa}, \ref{eq:intro:deltaC}), the characteristic width of the resonance $\delta_{c}$ is zero in this case, which indicates that the resonance is always narrow no matter how dilute the system is.

\begin{figure}[hhtb]
	\centering
	         \subfloat[$E_{F}<\tilde\delta$]{\label{fig:narrowFR:aboveSea}\includegraphics[width=.2\textwidth]{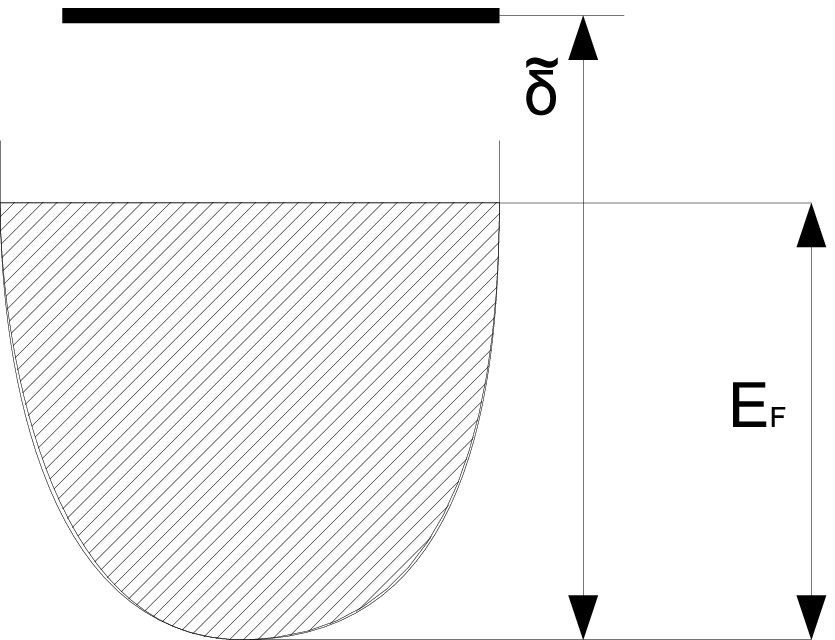}}\quad
		\subfloat[$0<\tilde\delta<E_{F}$]{\includegraphics[width=.30\textwidth]{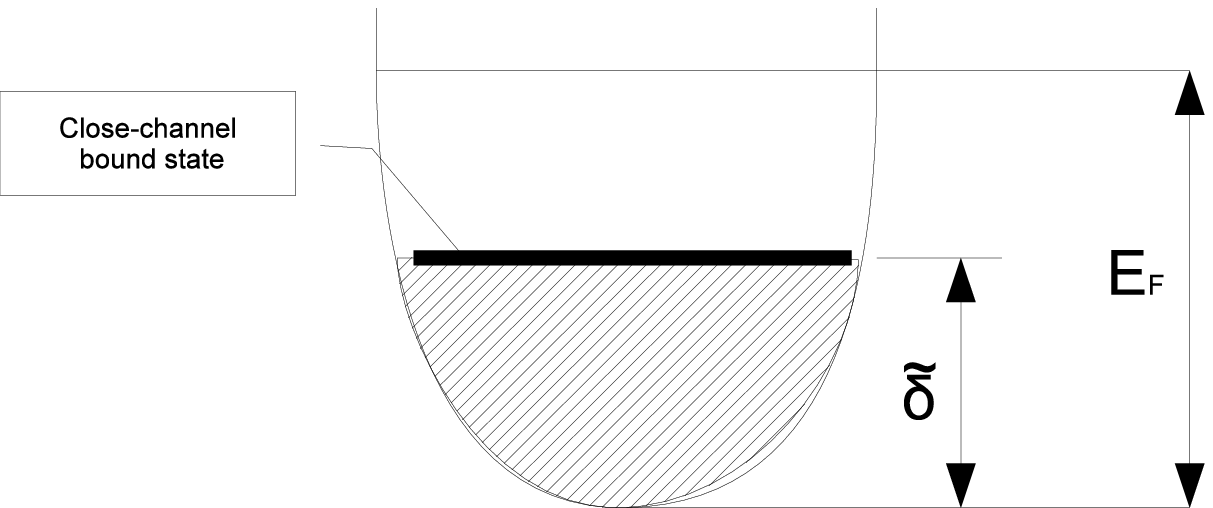}\label{fig:narrowFR:inSea}}\quad
		\subfloat[$\tilde\delta<0$]{\label{fig:narrowFR:belowSea}\includegraphics[width=.20\textwidth]{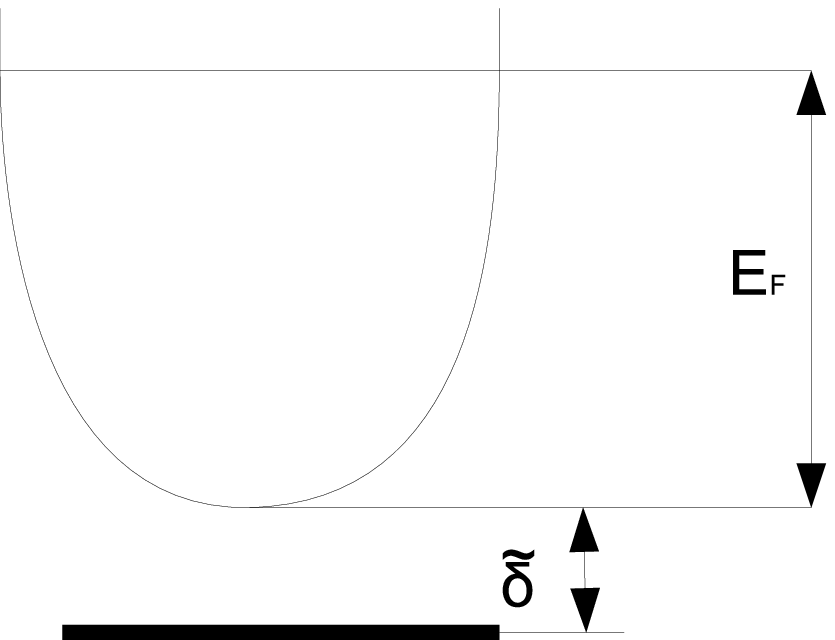}}
	\caption{Extremely narrow resonance\label{fig:narrowFR}}
	\small{The shaded area is occupied by atoms. }
	%\parbox{0.7\textwidth}{\small{  In Fig. \subref{fig:narrowFR:inSea} chemical potential would be close to the closed-channel bound state level (besides small shift due to the open-channel intra-channel coupling) and the ``Fermi sea'' above is empty. }}
\end{figure}
Here we assume that the atoms in the open-channel maintains a relatively clear Fermi surface, like the Fermi liquid or the BCS.  The many body situation is  quite straightforward. In general, three different situations exist, as Figs. \ref{fig:narrowFR:aboveSea}, \ref{fig:narrowFR:inSea} and \ref{fig:narrowFR:belowSea}.  When the closed-channel bound state level is above the Fermi sea ($E_{F}<\tilde\delta$, Fig. \ref{fig:narrowFR:aboveSea}), all atoms are in the open-channel and no atoms are in the closed-channel.  Therefore, it is a genuine  single-channel problem (only open-channel) with the original bare open-channel interaction.   There is no need to considering the inter-channel Pauli exclusion.  When the closed-channel bound state level is below the Fermi sea ($\tilde\delta<0$, Fig. \ref{fig:narrowFR:belowSea}), all atoms are in the closed-channel and no atoms are in the open-channel.  It is a simple di-atom molecule gas. There is no need to consider the inter-channel Pauli exclusion either.  The interesting situation is when the closed-channel bound state level is in the Fermi sea  ($0<\tilde\delta<E_{F}$, Fig. \ref{fig:narrowFR:inSea}).  The Fermi sea is then filled from the bottom (zero energy) all the way to the closed-channel bound state level in the open-channel.  Then all the rest atoms go into the closed-channel.  We do need to consider the inter-channel Pauli exclusion because atoms exist in both channel.

We can imagine the two channels are decoupled and each has $N^{(o)}$ ($N^{(c)}$ atoms.  It is not hard to find the wave function in each case.  The inter-channel Pauli-exclusion effect should be in the order of the overlap of these two wave functions. In the open-channel, the momentum space is filled  with occupation $1$ from the bottom up to the chemical potential $\mu$ (or $\tilde\delta$). In the closed-channel, as we discussed in the introduction, the occupation is roughly $1/E_b$ at the low momentum ($\lesssim{}E_F$), where $E_b$ is the binding energy of the closed-channel bound state.   So this effect should be roughly equal $E_F/{E_b}$.

\section{Model set-up and the Hubbard-Stratonovich transformation}

One way to approach this problem is to extend BCS ansatz and then find the set of parameters that optimizes the free energy.  We can write down a  three species BCS ansatz as 
\begin{equation}\label{eq:pathInt2:ansatz}
\ket{\Psi}=\prod_\vk\br{u_\vk+v_\vk{}a^\dg_\vk{}b^\dg_{-\vk}+w_\vk{}a^\dg_\vk{}c^\dg_{-\vk}}\ket{0}
\end{equation} 
with normalization, $\abs{u_\vk}^2+\abs{v_\vk}^2+\abs{w_\vk}^2=1$.   Here ``$a$'' is the common species of two channels. $(a,b)$ is the open channel and $(a,c)$ is the closed channel.  $\ket{0}$ is the vacuum state of fermions. This ansatz can be derived from a many-body coherent state of the two-body state as we did in the single-channel (Eq. \ref{eq:intro:BCScoherent}), 
\begin{equation}\label{eq:pathInt2:ansatzCoh}
\ket{\Psi}\propto\exp{}\mbr{\sum_{\vk}(\phi^{(ab)}_\vk{}a^\dg_\vk{}b^\dg_{-\vk}+\phi^{(ac)}_\vk{}a^\dg_\vk{}c^\dg_{-\vk})}\ket{0}
\end{equation} 
$\phi^{(ab)}$ and $\phi^{(ac)}$ are related to $u_\vk$, $v_\vk$ and $w_\vk$ as  
\begin{gather}
\phi_{\vk}^{(ab)}=\frac{v_{\vk}}{u_{\vk}}\\
\phi_{\vk}^{(ac)}=\frac{w_{\vk}}{u_{\vk}}\label{eq:pathInt2:phiWU}
\end{gather}
 We can then proceed to find the set of $(u_{\vk},v_{\vk},w_{\vk})$ that optimizes the free energy. This method offers good intuition of the state  with an explicit connection to two-body wave functions, $\phi^{(ab)}$ and $\phi^{(ac)}$, .  However,  it is not easy to find these parameters from optimization process. Furthermore, it is hard to extend this method beyond the mean-field to study phenomena such as collective modes.  We briefly discuss this approach in Appendix \ref{ch:mean}. Instead, in this chapter, we use the path integral method, which turns out to be  more convenient for this problem.   The Hubbard-Stratonovich transformation provides a powerful tool for studying non-trivial degrees of freedom (order parameters) in the system.  It is more or less equivalent to other approaches at the mean-filed level. But it has great advantage to be easily extended to explore the fluctuation over the mean-field result.  

For a two-channel problem, we write down the Hamiltonian \footnote{Here and hereafter in the chapter, we take $\hbar=1$.} as
\begin{equation}\label{eq:pathInt2:ham2}
\begin{split}
H-\mu{N}&=\int{d^{d}r}\bigg\{\sum_{j=(a,b,c)}\bar\psi_{j}\mbr{\nth{2m}(-i\nabla)^{2}-\mu+\eta_{j}}\psi_{j}\\
	&\qquad-U\bar\psi_{a}(r)\bar\psi_{b}(r)\psi_{b}(r)\psi_{a}(r)-V\bar\psi_{a}(r)\bar\psi_{c}(r)\psi_{c}(r)\psi_{a}(r)\\
	&\qquad-\mbr{Y\bar\psi_{a}(r)\bar\psi_{b}(r)\psi_{c}(r)\psi_{a}(r)+h.c.}
	\bigg\}
\end{split}
\end{equation}
Here $\eta_{j}$ is the Zeeman energy of the specific hyperfine species.  $a,b,c$ stands for the three hyperfine species defined as  same as in the ansatz approach (Eq. \ref{eq:pathInt2:ansatz}).  All the interactions ($U$, $V$, $Y$), are contact type, this simplifies the Hubbard-Stratonovich transformation considerably.  It is plausible as we only study low-energy phenomena for the short-range potential. 

A contact interaction is  flat in momentum space. In other words, it pairs not only zero center-of-mass momentum pairs, but also finite center-of-mass momentum pairs. This is different from the paring interaction used in the original BCS work\cite{BCS}, which only pairs zero center-of-mass momentum, i.e. opposite-momentum, pairs.  Nevertheless, the saddle point (mean-field) solution settles at the zero center-of-mass momentum pairing; therefore the mean-field solution coincides with that of  variation method derived from pairing  only opposite momentum atoms.  On the other hand, the collective mode (fluctuation of order parameters) emerges naturally from the included non-zero center-of-mass momentum interaction.  We do not need to introduce more general interaction terms beyond the simple opposite momentum pairing (e.g. the Coulomb interaction in \cite{AndersonBCS}) to study collective modes.   

Three different types of Hilbert spaces are used in this chapter.   The first one is the  (infinite dimension) coordinate space and its reciprocal momentum space.   The second one is the  (3 dimension) space of hyperfine spin, $a,\,b,\,c$.   These two are atom-based.  The full many-body Hilbert space is $N$-power (also direct-product type) of the direct product of these two.  As discussed previously, in some cases, instead of a 3-dimension hyperfine spin space, we use the third type of spaces, open- and closed-channel, $(a,b),\,(a,c)$ (2 dimension).  Strictly speaking, this one is just a subspace of the direct product of a pair hyperfine spin spaces.  Nevertheless,  it is sufficient within our model because only these two combinations (channels) of a pair are considered.    In principle, fermion fields $\psi$ and $\bar\psi$ (and $\Delta$ and $\bar\Delta$ defined according to $\psi$ and $\bar\psi$),  both Grassmann numbers, are independent to each other  and  not related as complex conjugate\footnote{Complex conjugate is not a well-defined concept for Grassmann algebra}.  This is marked by using ``$\bar{\;}$''(bar) instead of the normally used ``$^{\dg}$''(dagger) sign. In this chapter, a $3\times3$ matrix always refer to hyperfine spins; while a $2\times2$ matrix always refers to  open- and closed-channel.  And $^{}\dg$ sign  is reserved only for hermitian conjugate of these two types of matrices. 
%We can introduce a unitary transformation $Q$ (mixing two channels) to diagonalize the interaction matrix into diagonal matrix $A$.
%\begin{equation}
%\begin{split}
%Q^{\dg}AQ=\mtrx{U&Y\\Y^{*}&V}\equiv{}\tilde{U}\\
%\tilde{U}\equiv\mtrx{U&Y\\Y^{*}&V}=Q\mtrx{A_{11}&0\\0&A_{22}}Q^{\dg}
%\end{split}
%\end{equation}

Introduce two channels into the vector form.    $(\psi\psi)$  is a column vector and $(\bar\psi\bar\psi)$ is a row vector.
\begin{equation*}
(\bar\psi\bar\psi)=\mtrx{\bar\psi_{a}\bar\psi_{b}&\bar\psi_{a}\bar\psi_{c}}
\qquad(\psi\psi)=\mtrx{\psi_{b}\psi_{a}\\\psi_{c}\psi_{a}}
\end{equation*}
The two-body interaction can then be written as a ($2\times2$)  hermitian  matrix  $\tilde{U}$ in the channel space
\begin{equation}
\tilde{U}\equiv{}\mtrx{U&Y\\Y^{*}&V}
\end{equation}
We can now write the Hamiltonian in a more compact form
\begin{equation}
H=\int{d^{d}r}\bigg\{\sum_{j={a,b,c}}\bar\psi_{j}\mbr{\nth{2m}(-i\nabla)^{2}-\mu+\eta_{j}}\psi_{j}
 	-(\bar\psi\bar\psi)\tilde{U}(\psi\psi)
\end{equation}
The finite-temperature action is 
\begin{equation}\label{eq:pathInt2:actionFermi}
S(\bar\psi,\psi)=\int^{\beta}_{0}d\tau\int{d^{d}r}\mbr{\sum_{j}\bar\psi_{j}(\partial_\tau-\nth{2m}\nabla^{2}-\mu+\eta_{j})\psi_{j}
-(\bar\psi\bar\psi)\tilde{U}(\psi\psi)}
\end{equation}

Similar as in Chapter \ref{sec:pathInt}, we can perform the Hubbard-Stratonovich transformation here.   Introduce auxiliary fields (functional variables), $(\Delta_{1},\Delta_{2})$, as a 2-component vector   and start from the ``fat identity'', where  all the integral constant is absorbed into the measure of functional integral of $\bigD(\Delta,\bar\Delta)$ \cite{Altland}.
\begin{equation}\label{eq:pathInt2:identity}
1=\int{\bigD(\Delta,\bar\Delta)}\exp(-\int{dx}\Delta^{\dg}\tilde{U}^{-1}\Delta)
\end{equation}
\[
\Delta^{\dg}=(\bar\Delta_{1},\bar\Delta_{2})\qquad\Delta=\begin{pmatrix}\Delta_{1}\\\Delta_{2}\end{pmatrix}
\]
here $x$ is four-coordinate, $(\vr, \tau)$,  $\int{dx}=\int^{\beta}_{0}d\tau\int{d^{d}\vr}$. 

We can make a shift in $\Delta$
\begin{equation}\label{eq:pathInt2:DeltaPhi}
\Delta\longrightarrow\Delta-\tilde{U}(\psi\psi)
\end{equation}
Write it explicitly into the matrix form
\begin{equation*}
\mtrx{\Delta_{1}\\\Delta_{2}}\longrightarrow
	\mtrx{\Delta_{1}\\\Delta_{2}}-
	\mtrx{U&Y\\Y^{*}&V}
	\mtrx{\psi_{b}\psi_{a}\\\psi_{c}\psi_{a}}
\end{equation*}
\begin{equation*}
\mtrx{\bar\Delta_{1},\bar\Delta_{2}}\longrightarrow
	\mtrx{\bar\Delta_{1},\bar\Delta_{2}}-
	\mtrx{\bar\psi_{a}\bar\psi_{b}&\bar\psi_{a}\bar\psi_{c}}
	\mtrx{U&Y^{*}\\Y^{}&V}
\end{equation*}
First of all, notice that the new auxiliary field $\Delta$ is related to the mixture of two channels by a $2\times2$ matrix $\tilde{U}$.
Also note that  $\bar{\Delta}_{i}$ is not the complex conjugate of $\Delta_{i}$ in general as they are related to Grassmann fields $\psi\psi$ and $\bar\psi\bar\psi$ which are not complex conjugate to each other. But it can be verified later that, to the  mean-field (saddle point) level as well as to the simple  (phase-fluctuation) Gaussian level about collective modes, $\bar{\Delta}_{i}$ and ${\Delta}_{i}$ are indeed complex conjugate (and real for saddle point).  Hence, for the simplicity, we will treat them as such hereinafter and use $\bar\Delta$ and $\Delta^{*}$ more or less arbitrarily.  The other point to notice is that, $\Delta(\vr,\tau)$ carries the the same coordinates as $\psi(\vr,\tau)\psi(\vr,\tau)$ (or frequency-momentum coordinates in the reciprocal space) because there is only the contact interaction.  Now the ``fat identity'' Eq. \ref{eq:pathInt2:identity} becomes 
\begin{equation}
1=\int{\bigD(\Delta_{j},\bar\Delta_{j})}\exp\big\{-\int{dx}
	[\Delta^{\dg}\tilde{U}^{-1}\Delta-(\bar\psi\bar\psi)\Delta-\bar\Delta(\psi\psi)+(\bar\psi\bar\psi)\tilde{U}(\psi\psi)]\big\}
\end{equation}
The above equation use the fact $\tilde{U}$ is hermitian, so $\tilde{U}^{\dg}\tilde{U}^{-1}=\tilde{U}^{-1}\tilde{U}=I$.
It can be rearranged as 
\begin{equation}
\exp[\int{dx}(\bar\psi\bar\psi)\tilde{U}(\psi\psi)]
=\int{\bigD(\Delta,\bar\Delta)}\exp\big\{-\int{dx}
	[\Delta^{\dg}\tilde{U}^{-1}\Delta-(\bar\psi\bar\psi)\Delta-\bar\Delta{}(\psi\psi)]\big\}
\end{equation}
This is now ready to be applied to the original action in Eq. \ref{eq:pathInt2:actionFermi}, 
\begin{equation}\label{eq:pathInt2:actionMix}
S_{\tau}(\bar\Delta,\Delta,\bar\psi_{i},\psi_{i})=\int^{\beta}_{0}d\tau\int{d^{d}r}\bbr{\sum_{j}\bar\psi_{j}(\partial_\tau-\nth{2m}\nabla^{2}-\mu+\eta_{j})\psi_{j}
+[\Delta^{\dg}\tilde{U}^{-1}\Delta-(\bar\psi\bar\psi)\Delta-\bar\Delta{}(\psi\psi)]}
\end{equation}
We can introduce  a spinor similar to the Nambu spinor representation in the single-channel superconductivity.  
\begin{equation}
\bar\Psi=\mtrx{\bar\psi_{a}&\psi_{b}&\psi_{c}}\qquad\Psi=\mtrx{\psi_{a}\\\bar\psi_{b}\\\bar\psi_{c}}
\end{equation}
The action can then be rewritten in a more compact form with respect to $\Psi$ and $\bar\Psi$
\begin{equation}\label{eq:pathInt2:actionMixCompact}
S(\bar\Delta,\Delta,\bar\psi_{i},\psi_{i})=\int^{\beta}_{0}d\tau\int{d^{d}r}
	\mbr{\Delta^{\dg}\tilde{U}^{-1}\Delta-\bar\Psi\mathcal{G}^{-1}\Psi}
\end{equation}
where 
\begin{equation}
\mathcal{G}^{-1}=
\begin{pmatrix}\label{eq:pathInt2:nGDelta}
-\partial_{\tau}+\nth{2m}\nabla^{2}+\mu-\eta_{a}&\Delta_{1}&\Delta_{2}\\
\bar\Delta_{1}&-\partial_{\tau}-\nth{2m}\nabla^{2}-\mu+\eta_{b}&0\\
\bar\Delta_{2}&0&-\partial_{\tau}-\nth{2m}\nabla^{2}-\mu+\eta_{c}
\end{pmatrix}
\end{equation}
Rewriting  $\nG$ in the frequency-momentum space, we find $\nG$   decoupled in frequency and momentum. 

\begin{equation}\label{eq:pathInt2:nGDeltaK}
\mathcal{G}^{-1}=
\begin{pmatrix}
i\omega_{n}-\xi_{k}-\eta_{a}&\Delta_{1}&\Delta_{2}\\
\bar\Delta_{1}&i\omega_{n}+\xi_{k}+\eta_{b}&0\\
\bar\Delta_{2}&0&i\omega_{n}+\xi_{k}+\eta_{c}
\end{pmatrix}
\end{equation}
here\footnote{In principle, there are two chemical potentials, $\mu_{a}$ and $\mu_{b,c}$ because $a$ does not convert to $b$ or $c$.  But we omit this for simplicity and absorb all the difference into $\eta_{i}$.} $\xi_{k}=\nth{2m}k^{2}-\mu$. The diagonal elements of the second column/row (${b}$) and the third column/row (${c}$) corresponding to negative energy, because of the particular opposite choice in the  spinor ($\bar\Psi_{2}\Psi_{2}=\psi_{b}\bar\psi_{b}$ and $\bar\Psi_{3}\Psi_{3}=\psi_{c}\bar\psi_{c}$).  The non-diagonal elements mix $\psi_{a}$ with $\bar{\psi}_{b,c}$  and therefore lead to a number-non-conserved theory.  

%$\nG$ can be further simplified by introducing mixture within two channels.
%\begin{equation}\label{eq:pathInt2:Ddef}
%D\equiv\mtrx{D_{1}\\D_{2}}=Q^{\dg}\Delta
%\end{equation}
%\begin{equation*}
%\bar\Delta=\bar{\Delta}\,Q^{\dg}\qquad\Delta=Q\,D
%\end{equation*}
%It is not difficult to see that $D$ actually describe the off-diagonal coupling in fermionic field $\Psi$ and is the counterpart of order parameter, gap, in single-channel problem instead of $\Delta$ here.  We will see it  more clearly when  discussing mean-field solution. 

If we in addition assume $\eta_{a}=\eta_{b}=0$, $\eta_{c}=\eta$ (i.e., use $\eta$ as the absolute Zeeman energy difference of two channels)  , in frequency-momentum space, 
\begin{equation}\label{eq:pathInt2:nG}
\mathcal{G}^{-1}=i\omega_{n}I-
\begin{pmatrix}
\xi_{k}&-\Delta_{1}&-\Delta_{2}\\
-\bar{\Delta}_{1}&-\xi_{k}&0\\
-\bar{\Delta}_{2}&0&-(\xi_{k}+\eta)
\end{pmatrix}
\end{equation}

%and the first term $\bar\Delta{}A^{-1}\Delta$ in action Eq. (\ref{eq:pathInt2:actionMix}) becomes
%\[
%\bar\Delta{}A^{-1}\Delta=\bar{\Delta}Q^{\dg}A^{-1}QD=\bar{\Delta}\tilde{U}^{-1}D
%\]

%We can then change the functional variable into $D(\bar{\Delta})$ 
%\begin{equation}\label{eq:pathInt2:actionMixD}
%S(\bar{\Delta},D,\bar\psi_{i},\psi_{i})=\int^{\beta}_{0}d\tau\int{d^{d}r}
%	\mbr{\bar{\Delta}\tilde{U}^{-1}D-\bar\Psi\mathcal{G}^{-1}\Psi}
%\end{equation}
The action in Eq. \ref{eq:pathInt2:actionMixCompact} is  bilinear to $\Psi$ and we can integrate  out $\Psi$ ($\bar\Psi$) formally
\begin{equation}\label{eq:pathInt2:actionD}
S(\bar{\Delta},\Delta)=\int{dx}\br{\bar{\Delta}\tilde{U}^{-1}\Delta-\tr\ln\nG}
\end{equation}
Note that at this stage $\Delta(\vr,\tau)$  is not necessarily homogeneous in space or pseudo-time as in the mean-field result.

\section{Diagonalization of the Green's function\label{sec:diagonalGreen}}
Eq. \eqref{eq:pathInt2:actionD} looks fairly simple and compact.  Nevertheless, it has all the physics in it and is not as simple as it looks.  The major problem comes from the term $\tr\ln\nG$, which includes  logarithm and trace over an infinite-dimension matrix.   All these operations are fairly straight-forward if we can diagonalize the  Green's function (or its inverse) in a proper basis.      It is not hard to see $\nG$ is already decoupled in the frequency-momentum space (Eqs.  \ref{eq:pathInt2:nGDeltaK}, \ref{eq:pathInt2:nG}).  It is however mixed in the $3\times3$ hyperfine-spin space.  The rest of this section is dedicated to diagonalize this $3\times3$ matrix in the hyperfine-species space.  Note that in principle, the following discussion is not limited for constant $\Delta$, but also applies to inhomogeneous $\Delta(\omega_{n},\vk)$ as well because the $3\times3$ hyperfine space is independent to the frequency-momentum space.   Here we use the approach in Sec. \ref{sec:diagonalizeGreen1} to diagonalize it.   
In current problem, we need to diagonalize a $3\times3$ matrix (Eq. \ref{eq:pathInt2:nG}), in other words, we need to figure out the Bogoliubov canonical transformation over which the Hamiltonian/action is diagonalized.   Eigen-problem of the $3\times3$ matrix involves solving a cubic equation. An exact solution exists in principle.  However,  it offers little intuition to write down the exact result. Instead,  the spectrum from the broad-resonance, where the only effect of the closed-channel is to modify the effective interaction of the open-channel, serves a reasonable lowest order approximation. We proceed to find the next order of correction over it (See Appendix \ref{sec:diagonalize} and \ref{sec:pathApp:consistency}).

%Within the assumption that the spectrum  deviates not too much from the na\"{i}ve broad-resonance solution, 
We  can break down the unitary transformation into two steps $T$ and $L$. 
\begin{equation}\label{eq:pathInt2:B}
B_{\omega_{n},\vk}=L_k^{\dg}T_k^{\dg}G_{\omega_{n},\vk}^{-1}T_kL_k
\end{equation} 
Here $B_{k}$ is the diagonal matrix; $T$ and $L$ are both unitary transformation.  We take $T$ as the canonical transformation at the broad resonance, i.e., when we can ignore the inter-channel Pauli exclusion. 
\begin{equation}\label{eq:pathInt2:T}
T_k=\mtrx{u_k&v_k&0\\-v_k&u_k&0\\0&0&1}
\end{equation}	
where $u_{k}$ and $v_{k}$ are defined in a similar fashion as in the single-channel BCS  problem
\begin{gather}
v_{\vk}^{2}\equiv1-u_{\vk}^{2}\equiv\nth{2}\br{1-\frac{\xi_{\vk}}{E_{\vk}}}\\
E_{\vk}\equiv(\xi_{\vk}^{2}+\Delta_{1}^{2})^{1/2}
\end{gather}
Note that here $v_{\vk}^{2}$  does not carry the physical meaning of the occupation number of the (open-channel) atoms, and $E_{\vk}$   does not stand for fermionic excitation spectrum as in Chapter \ref{sec:intro:1channel} or Appendix \ref{ch:mean}.   They carry such meaning only  at the broad resonance. In the narrow resonance, as we currently discuss, they are the zeroth order approximates of such quantities. 

In the broad resonance, the closed-channel can be integrated out at the two-body level and  only the BCS pairing in the open channel needs to be considered at the many-body level. Matrix $T$, however, is enough to diagonalize $G^{-1}$ and $L$ is simply an identity matrix.  %Here we will try to approximate it to the first order correction due to Pauli exclusion between two-channel.  
In the narrow resonance, $T$ cannot diagonalize $G^{-1}$ because of the inter-channel Pauli exclusion between channels.  Consequently, $L$ stands the extra correction due to Pauli exclusion in the canonical transformation. Apply $T$ onto $G^{-1}$, we have 
\begin{equation}\label{eq:pathInt2:G2}
T_k^{\dg}G_{\omega_{n},\vk}^{-1}T_k=i\omega_nI+\mtrx{-E_k&0&u_k\Delta_2\\0&+E_k&v_k\Delta_2\\u_k\Delta_2&v_k\Delta_2&+\xi_k+\eta}
\end{equation}
We regard the off-diagonal elements as perturbation because  we  only seek the solution around the BCS wave function ($T$ transform). 
Introduce a dimensionless scale $\zeta$,
\begin{equation}\label{eq:pathInt2:zetaDef}
\boxed{\zeta=\frac{\Delta_{2}^{2}}{\Delta_{1}\eta}}
\end{equation}
Here both $\Delta_{1}$ and $\Delta_{2}$ are their mean-field (saddle point) values.  It can be verified that $\zeta\ll1$ (See Appendix \ref{sec:pathApp:consistency}).  
This matrix can then be diagonalized with  the unitary transformation $L_{\vk}$ within the first order of $\zeta$  (see Appendix \ref{sec:diagonalize} for details of calculation.)
\begin{equation}\label{eq:pathInt2:Bapprox}
\begin{split}
B_{\omega_{n},\vk}&=i\omega_{n}I-
	\mtrx{E_{1}{}_{\vk}&0&0\\0&-E_{2}{}_{\vk}&0\\0&0&-E_{3}{}_{\vk}}\\
%	&\approx{}i\omega_{n}I-
%	\mtrx{E_{\vk}+{}u_{\vk}^{2}\zeta&0&0\\
%	0&-\br{E_{\vk}-{}v_{\vk}^{2}\zeta}&0\\0&0&-\br{\xi_{\vk}+\eta-\frac{\zeta}{2}}}
%	&=
%	\mtrx{i\omega_{n}-E_{\vk}&0&0\\0&i\omega_{n}+E_{\vk}&0\\0&0&i\omega_{n}+\eta}
%	+\mtrx{-\frac{D_{1}^{2}}{\eta}&0&0\\0&-\frac{D_{2}^{2}}{\eta}&0\\0&0&+\frac{D_{1}^{2}+D_{2}^{2}}{2\eta}}\\
%	&\equiv{}B^{(0)}_{\vk}+B^{(1)}_{\vk}
\end{split}	
\end{equation}
The dispersion spectrum of fermions is
\begin{align}\label{eq:pathInt2:xiExpand}
E_{1\vk}&\equiv{}E_{\vk}+\gamma_{1\vk}\approx{}E_{\vk}+\frac{\Delta_{2}^{2}u_{\vk}^{2}}{\xi_{\vk}+\eta}
\approx{}E_{\vk}+u_{\vk}^{2}\zeta\frac{\eta}{\xi_{\vk}+\eta}\Delta_{1}
\approx{}E_{\vk}+u_{\vk}^{2}\Delta_{1}\zeta\\
E_{2\vk}&\equiv{}E_{\vk}+\gamma_{2\vk}\approx{}E_{\vk}-\frac{\Delta_{2}^{2}v_{\vk}^{2}}{\xi_{\vk}+\eta}
\approx{}E_{\vk}-v_{\vk}^{2}\zeta\frac{\eta}{\xi_{\vk}+\eta}\Delta_{1}
\approx{}E_{\vk}-v_{\vk}^{2}\Delta_{1}\zeta\label{eq:pathInt2:xiExpand2}\\
E_{3\vk}&\equiv{}\xi_{\vk}+\eta+\gamma_{3\vk}\approx{}\xi_{\vk}+\eta-\frac{\Delta_{2}^{2}}{2(\xi_{\vk}+\eta)}
\approx{}\epsilon_{\vk}+\eta-\frac{\zeta}{2}\frac{\eta}{\xi_{\vk}+\eta}\Delta_{1}
\approx{}\epsilon_{\vk}+\eta-\frac{\zeta}{2}\Delta_{1}
\label{eq:pathInt2:xiExpand3}
\end{align}
The last step of each equations is valid only for low momentum ($\sim{k_{F}}$) where $\eta\gg\xi_{\vk}$. 
Here we choose the sign convention to make  $E_{1,2,3}$  positive in their zeroth order.  Similarly as in Eq.  \ref{eq:pathInt2:nGDeltaK}, the second and third diagonal elements are negative because in the spinor representation, we choose $\bar{\psi}_{b}$ and $\bar{\psi}_{c}$ for $\Psi$, which gives extra negative signs for quantity such as $\bar\Psi\Psi$ in Eq. \ref{eq:pathInt2:actionMixCompact}.\footnote{Recall that $\psi_{i}$ and $\bar\psi_{i}$ are Grassmann fields; thereore $\psi_{i}\bar\psi_{i}=-\bar\psi_{i}\psi_{i}$. }  These negative signs disappear when we restore them to the normal order $\bar\psi\psi$ in Sec. \ref{sec:pathInt2:bog}.  %Therefore, the Fermi sea of the redefined Bogoliubov quasiparticle is always filled up without extra particle-hole remapping.  

One interesting  feature of this solution  is that the corrections do not disappear for zero inter-channel coupling, $Y=0$, which we discussed qualitatively in Sec. \ref{sec:pathInt2:extremelyNR}.     This is because the corrections are due to the inter-channel Pauli exclusion between two channels, which continues to exist even when there is no inter-channel coupling.  In fact, the inter-channel coupling, $Y$, contributing mostly energetically, modifies the effective interaction in the open-channel and affect the open-channel order parameters,  $\Delta_{1}$, greatly.  By taking the broad resonance result as the zeroth order, we have taken effects of $Y$ into consideration.  On the other hand, the inter-channel Pauli exclusion or statistics is left out as the higher order correction as illustrated above. Furthermore, from Appendix \ref{sec:pathApp:consistency} (Eq. \ref{eq:pathApp:zetaEs}), we know $\zeta\sim\frac{k_F}{\kappa}$, it is just the square root of our estimate  $E_F/E_b$ ($E_b=\hbar^2\kappa^2/2m$) in Sec. \ref{sec:pathInt2:extremelyNR}.

The extra factor of unitary transformation is 
\begin{equation}\label{eq:pathInt2:L1}
L_{\vk}\approx{}I+
\mtrx{0&-\frac{\Delta_{1}{}\Delta_{2}{}}{4E^{2}_{\vk}}&u_{\vk}\\
\frac{\Delta_{1}{}\Delta_{2}{}}{4E^{2}_{\vk}}&0&v_{\vk}\\
-u_{\vk}&-v_{\vk}&0
}\frac{\Delta_{2}{}}{\eta}
\equiv{}I+\delta_{k}\qquad
L^{\dg}_{\vk}=I-\delta_{\vk}
\end{equation}
Here we use $u_{\vk}v_{\vk}=\Delta_{1}/2E_{\vk}$.    Note that $L$ and $L^{\dg}$ are unitary only to the first order of $\Delta_{i}/\eta$.
Now it is easy to express the Green's function as
\begin{equation}
G_{\omega_{n},\vk}=T_{\vk}L_{\vk}B_{\omega_{n},\vk}^{-1}L_{\vk}^{\dg}T_{\vk}^{\dg}
\end{equation}
This is ready to be expanded over the perturbation in order of  $\zeta$ or $\Delta_{i}/\eta$.  It is easy to see that all $\omega_{n}$ dependence concentrates on $B_{\omega_{n},\vk}$, which is linear in $\omega_{n}$  and simplifies the Matsubara frequency summation considerably.   
\begin{subequations}\label{eq:pathInt2:Gexpand}
\begin{gather}
G_{\omega_{n},\vk}\approx{}T_{\vk}B_{\omega_{n},\vk}^{-1}T_{\vk}^{\dg}+T_{\vk}\delta_{\vk}B_{\omega_{n},\vk}^{-1}T_{\vk}^{\dg}
	-T_{\vk}B_{\omega_{n},\vk}^{-1}\delta_{\vk}T_{\vk}^{\dg}
	\equiv{}G_{\omega_{n},\vk}^{(0)}+G_{\omega_{n},\vk}^{(1)}\\
	G_{\omega_{n},\vk}^{(0)}=T_{\vk}B_{\omega_{n},\vk}^{-1}T_{\vk}^{\dg}\\
	G_{\omega_{n},\vk}^{(1)}=T_{\vk}\delta_{\vk}B_{\omega_{n},\vk}^{-1}T_{\vk}^{\dg}
	-T_{\vk}B_{\omega_{n},\vk}^{-1}\delta_{\vk}T_{\vk}^{\dg}
\end{gather}
\end{subequations}

%We only need to keep the zeroth order term for $B_{\vk}^{-1}$ for first order expansion.  
\section{Mean field equations \label{sec:pathInt2:meanfield}}
Use the same techniques for derivatives as Eq. (\ref{eq:pathInt:diffTr}), we derive two saddle point equations for $\Delta_{1}$ and $\Delta_{2}$ from Eq. \eqref{eq:pathInt2:actionD},
 \begin{align}
\frac{\delta}{\delta{}\Delta_{1}}:&\qquad&
(\tilde{U}^{-1})_{11}\bar{\Delta}_{1}+(\tilde{U}^{-1})_{21}\bar{\Delta}_{2}-\tr\mbr{{G_{0}}\cdot\cmtrx{0&1&0\\0&0&0\\0&0&0}}=0
\label{eq:pathInt2:mf01}\\
\frac{\delta}{\delta{}\Delta_{2}}:&\qquad&
(\tilde{U}^{-1})_{12}\bar{\Delta}_{1}+(\tilde{U}^{-1})_{22}\bar{\Delta}_{2}-\tr\mbr{{G_{0}}\cdot\cmtrx{0&0&1\\0&0&0\\0&0&0}}=0
\label{eq:pathInt2:mf02}
 \end{align}

 Taking $\Delta$ as real and constant,\footnote{\label{foot:pathInt2:real}When $Y$ is not real, $\Delta_{1}$ and $\Delta_{2}$ cannot be both real even at the mean field level.  Nevertheless, we can require one  real, then the other will have a phase just to compensate the phase in $Y$.  The final conclusion can be verified  still valid.   }
  %\footnote{Actually $D_{2}{_{\vk}}$ cannot be constant at high momentum.  However, for the momentum we are interested, i.e. the momentum lower or in the order of Fermi momentum, it slowly varies.  Therefore  it is reasonable to take it as constant.}     
  we can find the mean field result. Eq. (\ref{eq:pathInt2:nG}) can be inverted to get $G$.  The inversion is quite tedious, but fortunately, we only need two elements of the $G$ matrix ($G_{0\, (21)}$ and $G_{0 \,(31)}$).  The final mean-field equations are (For simplicity,  both $\Delta_{i}$'s are taken as real. Please see Appendix \ref{sec:pathInt2:deriveMF} for detail) 
  \begin{equation}\label{eq:pathInt2:mf}
\mtrx{\Delta_1\\\Delta_2}=\mtrx{U&Y\\Y^{*}&V}\sum_{\vk}\mtrx{h_{1\vk}\\h_{2\vk}}
\end{equation}
  where 
  \begin{gather}
  h_{1\vk}=\av{\psi_{a,-{\vk}}\psi_{b,+{\vk}}}
  =\Delta_{1}\frac{E_{1\,\vk}+\xi_{\vk}+\eta}{(E_{1\,\vk}+E_{2\,\vk})(E_{1\,\vk}+E_{3\,\vk})}\label{eq:pathInt2:h1}\\
  h_{2\vk}=\av{\psi_{a,-{\vk}}\psi_{c,+{\vk}}}
  =\Delta_{2}\frac{E_{1\,\vk}+\xi_{\vk}}{(E_{1\,\vk}+E_{2\,\vk})(E_{1\,\vk}+E_{3\,\vk})}\label{eq:pathInt2:h2}
  \end{gather}

Comparing Eq. \ref{eq:pathInt2:mf} with the gap equation for the single-channel crossover, we  see that $\Delta_{1,2}$ are  direct counterpart of the order parameter $\Delta$ in the single-channel problem. 
$h_{1\vk}$ and $h_{2\vk}$ are the equal-time  expectation of the anomalous Green's function for the open- and closed-channel. On the other hand, they correspond the macroscopic eigen-function of the two-body density matrix as described by Zhang and Leggett\cite{ZhangThesis,shizhongUniv} (see Sec. \ref{sec:intro:as}). For the purpose of many calculations, they are just like the two-body wave function. And indeed, they coincide with the two-body wave function at high momentum.  We will reference them often as the ``two-body correlation'' in the following.  
If we simply take $E_{i}$ to the lowest order of $\zeta$, and ignore the closed-channel, it is easy to identify $h_{1\vk}\approx{\Delta_{1}}/(2E_{\vk})$ as the many-body wave function $F_{\vk}$ in the single channel BEC-BCS crossover problem.

At high-momentum, both $h_{1\vk}$ and $h_{2\vk}$ behave as $1/\epsilon_{\vk}$ which makes the summation (or the converted integral) diverges in 3D.  This divergence can be mitigated by setting a high-momentum cutoff in integral or recognizing the decay of interaction  in high momentum.  It is important to recognize that this divergence is not unique in many-body and exists in the two-body physics, where  high-momentum component of wave function is  $1/\epsilon_{\vk}$ asymptotically as well.  In the next section, we proceed to  remove divergence of summation in $h_{1\vk}$ and $h_{2\vk}$ by noting the same divergence in the two-body wave function and manipulating accordingly. 
%It is interesting to look at Eq. \ref{eq:pathInt2:h2} more carefully, in BCS side, $\mu\approx{}E_{F}$, at low momentum ($k<k_{F}$), $\xi_k<0$ and $h_{2\,\vk}$ is close to 0; at higher momentum where $\epsilon_{\vk}>\mu$, we have $E_{2\,\vk}\approx\xi_{\vk}$, $E_{3\,\vk}\approx\xi_{\vk}+\eta$.  All these lead to $  h_{2\vk}\approx\frac{D_2}{(2\epsilon+\eta)}$, which coincides with two-particle wave function for such range ( momentum $k$ smaller than inverse of potential range $1/a_{c}$ as well as  characteristic of closed-channel binding energy ``$\kappa$'', but larger than other scale, such as $k_{F}$, $1/a_{s}$).
%;  at very high momentum where $\epsilon_{\vk}>\eta, 1/a_c$, $D_2$ can no longer be treated as a constant and decays with energy, its specific form is determined by the specific short-range shape of potential. We expect the high-momentum normalization follows the middle-momentum normalization when comparing to two-body bound state ($D_2$ here is the normalization factor.  See sec \ref{sec:intro:as}).  Only in the low-momentum, it differs from the two-body wave function and that is where 
 
\subsection {Renormalization of the mean field equations\label{sec:pathInt2:ren}}
The mean-field equations (Eq. \ref{eq:pathInt2:mf}) can be rewritten as 
\begin{align}
\Delta_{1\vp}&=\sum_{\vk}{}U_{\vp\vk}h_{1\vk}+\sum_{\vk}{}Y_{\vp\vk}h_{2\vk}\label{eq:pathInt2:mfopen}\\
\Delta_{2\vp}&=\sum_{\vk}{}Y_{\vp\vk}h_{1\vk}+\sum_{\vk}{}V_{\vp\vk}h_{2\vk}\label{eq:pathInt2:mfclose}
\end{align}
The first thing to notice is that we restore the momentum dependence of the interaction as well as of $\Delta_{1\vp}$ and $\Delta_{2\vp}$.     Here $\Delta_{1\vp}$ and $\Delta_{2\vp}$ vary slowly in low momentum and  it is a good approximation to take them as constant if we only study the low momentum properties.  We introduce the momentum-dependence in order to gain proper convergence in summation in high momentum. As pointed out previously, both $h_{1\vk}$ and $h_{2\vk}$ approach $1/\epsilon_{\vk}$ in high momentum asymptotically.  This  makes  all summations diverge at high momentum when they are converted to integral in 3D if we take the interaction as contact and pull them out of the summation.  However, we note that the real potential is not the contact-type and decays in high momentum.  One way to remove the divergence is to  introduce  a cutoff at high-momentum while keeping the interaction coefficient constant.  However, an arbitrary cutoff is  undesirable for a theory.  Alternatively, we can remove the divergence by restoring the momentum dependence in the  interaction and recognizing its decay at high-momentum.    The problem here is that those microscopic bare interactions are hard to pin down.      More readily available and observable is the two-body low-energy effective scattering matrix, $T$, which effectively integrates out the high-momentum component of the bare interaction.             In the single channel problem, (Sec. \ref{sec:pathInt:meanfield}),  we  introduce a relation involving the two-body s-wave scattering length $a_{s}$, (or $T_{0}$), Eq. \eqref{eq:pathInt:as}, which has the similar divergence in high momentum.  We subtract it from the summation and the remaining difference does not diverge in high momentum.  A cutoff or bare momentum-dependent interaction is then no longer necessary.  In principle, we can do the same in two channels and introduce the two-channel equivalent of the s-wave scattering length, a $2\times2$ scattering amplitude matrix.  However, this $2\times2$ matrix is awkward in definition and not clear for the physical meaning.   Therefore, we adopt a two-step process with better physical intuition behind it.   

Besides the assumption of short-range potentials, we make another assumption:  the  two-body closed-channel bound-state at resonance, $\phi_{0}$, is much smaller in size ($a_{c}$) than the interparticle distance ($a_{0}$), although it is  larger than the potential range ($r_{c}$),    $r_{c}\ll{}a_{c}\ll{}a_{0}$.  The problem would be a genuine 3-species problem and requires other techniques if the closed-channel bound-state is about the size  or even larger than the interparticle distance. As discussed at the beginning of this chapter, this assumption guarantees that the closed-channel two-body correlation $h_{2\vk}$ follows its two-body bound-state wave function in high momentum, and  we can write it as 
\begin{equation}\tag{\ref{eq:pathInt2:hphif}}
h_{\vk}\sim\phi_{0\vk}f(\vk)
\end{equation}
with the factor $f(\vk)$ encapsulating all the many-body effects. This replacement also helps to  cure the divergence involving $h_{2\vk}$ because $\phi_{0\vk}$ decays fast enough in high momentum and gives no singularity when integrated with the momentum-dependent interaction.   After $h_{2\vk}$ is taken care of, we proceed to remove the singularity of integral involving the open-channel two-body correlation $h_{1\vk}$ using the regular method by subtracting the open-channel low-energy two-body interacting kernel, or the open-channel s-wave scattering length, $a_{s}^{(o)}$, which involves  the same high-momentum divergence as $h_{1\vk}$.

Let us put the above qualitative statement into concrete mathematical formulas.   As discussed before, we start with the closed-channel correlation $h_{2\vk}$ (Eq. \ref{eq:pathInt2:h2}). In the lowest order of ${\zeta}$, we have $\frac{E_{1\,\vk}+\xi_{\vk}}{E_{1\,\vk}+E_{2\,\vk}}\approx{}u_{\vk}^{2}$, $h_{2\vk}$ can be rewritten into such form
\begin{equation}\label{eq:pathInt2:h2D2}
 h_{2\vk}\approx\frac{\Delta_{2\vk}}{(E_{1\,\vk}+E_{3\,\vk})}u_{\vk}^{2}
\end{equation}

%Within $r_{c}$, the  potential determine the specific shape of the wave-functions; while, at free region, $\mathcal{D}$ ($r\gg{r_{c}}$ see Sec. \ref{sec:intro:as}), system is ``free'', all the details of potential are encapsulated in a few parameters (for example, s-wave scattering length, $a_{s}$).       %    the behavior should be quite universal for these bound states except an overall normalization factor. 
  %Across the crossover, the internal structure of correlation ($h_{2\vk}$) at medium and high momentum ($k\gg{k_{F}}$) follows the two-body wave function ($\phi_{0\vk}$) does not change much, while the overall normalization does vary along tuning.   At low momentum, we can introduce a factor for  many-body effect as illustrated in Eq. \ref{eq:pathInt2:hphif}.   Eq. \ref{eq:pathInt2:h2} actually follows such form fairly well in the medium and low momentum.  In the lowest order, we have $u_{\vk}^{2}\approx\frac{E_{1\,\vk}+\xi_{\vk}}{E_{1\,\vk}+E_{2\,\vk}}$ and it serves as the many-body modifier.  At high momentum ($>1/r_{c}$), the mean field result is not really reliable as showed in the divergence of integral. $D_{2\vk}$ at high momentum is no longer a constant, but  has specific form determined by the specific form of  potential. 

%In three-species many-body problem, the above conclusion is still valid except at low-momentum (around or below Fermi energy) where Pauli exclusion between two channels is severe. 
Here $\nth{(E_{1\,\vk}+E_{3\,\vk})}\approx\nth{2\xi_{\vk}+\eta}$ is just the same as two-body wave function $\phi_{0}$  besides normalization (when $k\ll\kappa$); while the extra factor $u_{\vk}^{2}$ is the many-body factor $f(\vk)$ in Eq. \ref{eq:pathInt2:hphif} which describes the Pauli exclusion between two channels.  At the low-momentum, the open-channel weight is large, the phase space left for the closed-channel is limited, which is shown mathematically as a small $u_{\vk}^{2}$ factor.   At higher momentum ($k_{F}\ll{k}\ll1/r_{c}$),  there is almost nothing in the open-channel,  indicated by $u_{\vk}^{2}\approx1$ and thus the closed-channel wave-function just follows its two-body counterpart with a different normalization factor.   It is not hard to see that $\Delta_{2}$ is closely related to the ``\emph{integrated contact intensity}'', $C$, in Tan's work about universality (\cite{Tan2008-1,Tan2008-2}).  In his work, Tan concluded that  the high-end of the relative-momentum distribution asymptotically approaches  $C/k^{4}$.  Note that the high-end in his paper means momentum lower than $1/r_{c}$, but higher than any other scales ($1/a_{s}$, $1/a_{0}$).  In such scale, $u_{k}^2\approx1$, %and especially, for close-to-threshold bound-state, $E_{b}\ll{}\frac{\hbar^{2}}{mr_{c}^{2}}$, so 
$C=\frac{\Delta_{2}^{2}}{4m^{2}}$ for the closed-channel.   
At even higher momentum ($>1/r_{c}$), the mean-field / saddle-point solution no longer applies.  And the two-body correlation, $h_{2\vk}$, should just follow two-body wave function (See Sec. \ref{sec:intro:as}).  By all the above arguments, we replace $\frac{\Delta_{2}}{(E_{1\,\vk}+E_{3\,\vk})}$ with a normalization factor, $\alpha$, and the two-body wave function $\phi_{0\vk}$ in Eq. \ref{eq:pathInt2:h2D2},
\begin{equation}\label{eq:pathInt2:hphi}
h_{2\vk}=\alpha\phi^{}_{0\vk}u_{\vk}^{2}
\end{equation}
where $\phi_{0}$ is the normalized  solution of  two-body \sch equation, $\sum_{\vk}\abs{\phi_{0\vk}}^{2}=1$.
\begin{equation}\label{eq:pathInt2:phi}
-E_{b}^{(0)}\phi_{0\,\vp}=2\epsilon_{\vp}\phi_{0\,\vp}-\sum_{\vk}V_{\vp\vk} \phi_{0\,\vk}
\end{equation}
By replacing $h_{2\vk}$ with the two-body wave function at high momentum, we ignore the Pauli exclusion between different closed-channel bound-states, $\phi_{0}$, while keeping Pauli exclusion between channels through the factor $u_{k}^{2}$.  The higher order correction in this replacement is probably comes in as nonliner relationship between $h_{2\vk}$ and $\phi_{0\vk}$ or extra terms $\phi_{i}$ in expansion of $h_{2\vk}$. However, as we will show, $h_{2\vk}$ is always  much smaller than 1. These higher order correction is relatively minor and only modifies  the conclusion quantitatively without affecting the result qualitatively. 
Furthermore, this replacement also solves the renormalization problem automatically because the two-body wave function decays fast enough at high momentum combined with the momentum-dependent interactions and therefore we have no more  divergence in integration.  

Eq. \ref{eq:pathInt2:hphi} can also be understood in terms of BCS-type anstaz $\prod_\vk\br{u_\vk+v_\vk{}a^\dg_\vk{}b^\dg_{-\vk}+w_\vk{}a^\dg_\vk{}c^\dg_{-\vk}}\ket{0}$ (Eq. \ref{eq:pathInt2:ansatz}) which connects to coherent states of the two-body wave function (Eq. \ref{eq:pathInt2:ansatzCoh}), $\phi^{(ab)}+\phi^{(ac)}$.   It is not hard to show  $h_{2\vk}=\av{a_{\vk}c_{-\vk}}=u_{\vk}w_{\vk}$ for this ansatz. Using the relation $\phi^{(ac)}_{\vk}=w_{\vk}/u_{\vk}$(Eq. \ref{eq:pathInt2:phiWU}),  we have
\begin{equation}
h_{2\vk}=u_{\vk}w_{\vk}=u_{\vk}^{2}(\frac{w_{\vk}}{u_{\vk}})\propto{}u_{\vk}^{2}\phi^{(ac)}_{\vk}
\end{equation}
 In a two-body problem of the Feshbach resonance, we expect $\phi^{(ac)}\propto\phi_{0}$  (see Chapter \ref{sec:intro:twobody}).
 And we have 
 \begin{equation}
 h_{2\vk}\propto\phi_{0\vk}u_{\vk}^{2}
 \end{equation}
just like Eq. \ref{eq:pathInt2:hphi}.

%The two-body wave function for isolated closed-channel spreads in large momentum space and has small weight in low momentum (below $k_{F}$), and therefore, we expected $u_{\vk}^{2}$ has small correction only.  $D_{2}$ is the product of two factors: 1, the normalization ($A$) factor as in two-body wave function normalized to 1(Appendix \ref{sec:pathInt2:short-range}); 2, total number of particles in closed-channel.  

%As discussed, $1/(2\xi_{\vk}+\eta)$ is the form of two-body closed-channel wave function, $\phi_0$, at low momentum.  If we ignore the Pauli exclusion between closed-channel bound states, the closed-channel wave function $h_{2\vk}$ can be written as .
%\begin{equation}
%h_{2\vk}=\alpha\phi^{(0)}_{\vk}u_{\vk}^{2}+o(\frac{E_{F}}{\eta})
%\end{equation}
% where $\phi^{(0)}$ follows two-body \sch equation
%\begin{equation}\label{eq:pathInt2:phi}
%-E_{b}^{(0)}\phi_{0\,\vp}=\epsilon_{\vp}\phi_{0\,\vp}-\sum_{\vk}V \phi_{0\,\vk}
%\end{equation}

To  the first order of $\zeta$, $\phi_{0\vk}\sim\nth{\kappa^{2}+k^{2}}$, where $\kappa$ is the momentum scale related to the binding energy or the tuning, $\kappa^{2}/2m=E_{b}\approx\eta$.     In the many-body relevant region where momentum is not significantly larger than the Fermi momentum, this quantity is actually very small and approximately a constant, $\nth{\kappa^{2}}$, because $\kappa{}\gg{}k_{F}$.  The part where the many-body factor $u_{\vk}^{2}$ significantly alters the wave function is actually rather small comparing to the whole spread of the wave function over the momentum space, in the order of $\kappa$ ($E_{b}=\hbar^{2}\kappa^{2}/2m$).  Within the range where $u_{\vk}^{2}$ differs from 1 significantly, the wave function $\phi_{0\vk}$ is very small.  (See Appendix \ref{sec:pathApp:consistency})
\begin{equation*}
\abs{\phi_{0\,k=0}}^{2}N\sim\br{\sqrt{\frac{\kappa}{\mathcal{V}_{0}}}\nth{\kappa^{2}}}^{2}\frac{\mathcal{V}_{0}}{a_{0}^{3}}=\nth{(a_{0}\kappa)^{3}}\gg1
\end{equation*}
 This fact is valid across the full range of the crossover.  At some regions of crossover, $\Delta_{2\vk}$ (or $\alpha$) may be large and the total closed-channel weight is comparable or even larger than that of the open-channel. However, for each energy level in low  momentum, the factor $\phi_{0\vk}$ is so small that the product of $h_{2\vk}$ is still much smaller than $1$.  
In summary, \emph{the full closed-channel (including normalization) is always small in the low momentum region  even when the total closed-channel weight is large.}  This justifies our following perturbative treatment on the inter-channel Pauli exclusion between two channels. 

%Let us look at the second equation Eq. \ref{eq:pathInt2:mfclose} first, as discussed before, $h_{2}$ is similar to $\phi^{(0)}$ in two-body wave function. More specifically

Not far away from the resonance point, absolute detuning $\eta$ is always close to binding energy $E_{b}^{(0)}$ of the closed-channel bound-state, $\phi_{0}$.   
Using Eq. \ref{eq:pathInt2:h2D2}, we can write 
\begin{equation*}
\Delta_{2\vk}=h_{2\vk}\frac{(E_{1\vk}+E_{3\vk})}{u_{\vk}^{2}}\approx{}h_{2\vk}\frac{(E_{\vk}+\xi_{\vk}+\eta)}{u_{\vk}^{2}}
\end{equation*}
Combining the above equation with Eq. \ref{eq:pathInt2:mfclose}, we have 
\begin{equation*}
h_{2\vk}\frac{(E_{\vk}+\xi_{\vk}+\eta)}{u_{\vk}^{2}}=\sum_{\vk'}{}Y_{\vk\vk'}h_{1\vk'}+\sum_{\vk'}V_{\vk\vk'}h_{2\vk'}
\end{equation*}
We replace $h_{2\vk}$ with $\alpha{}u_{\vk}^{2}\phi_{\vk}$ as in Eq. \ref{eq:pathInt2:hphi}.  The above equation becomes
\begin{align*}
\alpha{}\phi_{\vk}{(E_{\vk}+\xi_{\vk}+\eta)}
	&=\sum_{\vk'}{}Y_{\vk\vk'}h_{1\vk'}+\alpha{}\sum_{\vk'}V_{\vk\vk'}u_{\vk'}^{2}\phi_{\vk'}\\
	&=\sum_{\vk'}{}Y_{\vk\vk'}h_{1\vk'}-\alpha{}\sum_{\vk'}V_{\vk\vk'}v_{\vk'}^{2}\phi_{\vk'}
		+\alpha{}\sum_{\vk'}V_{\vk\vk'}\phi_{\vk'}
\end{align*}
The last term can be rewritten as $\sum_{\vk}V_{\vp\vk} \phi_{0\,\vk}=(2\epsilon_{\vp}+E_{b})\phi_{0\,\vp}$ using the two-body \sch equation of the isolated closed-channel Eq. \ref{eq:pathInt2:phi}.  We move the last two terms from r.h.s to l.h.s, and using $\xi_{\vk}=\epsilon_{\vk}-\mu$
\begin{equation*}
\alpha{}\phi_{\vk}\br{-E_{b}+\eta-2\mu+E_{\vk}-\xi_{\vk}+\sum_{\vk'}V_{\vk\vk'}v_{\vk'}^{2}\phi_{\vk'}}
=\sum_{\vk'}{}Y_{\vk\vk'}h_{1\vk'}\end{equation*}
Multiply both sides with $\phi^{*}_{\vk}$ and integrate over the momentum,
\begin{equation}
\alpha\mbr{-E_{b}+\eta-2\mu+\sum_{\vk}{\phi_{\vk}^{*}}{(E_{\vk}-\xi_{\vk})}\phi_{\vk}
	+\sum_{\vk\vk'}{\phi_{\vk}^{*}}{v_{\vk'}^{2}V_{\vk\vk'}}\phi_{\vk'}}
	=\sum_{\vk\vk'}{\phi_{\vk}^{*}}{Y_{\vk\vk'}}{h_{1\vk'}}
\end{equation}
 We rewrite this equation in the form
\begin{equation}\label{eq:pathInt2:alpha}
\alpha=\frac{\sum_{\vk\vk'}{\phi_{\vk}^{*}}{Y_{\vk\vk'}}{h_{1\vk'}}}{\br{-E_{b}+\eta-2\mu-\lambda_{1}}}
\end{equation}
\begin{equation}\label{eq:pathInt2:lambda1}
\lambda_{1}(\eta)\equiv-\sum_{\vk}{\phi_{\vk}^{*}}{(E_{\vk}-\xi_{\vk})}\phi_{\vk}
	-\sum_{\vk\vk'}{\phi_{\vk}^{*}}{v_{\vk'}^{2}V_{\vk\vk'}}\phi_{\vk'}
\end{equation}
Comparing this to the two-body problem Eq. \ref{eq:intro:closeCoeff}, detuning is shifted by the many-body effects $\mu$ (mostly due to the Pauli exclusion within the open-channel) and $\lambda_{1}$ (mostly due to the Pauli exclusion between channels). 
$\lambda_{1}$ depends on detuning $\eta$ (through $E_{\vk}$, $v_{\vk}$), but the dependence is rather weak because the integration is over mostly the short-range quantities and  insensitive to the detuning. We will discuss more detail about it later in Sec \ref{sec:pathInt2:lambda}. 
Using Eq. \ref{eq:pathInt2:alpha}, we can express the $\alpha$ in $h_{2\vk}$ (Eq. \ref{eq:pathInt2:hphi}), and then plug it into Eq. \ref{eq:pathInt2:h1}
\begin{equation*}
\Delta_{1\vp}=\sum_\vk{}U_{\vp\vk}{h_{1_{\vk}}}+
	{\sum_{\vp'} Y_{\vp\vp'}\alpha{\phi_{\vp'}}u_{\vp'}^{2}}
				=\sum_\vk{}U_{\vp\vk}{h_{1_{\vk}}}+
	\frac{\sum_{\vk\vk^{'}\vp'} Y_{\vp\vp'}{\phi_{\vp'}}{\phi_{\vk'}^{*}}{Y_{\vk\vk'}}u_{\vp'}^{2}{{h_{1\vk}}}}
		{\br{-E_{b}+\eta-2\mu-\lambda_{1}}}
\end{equation*}
Comparing this with the two-body problem, we can see that the detuning part (denominator of the second term) is shifted by $2\mu+\lambda_{1}$ and there is an extra $u_{\vp'}^{2}$ term introduced as the many-body effect.  Nevertheless, none of these affect the high-momentum behavior. Therefore, the equation can be renormalized exactly as in the single-channel problem by introducing the long-wave-length s-wave scattering length $a_{s}$.  We rewrite the above equation
\begin{equation*}
\begin{split}
\Delta_{1\vp}=&\sum_{\vk}\br{U_{\vp\vk}+
	\frac{\sum_{\vk^{'}\vp'} Y_{\vp\vp'}{\phi_{\vp'}}{\phi_{\vk'}^{*}}{Y_{\vk\vk'}}}
		{\br{-E_{b}+\eta-2\mu-\lambda_{1}}}}{{h_{1\vk}}}\\
	&-\frac{\sum_{\vk\vk{'}\vp'} Y_{\vp\vp'}{\phi_{\vp'}}{\phi_{\vk'}^{*}}{Y_{\vk\vk'}}v_{\vp'}^{2}{h_{1\vk}}}
		{\br{-E_{b}+\eta-2\mu-\lambda_{1}}}{}
\end{split}
\end{equation*}
The second term in the r.h.s. has no divergence at high momentum in 3D due to the extra $v_{p'}^{2}$ factor. Actually the factor, $v_{p'}^{2}$, decreases quickly over a small range in the order of  ``gap'' $\Delta_{1\vp}$; therefore, the summation in the second term is essentially only over low-momentum, and is very small.   In addition, considering the short-range nature for $Y_{\vp\vp'}$, this term varies slowly over momentum, $\vp$.  

Multiply both side with $(1+TG)$,  where $T$ is the scattering matrix for the open-channel, and $G=(\omega-H_{0})^{-1}$ is the Green's function for a free pair in the open-channel.\footnote{Here we use the relation of the scattering $T$-matrix, the free pair Green's function $G=(\omega-H_{0})^{-1}$ and the bare interaction $V=-U_{\text{eff}}$.
\begin{equation*}
T=V+TGV=-(1+TG)U_{\text{eff}}
\end{equation*}
 }
\begin{equation*}
(1+TG)\Delta_{1}=-Th_{1}-\lambda_{2}
\end{equation*}
\begin{equation}\label{eq:pathInt2:lambda2}
\lambda_{2\vp}\equiv\sum_{\tilde\vp}(1+TG)_{\vp\tilde\vp}\frac{\sum_{\vk\vk{'}\vp'} Y_{\tilde\vp\vp'}{\phi_{\vp'}}{\phi_{\vk'}^{*}}{Y_{\vk\vk'}}v_{\vp'}^{2}{h_{1_{\vk}}}}
		{\br{-E_{b}+\eta-2\mu-\lambda_{1}}}{}
		=\alpha\sum_{\tilde\vp\,\vp'}(1+TG)_{\vp\tilde\vp} Y_{\tilde\vp\vp'}{\phi_{\vp'}}v_{\vp'}^{2}
\end{equation}

In principle, $\lambda_{2\vp}$ depends on momentum, however, like $\Delta_{1\vp}$, it is approximately constant at low momentum.
We are only interested in the low momentum/frequency properties as the system is cold, dilute and governed by only short-range interactions. The scattering matrix $T$ is approximately its s-wave zero-energy value ${4\pi{\tilde{a}_{s}(\mu,\lambda_{1})}}/{m}$ although $\tilde{a}_{s}(\mu,\lambda_{1})$ is the s-wave scattering length of the shifted (by $2\mu+\lambda_{1}$) detuning. The s-wave scattering length, $a_s$, in the two-body problem without many-body shift $2\mu+\lambda_{1}$ is simply Eq. \ref{eq:intro:asK}.  
\begin{equation}\tag{\ref{eq:intro:asK}}
a_{s}(\delta)=a_{bg}\br{1+\frac{\mathcal{K}}{\delta}}
\end{equation}
Here in the many-body context with the extra detuning $2\mu+\lambda_{1}$, $\tilde{a}_{s}$ is 
\begin{equation}\label{eq:pathInt2:asKshift}
\tilde{a}_{s}=a_{\text{bg}}(1+\frac{\mathcal{K}}{\delta-2\mu-\lambda_{1}})
\end{equation}
where $\mathcal{K}$ is introduced in Eq. \ref{eq:intro:kappa} of Chapter \ref{sec:intro:twobody}. 
The Green's function is simply $1/2\epsilon_{\vk}$ at zero energy.   Similarly,   $\Delta_{1\vp}$ and $\lambda_{2\vp}$ is approximately  constant at low momentum.  We drop the subscript $\vp$ of them, and pull $\Delta_{1}$ out of summation. 
\begin{equation}
\Delta_{1}=-T\sum(h_{1}+G\Delta_{1})-\lambda_{2}
=-\frac{4\pi{\tilde{a}_{s}(\mu,\lambda_{1})}}{m}\Delta_{1}\sum{}(\frac{E_{1\,\vk}+\xi_{\vk}+\eta}{(E_{1\,\vk}+E_{2\,\vk})(E_{1\,\vk}+E_{3\,\vk})}-\nth{2\epsilon_{\vk}})
	-\lambda_{2}
\end{equation}
Here the last step use the zero energy value of the zero-energy free pair Green's function $G(\omega=0)=(-2\epsilon_{\vk})^{-1}$. Dividing both sides with $\Delta_{1}$, we have the renormalized equation
\begin{equation}
1=-\mbr{\frac{4\pi{\tilde{a}_{s}(\mu,\lambda_{1})}}{m}\sum(\frac{E_{1\,\vk}+\xi_{\vk}+\eta}{(E_{1\,\vk}+E_{2\,\vk})(E_{1\,\vk}+E_{3\,\vk})}-\nth{2\epsilon_{\vk}})}
	-\frac{\lambda_{2}}{\Delta_{1}}
\end{equation}
Note that $\tilde{a}_{s}(\mu,\lambda_{1})$ corresponds to the two-body s-wave scattering length at detuning shifted by $2\mu+\lambda_{1}$.

\begin{figure}[htbp]
\begin{center}
\includegraphics[width=0.8\textwidth]{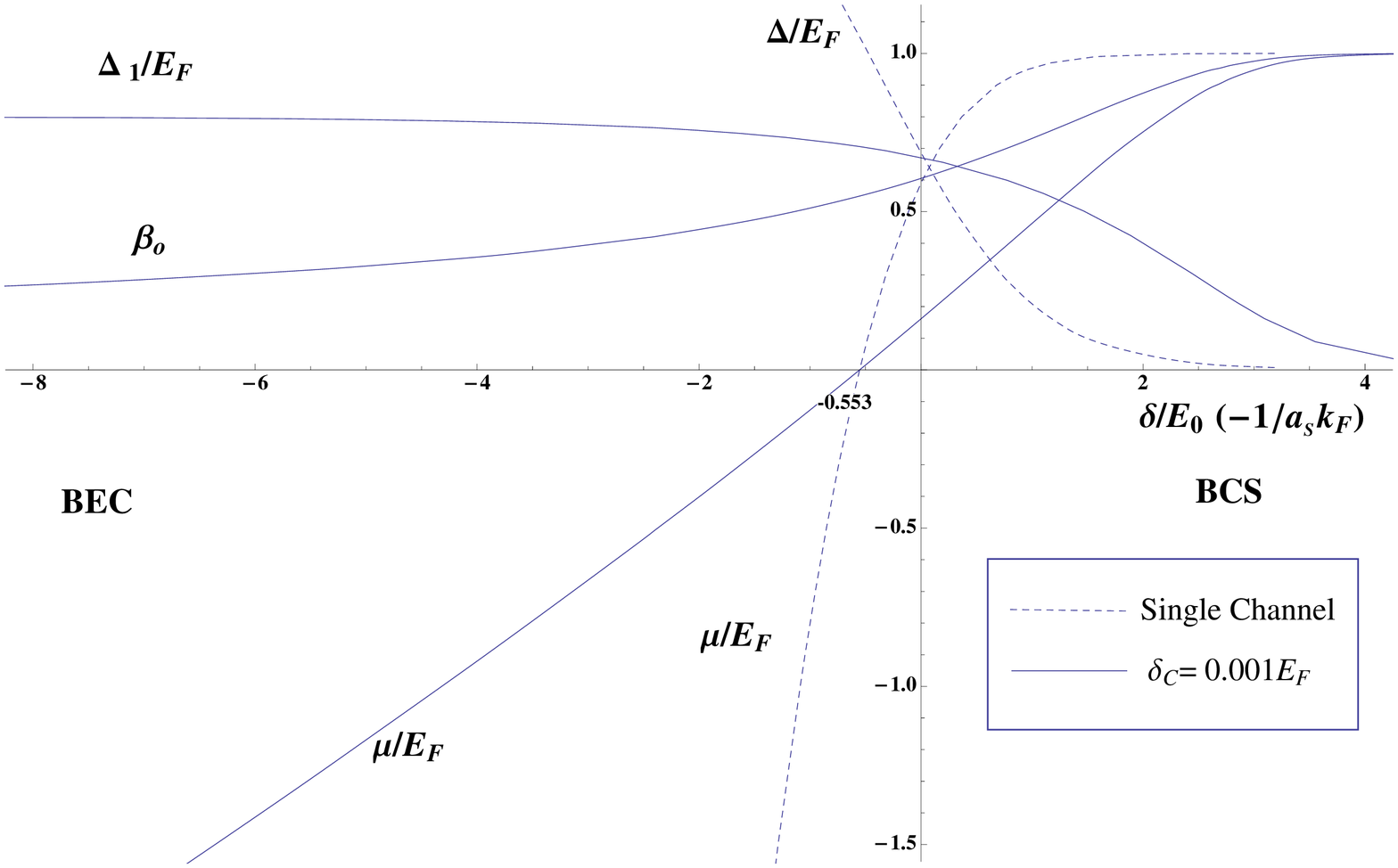}
\caption{ The narrow Feshbach resonance w/o the inter-channel Pauli exclusion  vs.  the single-channel model} 
\label{fig:pathInt2:narrow}
\parbox{0.8\textwidth}{\small We plot the chemical potential $\mu$, the open-channel gap $\Delta_{1}$ and the open-channel relative weight $\beta_{o}$  for the narrow resonance; and chemical potential and the gap for the single-channel model.     The gap in the open-channel $\Delta_{1}$ and the chemical potential $\mu$ are rescaled with the the Fermi energy of the total density, $E_{F}^{(tot)}$.  The $x$-axis is the detuning $\delta$ rescaled with $E_{0}$ (see Eq. \ref{eq:pathInt2:E0}) for the narrow resonance; while it is $-1/a_{s}k_{F}$ for the single-channel curves. We have taken $\delta_{c}=0.001E_{F}^{(tot)}$ for the narrow resonance figure. We used the detuning from the resonant point  for the $x$-axis in the narrow resonance instead of $-1/a_{s}k_{F}$ in the single-channel because the additional shift, $2\mu$, considering in Eq. \ref{eq:pathInt2:simplenarrowAs}.  Consequently, the chemical potential lines in both cases cross the $x$-axis at the same point where $\mu=0$.  }
\end{center}
\end{figure}

Now we can expand the first term in the parentheses to the first order of $\zeta$, using Eq. \ref{eq:pathInt2:xiExpand}. Keeping in mind $E_{\vk}\ll{\eta}$ at low momentum where summation is about, we have\footnote{Here we used $u_{\vk}^{2}-v_{\vk}^{2}=\frac{\xi_{\vk}}{E_{\vk}}$}
\begin{equation}\label{eq:pathInt2:gapRenorm}
1=-\mbr{\frac{4\pi{\tilde{a}_{s}(\mu,\lambda_{1})}}{m}\sum(\nth{2E_{\vk}}-\nth{2\epsilon_{\vk}}-\frac{\Delta_{2}^{2}\xi_{\vk}}{4(\xi_{\vk}+\eta){E_{\vk}^{3}}})}
	-\frac{\lambda_{2}}{\Delta_{1}}
\end{equation}
The correction term (the third term in parentheses) does not have divergence in summation of high-momentum. 
In summary, there are several difference of gap equation here comparing to single-channel problems:
\begin{enumerate}
\item\label{item:pathInt2:mu}The shift of $2\mu$ in detuning through $\tilde{a}_{s}$;
\item The extra shift of $\lambda_{1}$ (see Eq. \ref{eq:pathInt2:lambda1}) in detuning through $\tilde{a}_{s}$;
\item The extra term $-\frac{\lambda_{2}}{\Delta_{1}}$ in Eq. \ref{eq:pathInt2:gapRenorm};
\item The extra term $-\frac{\Delta_{2}^{2}\xi_{\vk}}{4(\xi_{\vk}+\eta){(E_{\vk})^{3}}}\approx-\frac{\xi_{\vk}\Delta_{1}}{4{E_{\vk}^{3}}}\zeta$ in the summation of Eq. \ref{eq:pathInt2:gapRenorm};
\end{enumerate}
Item \ref{item:pathInt2:mu} corresponds to a many-body effect  common to both three and four species problem.
\begin{unsure}
In a many-body system, most of the properties can obtained when considering what happens at the next extra particle or particle pair. In a two-body system,  we can pretend that the vacuum is the initial state and the next particle pair comes in from energy zero.  In a many-body system, this starting point is no longer the absolutely zero, but the chemical potential.  Excitations are counted from the chemical potential, $\mu$.  The energy detuning in Feshbach resonance is no exception.  In a many-body formula, the detuning is shifted by $2\mu$ (for a pair).  In the broad resonance, the Fermi energy is smaller comparing to the detuning during the relevant probing region.  Consequently, this shift is negligible.  In the narrow resonance, the Fermi energy, however, is substantial and a detuning of the chemical potential needs to be taken into account.  At the BCS side, $\mu$ is positive and close to the Fermi energy. This shifting just confirms the fact that most interesting phenomenon in a fermion system happens around Fermi surface instead of absolute zero.  With an extra positive shift $\mu$, detuning is reduced; therefore the crossover region is actually reached earlier (i.e. by larger detuning) than in two-body cases.  On the other hand, moving toward to the resonance point and BEC side, the chemical potential becomes smaller and finally flips sign to negative. Accordingly, the detuning is reduced lesser and lesser;  finally it is enhanced.  
%The Feshbach resonance in many-body cases is skewed toward BCS (positive detuning) side from two-body cases.  
Interestingly, the crossing point, $\mu=0$, is not shifted from a model with simple  s-wave scattering length directly from two-body physics if only considering this shift  (See Fig. \ref{fig:pathInt2:narrow}). This effect is also  studied extensively previously. (Sec. 6.2 of \cite{GurarieNarrow}).

The next three corrections are unique for three-species problem where the inter-channel Pauli exclusion cannot be neglected.  
Furthermore,  close look into $\lambda_1$ and $\lambda_2$ (Sec. \ref{sec:pathInt2:lambda}) reveals that they vary slowly across the full region of crossover and are functions of the density of atoms in the open-channel .  They describe  fundamental many-body effects, and do not have counterparts in a two-body problem.  
Both $\lambda_{1}$ (see Eq. \ref{eq:pathInt2:lambda1})  and $\lambda_{2}$ (see Eq. \ref{eq:pathInt2:lambda2})  involve overlap integrals between the open-channel wave function and the closed-channel wave function, (factor $v_{\vk}^{2}$ or $u_{\vk}^{2}$ describes mostly the open-channel wave function; while $\phi$ describes the closed-channel wave function). The larger the overlap of the two, the larger $\lambda_{1}$ and $\lambda_{2}$ are.  This has a very intuitive interpretation:  more overlap leads more severe inter-channel Pauli exclusion, which in turn leads to larger $\lambda_{1}$  and $\lambda_{2}$  that describe this effect between two  channels.  In our model, an open-channel wave function is  spread all over the real-space, (even at BEC-side, the real bound-state is very loosely bound comparing to the closed-channel bound state), while the closed-channel wave function is very sensitive to the binding energy, $E_{b}(\approx\eta)$.  When the closed-channel  bound-state is closer to the threshold, i.e.,  the binding energy  is smaller, the closed-channel bound-state is more spread out  in real space and has larger overlap with open-channel. Consequently,  $\lambda_{1}$ and $\lambda_{2}$ are larger in such cases.                           Nevertheless, $\lambda_1$ is much smaller than the  Fermi energy $E_{F}$, or the other shift, the chemical potential, $2\mu$.  So the shift is not very large and it is still all right to treat it as a perturbation.  The correction is   shifted with the change of the density.  In the many-body system, no dramatic jump in physical quantities happens at resonance due to the crossover nature, and it is hard to observe this shift over the resonance position though.  %Nevertheless, this effect can show up at two-body type experiments, where resonance position is easier to detect.  

\subsection{Evaluation and estimation of $\lambda_{1}$ and $\lambda_{2}$\label{sec:pathInt2:lambda}}
From the last section, we  see  that a lot of the inter-channel Pauli exclusion is encapsulated in two parameters $\lambda_{1}$ and $\lambda_{2}$.   Therefore, it is well-worthwhile to study them in more details.  $\lambda_{1}$ is defined as 
\begin{equation}\tag{\ref{eq:pathInt2:lambda1}}
\lambda_{1}(\eta)\equiv-\sum_{\vk}{\phi_{\vk}^{*}}{(E_{\vk}-\xi_{\vk})}\phi_{\vk}
	-\sum_{\vk\vk'}{\phi_{\vk}^{*}}{v_{\vk'}^{2}V_{\vk\vk'}}\phi_{\vk'}
\end{equation}
Use relationship $v_{k}^{2}=\frac{E_{\vk}-\xi_{\vk}}{2E_{\vk}}$, we can rewrite the above equation into 
\begin{equation}
\lambda_{1}(\eta)\equiv-\sum_{\vk}{\phi_{\vk}^{*}}v_{\vk}^{2}(2E_{k}\phi_{\vk}+\sum_{\vk'}V_{\vk\vk'}\phi_{\vk'})
\end{equation}
We can then replace the second term  in the parentheses using the two-body \sch equation of the isolated closed-channel Eq. \ref{eq:pathInt2:phi}, $\sum_{\vk}V_{\vp\vk} \phi_{\vk}=(E_{b}+2\epsilon_{\vp})\phi_{\vp}$.
\begin{equation*}
\lambda_{1}(\eta)=-\sum_{\vk}{\phi_{\vk}^{*}}v_{\vk}^{2}\mbr{2(E_{k}+\epsilon_{k})+E_b}\phi_{\vk}
	\approx\sum_{\vk}\abs{\phi_{\vk}}^2v_{\vk}^{2}\mbr{2\br{E_{k}+\epsilon_{k}}+\eta}
\end{equation*}
In the above summation, $v_{\vk}^2$ is only non-zero below or not much higher than the Fermi momentum.  In this range, $E_\vk,\epsilon_\vk\ll\eta$ and we can neglect $E_\vk+\epsilon_\vk$ comparing to $\eta$.  
Furthermore, $\phi_{\vk}$ varies slowly  in this range, because the closed-channel bound state is much smaller than the interparticle distance.    It is a good approximation to replace $\phi_{\vk}$ with $\phi_{\vk=0}\approx\frac{A}{\kappa^{2}}$ and to take it out of the summation.  
Put all these together, we can estimate $\lambda_{1}$ 
\begin{equation}\label{eq:pathInt2:lambda1es}
\lambda_{1}\approx-\eta\abs{\phi_{k=0}}^2\sum{}v_{\vk}^{2}=-\eta\abs{\phi_{k=0}}^2N_{o}	
\end{equation}
where $N_{o}$ is the total number of atoms in the open-channel. It is easy to see that $\abs{\lambda_{1}}$ is much smaller than $\eta$ as $\abs{\phi_{k=0}}^2N_{o}\ll1$.  Furthermore, using $\phi_{k=0}\approx\sqrt{\frac{8\pi\kappa}{\mathcal{V}_{0}}}\frac{1}{\kappa^{2}}$ (Eq. \ref{eq:pathInt2:phi2body} of Appendix \ref{sec:pathInt2:short-range}), we get $\lambda_{1}\sim\frac{a_{c}}{a_{0}}E_{F}\ll{}E_{F}$, where $a_{0}$ is the average interparticle distance and $a_{c}$ is the size of the close-channel bound-state.   We have showed that $\lambda_{1}$ is indeed a minor correction over shift $\mu$.
It is not easy to estimate $\lambda_{1}$ precisely from  microscopic parameters above because these parameters themselves are  in turn hard to  estimate precisely.  Nevertheless, a key observation is that $\lambda_1$ depends on the (open-channel) density ($n_{0}={\mathcal{V}_{0}^{-1}}\sum_{\vk}v_\vk^2$) linearly to the lowest order of $\frac{a_{c}}{a_{0}}$, 
\begin{equation}\label{eq:pathInt2:lamb1}
\lambda_1\approx\lambda_1^{(0)}n_{o}
\end{equation}  
We can perform the experiments at different densities and estimate the coefficient $\lambda_1^{(0)}$.

 $\lambda_2$ is defined as
\begin{equation}\tag{\ref{eq:pathInt2:lambda2}}
\lambda_{2\vp}\equiv\sum_{\tilde\vp}(1+TG)_{\vp\tilde\vp}\frac{\sum_{\vk\vk{'}\vp'} Y_{\tilde\vp\vp'}{\phi_{\vp'}}{\phi_{\vk'}^{*}}{Y_{\vk\vk'}}v_{\vp'}^{2}{h_{1_{\vk}}}}
		{\br{-E_{b}+\eta-2\mu-\lambda_{1}}}{}
		=\alpha\sum_{\tilde\vp\,\vp'}(1+TG)_{\vp\tilde\vp} Y_{\tilde\vp\vp'}{\phi_{\vp'}}v_{\vp'}^{2}
\end{equation}
Here the argument goes similar as in the $\lambda_{1}$ case, considering the short-range nature, both  $Y_{\tilde\vp\vp'}$ and $\phi_{\vp'}$ vary slowly and can be approximated by their value at $\vp'=0$ for the momentum below or around the Fermi momentum.  After pulling these two quantities out of the summation over $\vp'$, we only need to sum over   $v_{\vp'}^{2}$, which gives a familiar factor,  the (open-channel) density, $n_{o}$. In addition, $\alpha=\sqrt{n_{c}}$.  The factor $(1+TG)_{\vp\tilde\vp}$ integrate out the high momentum component and $\lambda_{2\vp}$ is close to a constant at low momentum $\vp$.  We have 
\begin{equation}\label{eq:pathInt2:lamb2}
\lambda_2\approx\lambda_2^{(0)}n_{o}\sqrt{n_{c}}
\end{equation}
  In experiments, $\lambda_{1}$ and $\lambda_{2}$ can be measured at different total densities.   The two coefficients, $\lambda_{1}^{(0)}$  and $\lambda_{2}^{(0)}$,  which do not depend on the density, can then be estimated according to Eqs. \ref{eq:pathInt2:lamb1} and \ref{eq:pathInt2:lamb2}.

%The Direct replacement of $1/(E_{1\vk}+E_{3\vk})$ by $\phi_{k}$ in Eq. \ref{eq:pathInt2:hphi} introduces certain error. They might lead to a non-linear correction on $\lambda_{i}$ about density $n$.  Nonetheless, we expect the higher order non-linearity is weak and not important for small corrections, $\lambda_{i}$.

\subsection{Number equations}
There is one  number equation for each channel,  
\begin{gather*}
\sum_{\omega_{n}, \vk}G_{22}e^{(-i\omega_n\delta_-)}=N_{open}\\
\sum_{\omega_{n},\vk}G_{33}e^{(-i\omega_n\delta_-)}=N_{close}
\end{gather*}
Note that the Matsubara summation  formally diverges and we need to put in a small negative part into the summation in order to prevent the divergence.  Here we put in a small  negative part instead of a positive part  because $\Psi_{2}$ and $\Psi_{3}$ are conjugate of real particles, $\Psi_2=\bar\psi_b$, $\Psi_3=\bar\psi_c$.  The Matsubara summation can be performed with the normal trick of multiplying a Fermi function to summand and deform the contour  (see Appendix \ref{sec:pathInt2:deriveMF} for more details).  For the summation at zero temperature, we just need to consider the positive roots, $E_{1\,\vk}$.  It is straightforward to find 
\begin{gather}
N_{\text{open}}=\sum_{\vk}\frac{(E_{1\,\vk}-\xi_{\vk})(E_{1\,\vk}+\xi_{\vk}+\eta)-\Delta_2^2}{(E_{1\,\vk}+E_{2\,\vk})(E_{1\,\vk}+E_{3\,\vk})}
\label{eq:pathInt2:numOpen}\\
N_{\text{closed}}=\sum_{\vk}\frac{(E_{1\,\vk}-\xi_{\vk})(E_{1\,\vk}+\xi_{\vk})-\Delta_1^2}{(E_{1\,\vk}+E_{2\,\vk})(E_{1\,\vk}+E_{3\,\vk})}
=\sum_{\vk}\frac{E_{1\,\vk}^2-E_{\vk}^2}{(E_{1\,\vk}+E_{2\,\vk})(E_{1\,\vk}+E_{3\,\vk})}\label{eq:pathInt2:numClose}
\end{gather}
Let us look at the equation of the closed-channel first, if we expand $E_{i\,\vk}$ using Eqs. \ref{eq:pathInt2:xiExpand}-\ref{eq:pathInt2:xiExpand3}, the lowest order is 
\begin{equation}\label{eq:pathInt2:closeD1}
N_{\text{closed}}\approx\sum_{\vk}\frac{\gamma_{1\,\vk}}{(E_{\vk}+\xi_{\vk}+\eta)}=\sum_{\vk}\frac{\Delta_{2}^2u_{\vk}^{2}}{(\xi_{\vk}+\eta)(E_{\vk}+\xi_{\vk}+\eta)}
\end{equation}
Here $\gamma_{1}$ is the correction for the fermionic correlation spectrum due to the inte-channel Pauli exclusion (Eq. \ref{eq:pathInt2:xiExpand}).  We  take the full value $\frac{\Delta_{2}^2u_{\vk}^{2}}{(\xi_{\vk}+\eta)}$, instead of the low-momentum value because the summation has substantial contribution from  high momentum.  This is consistent with Eq. \ref{eq:pathInt2:h2} and \ref{eq:pathInt2:h2D2} if we assume\footnote{As discussed before, in the closed-channel $h_{2\vk}\ll1$ over all momentum.  Furthermore, the summation goes over very large momentum range, the difference between $h_{2\vk}$ and the closed-channel particle number is only large below or around Fermi energy, but the total weight of the closed-channel atoms  in such a region is very small.  }  $N_{\text{closed}}\approx\sum{h_{2}^{2}}$.    From appendix \ref{sec:pathApp:consistency}, we know that the summand is much smaller than 1 in all regions. Particular, in the low momentum ($\lesssim{}k_{F}$), the summand is $\zeta\frac{\Delta_{1}}{\eta}u_{\vk}^{2}\ll1$. Nevertheless, the weight spread in a very large range of momentum ($\sim\eta$), so the resulting sum can be still in the order of total number $N$ at certain detuning. 
%We cannot simply neglect a term in summation one order higher in $\zeta$ in general.  

Note that the connection between $\Delta_{2}$ and the ``Contact'' in the theory of universality (\cite{Tan2008-1,Tan2008-2}).  The summand as  density of momentum is valid at momentum up to the scale $1/a_{c}$, and is mostly determined by two-body physics except an overall factor at ``high momentum'' ($k_{F}\ll{}k\ll{}1/r_{c}$).  Furthermore, the contribution from the very high momentum ($k\gg1/r_{c}$) to integral is small.  All the deviation there can be absorbed as a small correction in $\Delta_{2}$.  To zeroth order of $\zeta$, $E_{i\vk}\approx\xi_\vk$ and $u_\vk^2\approx1$ for the most of the integral domain.  This summation can be written as 
\begin{equation}\label{eq:pathInt2:closeD2}
N_{\text{closed}}\approx\sum_{\vk}\frac{\Delta_{2}^2}{(\xi_{\vk}+\eta)(2\xi_{\vk}+\eta)}
\end{equation}
This equation has only one unknown parameter $\Delta_{2}$.  Therefore, this equation can be used to estimate $\Delta_{2}$ from experiments.  The  estimation of  $\Delta_{2}$ from such procedure automatically includes the correction from  high momentum. 

%Similar to the argument in gap equation, the summand  above is mostly useful at energy below $\eta$, where the component is small everywhere comparing to 1, however, the summation runs over high-momentum ($\sim\eta$), where $D_2$ should no longer be regarded as constant and summation can gives $N_{close}$ comparable to total number. 
% \mycomment{ This might not be correct statement.  Maybe it is just fine as it properly converges. } This summation shows that the closed-channel state is small and therefore the summation runs up to high momentum, where mostly is determined by two-body physics.  

The open-channel number equation  can be expanded perturbatively as well, 
\begin{equation}\label{eq:pathInt2:openNum}
\begin{split}
N_{\text{open}}\approx&\sum_\vk\mbr{\frac{E_\vk-\xi_\vk}{2E_\vk}+\frac{\gamma_{1\vk}}{2E_\vk}
	-\frac{(E_\vk-\xi_\vk)(\gamma_{1\vk}+\gamma_{2\vk})}{4E_\vk^2}
	-\frac{(E_\vk-\xi_\vk)\gamma_{3\vk}}{2E_\vk(\xi_\vk+E_\vk+\eta)}
	-\frac{\Delta_{2}^{2}}{2E_\vk(\xi_\vk+E_\vk+\eta)}}\\
	\approx&\sum_\vk\mbr{\frac{E_\vk-\xi_\vk}{2E_\vk}+\frac{E_\vk-\xi_\vk}{2E_\vk}\frac{\Delta_{1}}{2(E_{\vk}+\xi_{\vk}+\eta)}\zeta -\frac{\Delta_{1}^{3}}{4E_\vk^{3}}\zeta
	}	
\end{split}
\end{equation}
All terms behave well  in 3D and do not need any renormalization.   It is like the number equation in the single channel with  a few correction in the order of $\zeta=\Delta_2^2/\eta\Delta_{1}$.  The second term, except the factor $\frac{E_\vk-\xi_\vk}{2E_\vk}=v_{\vk}^{2}$, looks like the summand in the closed-channel number equation, Eq. \ref{eq:pathInt2:closeD1},  which can sum up in the order of $N$.  Nevertheless, the $v_{\vk}^{2}$ factor, decreases quickly in high momentum, becomes close to zero for $k\gg{}k_{F}$, and limits the summation to only low momentum. This makes the sum  a small correction as we discussed before.    Similarly, $\frac{\Delta_{1}^{2}}{E_{\vk}^{2}}$ in the third term also limits the summation to low momentum and ensures that the sum is a small correction. At low momentum, this term is actually more important than the second term.

\section{Discussion of the mean-field solution\label{sec:pathInt2:mean2}}
As discussed before, the correction of the narrow Feshbach resonance can be taken into account in two steps.  First,  omitting the inter-channel Pauli exclusion, we only consider the chemical potential $\mu$ in the shift and  the extra counting of the closed-channel.  Then in the second step, we can correct the previous result with quantities originated from the inter-channel Pauli-exclusion unique in the three-species problem. 

In  the first step, the gap equation and the (open-channel) number equation are simplified to 
\begin{gather}
1=-\mbr{\frac{4\pi{\tilde{a}_{s}(\mu)}}{m}\sum(\nth{2E_{\vk}}-\nth{2\epsilon_{\vk}})}\label{eq:pathInt2:narrowGapS}\\
N_{\text{open}}=\sum_\vk\frac{E_\vk-\xi_\vk}{2E_\vk}\label{eq:pathInt2:narrowNumS}
\end{gather}
Here we only consider the shift of the  chemical potential $2\mu$ in   $\tilde{a}_s$ (Eq. \ref{eq:pathInt2:asKshift}),
\begin{equation}
\tilde{a}_{s}=a_{\text{bg}}(1+\frac{\mathcal{K}}{\delta-2\mu})\approx{}\frac{a_{\text{bg}}\mathcal{K}}{\delta-2\mu}
%=\frac{\sqrt{2\delta_{c}\hbar^{2}/m}}{\delta-2\mu}
\label{eq:pathInt2:simplenarrowAs}
\end{equation}
 In the second equation, we assume $a_{\text{bg}}\ll{}a_{0}$ ($a_{0}$ is the average interparticle distance) and we only study situations close to resonance, where $\mathcal{K}\gg{}\delta \,\text{or}\,\mu$ ($\delta$ is the detuning from the resonant point in two-body physics).  From the two-body analysis of the Feshbach resonance, we know the numerator in the Eq. \ref{eq:pathInt2:simplenarrowAs} is closely related to the characteristic scale $\delta_{c}$ defined in Eq. \ref{eq:intro:deltaC}
\begin{equation}\tag{\ref{eq:intro:deltaC}}
\delta_{c}\equiv\frac{\mathcal{K}^{2}}{\hbar^{2}/m_{r}a_{bg}^{2}}=\frac{(\mathcal{K}a_{\text{bg}})^{2}}{\hbar^{2}/m_{r}}
\end{equation}
\end{unsure}
We can rewrite the $\tilde{a}_{s}$ with respect to $\delta_{c}$
\begin{equation}\label{eq:pathInt2:asNarrowSim}
\tilde{a}_{s}=\frac{\sqrt{2\delta_{c}\hbar^{2}/m}}{\delta-2\mu}
\end{equation}

In the narrow resonance, the weight of the closed-channel becomes substantial quickly.  We can calculate the amplitude of the closed-channel according to  Eq. \ref{eq:pathInt2:alpha}
\begin{equation}\tag{\ref{eq:pathInt2:alpha}}
\alpha=\frac{\sum_{\vk\vk'}{\phi_{\vk}^{*}}{Y_{\vk\vk'}}{h_{1\vk'}}}{\br{-E_{b}+\eta-2\mu-\lambda_{1}}}
\end{equation}
  $Y$, as a short-range interaction, only picks up the short-range part of $h_{1}(r)$ and $\phi(r)$ in the above integral. The short-range part of $h_{1}$ is assumed to be proportional to the short-range part of the two-body wave function. Its normalization, however, comes from many-body physics. If we assume the zeroth order form, $h_{1}=\Delta_{1}/2E_{\vk}$,   Zhang has derived the form of  $h_{1}(\vr)$ in the real space, (Eq. (25) in \cite{shizhongSumRule})
  \begin{equation}
  h_{1}(r)=\frac{m\Delta_{1}}{4\pi\hbar^{2}}\frac{1-r/a_{s}}{r}
  \end{equation}
Comparing this one with the Bethe-Peierls boundary condition Eq. \ref{eq:intro:Bethe}, we find that $\frac{m\Delta_{1}}{4\pi\hbar^{2}}$ is the normalization factor.  And $\Delta_{1}$ is determined by the number equation for open-channel density $n_{0}$.  Similar as the two-body case (Eq. \ref{eq:intro:closeWeight}), we can write
\begin{equation}
\alpha^{2}=\frac{\abs{\sum_{\vk\vk'}{\phi_{\vk}^{*}}{Y_{\vk\vk'}}{h_{1\vk'}}}^{2}}{\br{-E_{b}+\eta-2\mu-\lambda_{1}}^{2}}
\end{equation}
Extracting the normalization factor of $h_{1}$, the integral in the numerator then is just the same as that in the two-body physics and can be rewritten using the expression of  $\mathcal{K}$ according to Eq. \ref{eq:intro:kappa}.  
\begin{equation}
\abs{\sum_{\vk\vk'}{\phi_{\vk}^{*}}{Y_{\vk\vk'}}{h_{1\vk'}}}^{2}=\br{\frac{m\Delta_{1}}{4\pi\hbar^{2}}}^{2}\frac{\hbar^{2}\mathcal{K}}{ma_{bg}}
%:
\end{equation}
Here we use $m_{r}=\nth{2}m$.
We further assume that the energy difference between the resonant point ($a_{s}\to\pm\infty$) and the level crossing point, where the uncoupled closed-channel bound state level is at the threshold of the open-channel, $\mathcal{K}$, is in fact much larger than $\mu$ or $\delta_{c}$, i.e. $\abs{-E_{b}+\eta}\gg2\mu,\,\lambda_{1}$. So we have $\abs{-E_{b}+\eta-2\mu-\lambda_{1}}\approx\mathcal{K}$ (note that here $\eta$ is equivalent to $\tilde\delta$ in the two-body Eq. \ref{eq:intro:shiftK}). We can  then write
\begin{equation}
{n_{c}}\approx\frac{m\Delta_{1}^{2}}{32\pi^{2}\hbar^{2}}\nth{\mathcal{K}a_{bg}}
\end{equation}
From Eq. \ref{eq:intro:deltaC} above, we can replace $\mathcal{K}a_{bg}$ with $\sqrt{2\delta_{c}\hbar^{2}/m}$ using Eq. \ref{eq:intro:deltaC}. After  dividing both sides with the total density $n_{tot}=\br{\frac{mE_{F}^{(tot)}}{\hbar^{2}}}^{3/2}\frac{\sqrt{2}}{3\pi^{2}}$, we find the weight for the closed-channel
\begin{equation}
\beta_{c}=\frac{n_{c}}{n_{tota}}\approx\frac{3\sqrt2}{128}\frac{\Delta_{1}^{2}}{\sqrt{{E_{F}^{(tot)}}{}^{3}\delta_{c}}}
\end{equation}
 $\beta_{c}$ is proportional to $\Delta_{1}^{2}$.  This form is consistent with the formula obtained by Zhang previously \cite{shizhongUniv}.   In Fig. \ref{fig:pathInt2:narrow}, we plot the open-channel weight 
 \begin{equation}
 \beta_{o}=1-\beta_{c}
 \end{equation}

We can see an example of the narrow resonance ($\delta_{c}=0.001E_{F}^{(tot)}$) in the Fig. \ref{fig:pathInt2:narrow}. We rescale the detuning with $E_{0}$ \footnote{We choose this scale in order to make sure that $\tilde{a}_{s}k_{F}$ would match  $a_{s}k_{F}$ in the single channel discussion in Chapter \ref{sec:intro:1channel} if there was no $-2\mu$ shift as in Eq. \ref{eq:pathInt2:simplenarrowAs} in order to comparing with the single channel case as in Fig. \ref{fig:pathInt:meanField}.  When $\delta/E_{0}=1$, $-\tilde{a}_{s}k_{F}=1$ if there was no extra $-2\mu$ shift.}
\begin{equation}\label{eq:pathInt2:E0}
E_{0}=\sqrt{2\delta_{c}\hbar^{2}/m}\cdot{}k_{F}=2\sqrt{\delta_{c}E_{F}}
\end{equation}
$\Delta_{1}$ saturates at about $0.8E_{F}$. At the BCS limit, $n_{o}\approx{}n_{tot}$, $\beta_{0}\approx1$.  $\Delta_{1}$ is exponentially small, there is only very small amount of atoms in the closed-channel.  Moving toward the resonance and then the BEC side, $\Delta_{1}$ increases and the closed-channel density increases as well.  At the universality point of the two-body physics, $\Delta_{1}$ is in the order of the Fermi energy, and $\beta_{o}$ is already substantially smaller than $1$ if the resonance is narrow.  
Because the maximum of the $n_{c}$ is the total density $n_{tot}$, we can see that $\Delta_{1}$ saturates.  The narrower the resonance, the sooner it saturates.  

%\begin{figure}[htbp]
%\begin{center}
%\includegraphics[width=0.8\textwidth]{simpleNarrowMuDt}
%\caption{The chemical potential $\mu$ and gap $\Delta$  over crossover with the Feshbach resonance of different $\delta_{c}$} 
%\label{fig:pathInt2:simpleNarrowMuDt}
%\parbox{0.8\textwidth}{\small  Here $E_{0}=a_{\text{bg}}\mathcal{K}k_{F}$.  $\delta_{c}={E_{0}^{2}}/{E_{F}}$. So the two dashed lines correspond to $\delta_{c}=10E_{F}$ and $100E_{F}$. Notice that $E_{F}$ and $k_{F}$ here are referred to the open-channel quantities. }
%\end{center}
%\end{figure}
%In Fig. \ref{fig:pathInt2:simpleNarrowMuDt}, $\delta_{c}={E_{0}^{2}}/{E_{F}}$, we see the two lines correspond to $\delta_{c}=10E_{F}$ and $100E_{F}$.  Notice that all scales in Fig. \ref{fig:pathInt2:simpleNarrowMuDt} is related the Fermi energy and the Fermi momentum of atoms only in the open-channel.  It is important to find the relation between the open-channel density and the closed-channel density.  
 We can understand the saturation of $\Delta_{1}$ from another respect.  Let us look at the open-channel gap equation Eq. \ref{eq:pathInt2:mfopen} 
\begin{equation}
\Delta_{1\vp}=\sum_{\vk}{}U_{\vp\vk}h_{1\vk}+\sum_{\vk}{}Y_{\vp\vk}h_{2\vk}\tag{\ref{eq:pathInt2:mfopen}}
\end{equation}
Not far away from the resonance, the second term dominates the first term.  And considering Eq. \ref{eq:pathInt2:hphi}, $h_{2\vk}=\alpha\phi^{}_{0\vk}u_{\vk}^{2}$, it is not hard to see that the closed-channel normalization factor $\alpha$ is proportional to $\Delta_1$. Moving toward the BEC end from the BCS end,  the closed-channel density increases as $\Delta_1$ increases and eventually dominates the open-channel density.  In fact,  when the closed-channel dominates, $\alpha\approx\sqrt{N}$; consequently $\Delta_1$ saturates and is directly proportional to the square root of the total density.  
\begin{equation}
\Delta_1\approx\sqrt{N}\sum_{\vk}{}Y_{\vp=0,\,\vk}\phi^{}_{0\vk}u_{\vk}^{2}\approx{\sqrt{n_{\text{tot}}}\Delta_{1,\text{sat}}^{(0)}}
\end{equation}
Here the coefficient $\Delta_{1,\text{sat}}^{(0)}$ is common for one specific Feshbach resonance.  It is possible to observe this saturation experimentally by measuring the fermionic excitation spectrum.

 %At such region, the open-channel density is very small.  Fig. \ref{fig:pathInt2:simpleNarrowMuDt} shows $\Delta_1/E_{F,\text{open}}$ increases toward BEC end.  Given the total density,  this increase is   not because $\Delta_{1}$ increases, but because the open-channel density decreases quickly.  Nevertheless, for the large negative detuning, where this formula applies, the chemical potential is most likely negative and the fermionic excitation modes are $\sqrt{\mu^{2}+\Delta_{1}^{2}}$ instead of simple $\Delta_{1}$.  Furthermore, the more important low-energy excitation modes in such systems are most likely to be the bosonic ones.    
 
  We have the zeroth order result, $\mu^{(0)}$,  $\Delta_1^{(0)}$ from the previous step.  Now we can write down mean-field equations   including the inter-channel Pauli exclusion effects,  expand them and look for the first order corrections,  $\mu^{(1)}$ and $\Delta_1^{(1)}$.
$\Delta_{2}$ is a first order quantity by itself,  and is related to the closed-channel density by Eq. \ref{eq:pathInt2:closeD1}. Therefore, we do not need to include $\Delta_{2}$ when looking for the first order corrections.

We start from the open-channel  gap equation (Eq. \ref{eq:pathInt2:gapRenorm}) and number equation (Eq. \ref{eq:pathInt2:openNum}) 
\begin{gather}
\frac{m}{4\pi{\tilde{a}_{s}(\mu,\lambda_{1})}}=-{\sum(\nth{2E_{\vk}}-\nth{2\epsilon_{\vk}})
+\sum\frac{\Delta_{2}^{2}\xi_{\vk}}{4(\xi_{\vk}+\eta){E_{\vk}^{3}}}}
-\frac{m}{4\pi{\tilde{a}_{s}(\mu,\lambda_{1})}}(\frac{\lambda_{2}}{\Delta_{1}})
	%\tag{ \ref{eq:pathInt2:gapRenorm}}
	\\
N_{\text{open}}\approx\sum{\br{\nth{2}-\frac{\xi_\vk}{2E_\vk}}+\sum\frac{E_\vk-\xi_\vk}{2E_\vk}\frac{\Delta_{1}}{2(E_{\vk}+\xi_{\vk}+\eta)}\zeta -\sum\frac{\Delta_{1}^{3}}{4E_\vk^{3}}\zeta}%\tag{\ref{eq:pathInt2:openNum}}
\end{gather}
We have rearranged these two equations to better compare them  with  the gap equation and the number equation used in the previous step, Eqs. \ref{eq:pathInt2:narrowGapS} and \ref{eq:pathInt2:narrowNumS}.        The main structure of the equations are the similar. In the gap equation (Eq. \ref{eq:pathInt2:gapRenorm}), the second and the third term in the r.h.s are due to the inter-channel Pauli exclusion.  In addition, $\tilde{a}_{s}$ in the l.h.s. also has a small correction due to the shift $\lambda_{1}$(Eq. \ref{eq:pathInt2:asNarrowSim} and Eq. \ref{eq:pathInt2:asKshift}).  
\begin{equation}
\nth{\tilde{a}_{s}^{{(0)}}(\mu^{(0)})}-\nth{\tilde{a}_{s}(\mu,\lambda_{1})}=\frac{1}{\sqrt{2\delta_{c}\hbar^{2}/m}}\lambda_1
\end{equation}
We can expand the $1/(2E_{\vk})$  with $\mu\to\mu^{(0)}+\mu^{(1)}$ and $\Delta_1\to\Delta_1^{(0)}+\Delta_1^{(1)}$, where the zeroth order terms $\mu^{(0)}$ and $\Delta_1^{(0)}$ satisfy gap equation in the previous step, Eq. \ref{eq:pathInt2:narrowGapS}.      
We then get the equation for  $\mu^{(1)}$ and $\Delta_1^{(1)}$. 

\begin{gather}
\frac{m}{4\pi}\frac{\lambda_1}{\sqrt{2\delta_{c}\hbar^{2}/m}}=-\nth{2}\sum\frac{\xi_{\vk}}{E_{\vk}^3}\mu^{(1)}
+\nth{2}\sum\frac{\Delta_1^{(0)}}{E_{\vk}^3}\Delta_1^{(1)}+\sum\frac{\Delta_{2}^{2}\xi_{\vk}}{4(\xi_{\vk}+\eta){E_{\vk}^{3}}}
-\frac{m}{4\pi{\tilde{a}^{{(0)}}_{s}(\mu,\lambda_{1})}}(\frac{\lambda_{2}}{\Delta_{1}})\\
\xi_{\vk}=\epsilon_{\vk}-\mu^{(0)}\\
E_{\vk}=\sqrt{\xi_{\vk}^2+\Delta_1^{(0)}{}^2}
\end{gather}
Similarly, we can expand the $\xi_{\vk}/(2E_{\vk})$ in the number equation  with $\mu\to\mu^{(0)}+\mu^{(1)}$ and $\Delta_1\to\Delta_1^{(0)}+\Delta_1^{(1)}$, where the zeroth order terms $\mu^{(0)}$ and $\Delta_1^{(0)}$ satisfy number equation in the previous step, Eq. \ref{eq:pathInt2:narrowNumS}.  We then find the other equation for  $\mu^{(1)}$ and $\Delta_1^{(1)}$. 
\begin{gather}
0=-\nth{2}\sum\frac{\Delta_1^{(0)}{}^2}{E_{\vk}^3}\mu_1^{(1)}-\nth{2}\sum\frac{\xi_{\vk}\Delta_1^{(0)}}{E_{\vk}^3}\Delta_1^{(1)}+
\sum\frac{E_\vk-\xi_\vk}{2E_\vk}\frac{\Delta_{1}^{(0)}}{2(E_{\vk}+\xi_{\vk}+\eta)}\zeta
 -\sum\frac{\Delta_{1}^{(0)}{}^{3}}{4E_\vk^{3}}\zeta
\end{gather}
Please refer to   Eqs. \ref{eq:pathInt2:lambda1} and \ref{eq:pathInt2:lambda2} as well as Sec. \ref{sec:pathInt2:lambda} for $\lambda_{1}$ and $\lambda_{2}$.  These two quantities describe the overlapping between two channels, i.e. the inter-channel Pauli exclusion.  Their values are hard to estimate without the detail knowledge of the potential and wave-functions.  Nevertheless, they can be derived from the experiments (Sec. \ref{sec:pathInt2:lambda}). Once we obtain them, we can  solve the above equations  for the first order correction  $\mu^{(1)}$ and $\Delta_1^{(1)}$. In the above equations, all correction terms either have $\zeta$ explicitly or in the similar order.   Given the non-singular nature of the crossover problem, we  expect the correction $\mu^{(1)}$ and $\Delta_1^{(1)}$ to be also in the order of $\zeta$.

There are three distinct length scales in the problem.  The range of the potential, $r_c$, is the smallest.  The intra-channel and inter-channel potentials are  essentially zero outside this range. Potential energy dominate kinetic energy in this range and the correlation (wave function) is totally governed by the potential and might have large oscillation.   The two-body correlation follows the two-body wave function in this range.  The second range is the size of the closed-channel bound-state, $a_c$.  The closed-channel bound state  has only negligible weight outside this range.  In this range, the many-body correlation still follow the two-body wave function in the closed-channel.  We assume that  the closed-channel bound state is close to threshold  and therefore has most weight in this region.  The inter-channel Pauli exclusion are mostly accounted in this  region.  Not surprisingly, this region contributes most in the integral of $\lambda_{1}$ and $\lambda_{2}$. When much larger than $a_c$, many-body effects are important. However, the closed-channel has only negligible weight in it and therefore only the open-channel needs to be considered.

\chapter{Excitation modes\label{ch:excitation}}
In the single channel crossover, fermionic modes are related to the two-body correlation of the original fermions; while the bosonic modes are related to the two-body correlation of the auxiliary new bosonic fields (order parameters $(\Delta_{1}, \Delta_{2})$).  Similarly to the single-channel case, most modes are gapped with minimum at  $\Delta_{1}$ and only one  Goldstone mode for the in-phase phase fluctuation of order parameters (bosonic fields) is gapless with linear dispersion at low energy. 
\section{Fermionic excitation modes and Bogoliubov transformation\label{sec:pathInt2:bog}}
If we limit ourselves in the mean field level, we can interpret the transformation $T_\vk{}L_\vk$ in Eq. \eqref{eq:pathInt2:B} as the Bogoliubov canonical transformation, while  the $3\times3$ fermionic correlation matrix $B$ in Eqs. (\ref{eq:pathInt2:Bapprox}-\ref{eq:pathInt2:xiExpand3}) gives us the spectrum of the  fermionic quasi-particle excitation. From Eq. \eqref{eq:pathInt2:actionMix} and Eq. \eqref{eq:pathInt2:B}, the action is diagonal for new fermions (quasi-particles) field 
\begin{equation*}
\Phi_\vk\equiv\mtrx{\phi_{\text{\Rmnum{1}},+\vk}\\\bar\phi_{\text{\Rmnum{2}},-\vk}\\\bar\phi_{\text{\Rmnum{3}},-\vk}}
=L^{\dg}_{\vk}T^{\dg}_{\vk}\mtrx{\psi_{a\,\vk}\\\bar\psi_{b\,-\vk}\\\bar\psi_{c\,-\vk}}
\end{equation*}
Putting the above transformation into the operator language and using $\mathcal{X}_{\text{I}}$, $\mathcal{X}_{\text{II}}$, and $\mathcal{X}_{\text{III}}$ to denote the new quasi-particles,  we have the relation 
\begin{equation}
\mtrx{\mathcal{X}^{}_{\text{\Rmnum{1}},+\vk}\\\mathcal{X}_{\text{II},-\vk}^{\dg}\\\mathcal{X}_{\text{III},-\vk}^{\dg}}
=L^{\dg}_{\vk}T^{\dg}_{\vk}  \mtrx{a_{\vk}\\b_{-\vk}^{\dg}\\c_{-\vk}^{\dg}}
\end{equation}   

\begin{equation}\tag{\ref{eq:pathInt2:T}}
T_k=\mtrx{u_k&v_k&0\\-v_k&u_k&0\\0&0&1}
\end{equation}
\begin{equation}\tag{\ref{eq:pathInt2:L1}}
L_{\vk}\approx{}I+
\mtrx{0&-\frac{\Delta_{1}{}\Delta_{2}{}}{4E^{2}_{\vk}}&u_{\vk}\\
\frac{\Delta_{1}{}\Delta_{2}{}}{4E^{2}_{\vk}}&0&v_{\vk}\\
-u_{\vk}&-v_{\vk}&0
}\frac{\Delta_{2}{}}{\eta}
\equiv{}I+\delta_{k}\qquad
L^{\dg}_{\vk}=I-\delta_{\vk}
\end{equation}
First of all, mixture of $a_{\vk}$ and $b^{\dg}_{-\vk}$ ($c^{\dg}_{-\vk}$) indicates the pairing of atoms of the opposite momentum. Furthermore,   mixture of creators and annihilators dictates the approach of the grand-canonical ensemble, i.e., the ground state is a number non-conserved state.   In addition, $L$ matrix cannot be separated into two channels, which indicate the mixture of two channels.   At the mean field level, both $\Delta_{1}$ and $\Delta_{2}$ are taken as real and constant.  The new Hamiltonian at this level is 
\begin{equation}
\hat{H}=f(\Delta_{1},\Delta_{2})+E_{1\,\vk}\mathcal{X}^{\dg}_{\text{I},+\vk}\mathcal{X}^{}_{\text{I},+\vk}
+E_{2\,\vk}\mathcal{X}^{\dg}_{\text{II},-\vk}\mathcal{X}^{}_{\text{II},-\vk}
+E_{3\,\vk}\mathcal{X}^{\dg}_{\text{III},-\vk}\mathcal{X}^{}_{\text{III},-\vk}
\end{equation}
And the spectrum is just like we calculated in Sec. \ref{sec:diagonalGreen}, Eqs. \ref{eq:pathInt2:xiExpand}-\ref{eq:pathInt2:xiExpand3}.  We quote here again
\begin{align}
E_{1\vk}&\equiv{}E_{\vk}+\gamma_{1\vk}\approx{}E_{\vk}+u_{\vk}^{2}\zeta\tag{\ref{eq:pathInt2:xiExpand}}\\
E_{2\vk}&\equiv{}E_{\vk}+\gamma_{2\vk}
\approx{}E_{\vk}-v_{\vk}^{2}\zeta\tag{\ref{eq:pathInt2:xiExpand2}}\\
E_{3\vk}&\equiv{}\xi_{\vk}+\eta+\gamma_{3\vk}\approx{}\epsilon_{\vk}+\eta+\frac{\zeta}{2}
\tag{\ref{eq:pathInt2:xiExpand3}}
\end{align}
Here $\gamma_{1}$, $\gamma_{2}$ and $\gamma_{3}$ are correction due to the inter-channel Pauli exclusion. 

This clearly shows that $\mathcal{X}^{\dg}_{\text{I},\vk}$, $\mathcal{X}^{\dg}_{\text{II},\vk}$, and $\mathcal{X}^{\dg}_{\text{III},\vk}$ are fermionic quasi-particle excitation modes with spectrum $E_{i\,\vk}$ and   $L^{\dg}_{\vk}T^{\dg}_{\vk}$ is  the Bogoliubov canonical transformation to transfer the normal fermionic modes into these elementary quasi-particle modes .  Here we see in the excitation, different species of opposite momentum, $(a,b)$ and $(a,c)$, mixed together to form the elementary excitation due to the paring in the ground state.  
  From Eqs. (\ref{eq:pathInt2:xiExpand}-\ref{eq:pathInt2:xiExpand3}), we see that the fermionic excitation modes basically follow the pattern in the broad resonance.  In the broad resonance where the closed-channel only modifies the interaction in the open-channel,  there are  three fermionic quasi-particle modes: two (degenerate) Bogoliubov quasi-particle modes  in the open-channel, $E_{\vk}=\sqrt{\xi^{2}_{\vk}+\Delta_{1}^{2}}$ as in BCS theory (gapped at $\Delta_{1}$ in the BCS-like states, $\mu>0$ and ${\sqrt{\mu^{2}+\Delta_{1}^{2}}}$ in the BEC-like states, $\mu<0$)  
; and one high fermionic excitation mode in the closed-channel, $\xi_{\vk}+\eta$, as in normal gas.  In the narrow resonance, first of all, the mean-field value of the gap $\Delta_{1}$ (or $\sqrt{\mu^{2}+\Delta_{1}^{2}}$) itself is modified as described in the last chapter (Sec. \ref{sec:pathInt2:mean2}).  Once that is taken into account, the above conclusion is approximately correct except high-order corrections in $\zeta$.   The originally double degenerate excitation modes, $E_{\vk}=\sqrt{\xi^{2}_{\vk}+\Delta_{1}^{2}}$ , now split by $\zeta\Delta_{1}$; while the third high excitation mode corresponding to the normal fermionic excitation in the closed-channel has a small correction in the same order.   On the other respect, as discussed earlier in Sec. \ref{sec:diagonalGreen},  corrections  due to the inter-channel Pauli exclusion do not vanish when the inter-channel coupling, $Y$, approaches zero.  

\section{Collective excitation modes}
Fermionic modes are derived from the correlation function $\nG$ of the fermion fields $\Psi$, and therefore are mostly single (quasi)particle like.  On the other hand, order parameters ($\Delta_{1}$, $\Delta_{2}$) are defined in terms of collective behavior of many fermion atoms.  Fluctuations of order parameters thus marked the collective excitation modes of the system. Here with a two-component order parameter, four independent modes exist:   two for magnitude variation of each $\Delta_i$,  internal phase between two $\Delta_i$, and the overall local phase $\theta(x)$ of $\Delta_1$ and $\Delta_2$.  The first three change the magnitude of action and therefore massive; while the last one leaves the action invariant and thus massless. 
We study two modes of the phase fluctuation.  The in-phase mode is the counterpart of the Anderson-Bogoliubov modes in the single channel problem; while the two channels introduce a new out-of-phase mode. 
 
  For the single-channel crossover, Sec. \ref{sec:collective1} considers all (magnitude and phase) modes of the order parameter fluctuation at the same time, and only calculates  the low sound-like part of the spectrum. In the two-channel case, a general analysis with all modes in becomes unwieldy.  
We instead focus on one mode a time. We isolate the change in one mode and leave others at zero.

\subsection{The in-phase phase fluctuation }
The action of $\Delta$, $S(\bar{\Delta}_i,\Delta_i)$ (Eqs. \ref{eq:pathInt2:nG}, \ref{eq:pathInt2:actionD}), is invariant if the phases of $\Delta_{1\,\vk}$ and $\Delta_{2\,\vk}$ rotate simultaneously. We therefore conclude that there exists a massless (Goldstone) mode corresponding to the local phase invariance. We first study the massless two-channel in-phase phase fluctuation.   Introduce the phase fluctuation $\theta$, 
\begin{equation*}
\Delta_{i}(x)\rightarrow{}\Delta_{i}e^{i2\theta(x)}\qquad{}
\bar{\Delta}_{i}(x)\rightarrow{}\bar{\Delta}_{i}e^{-i2\theta(x)}
\end{equation*}
This is equivalent to  phase  rotation of fermionic variable $\psi$
\begin{equation*}
\psi_{i}(x)\rightarrow{}\psi_{i}(x)e^{i\theta(x)}\qquad{}
\bar{\psi}_{i}(x)\rightarrow{}\bar{\psi}_{i}(x)e^{-i\theta(x)}
\end{equation*}
Again, the phase $\theta(x)$ is common for  both channels.   With a phase shift, we can rewrite the action (taken mean-field value $\Delta^{(0)}=(\Delta_1,\Delta_2)^\dg$) (here we  follow treatment of Nagaosa\cite{Nagaosa})
\begin{subequations}\label{eq:pathInt2:actionPhase}
\begin{align}
S[\theta,\bar\psi_{i},\psi_{i}]=&S_0[\bar\psi_{i},\psi_{i}]+S_1[\theta,\bar\psi_{i},\psi_{i}]+S_2[\theta,\bar\psi_{i},\psi_{i}]\\
S_0[\bar\psi_{i},\psi_{i}]=&\int{dx}
\Big\{\sum_{j}\bar\psi_{j}(\partial_\tau-\nth{2m}\nabla^{2}-\mu+\eta_{j})\psi_{j}\nonumber\\
&\quad+\Delta^{(0)}{}^{\dg}\tilde{U}^{-1}\Delta^{(0)}-(\bar\psi\bar\psi)\Delta^{(0)}-{\Delta^{(0)}}{}^{\dg}(\psi\psi)\Big\}\\
S_1[\theta,\bar\psi_{i},\psi_{i}]=&\int{dx}\sum_{j}\Big\{
   i\,\bar\psi_{j}(\partial_{\tau}\theta)\psi_{j}+\nabla\theta\cdot\nth{2mi}[\bar\psi_{j}\nabla\psi_{j}-(\nabla\bar{\psi}_{j})\psi_{j}]\Big\}\\
S_2[\theta,\bar\psi_{i},\psi_{i}]=&\int{dx}\sum_{j}\nth{2m}(\nabla\theta)^{2}\bar\psi_{j}\psi_{j}
\end{align}
\end{subequations}
Note that here $\Delta^{(0)}$ is the mean-field value of $\Delta_{i}$, a constant 2-component vector, no longer a functional variable.  Here we see that $S_{0}$ has the same form as before except it only takes the mean field value of $\Delta$ and it is described by the same correlation $G_{0}$ (Eq. \ref{eq:pathInt2:nG}).  We can regard $S_{1}$ and $S_{2}$ as perturbation for the so-called gradient expansion on $\nabla\theta$ \cite{Altland}.  It is then obvious that $S_{1}$ is in the first order while $S_{2}$ is in the second order regarding with the (time or space) derivative of $\theta$.  Use the same spinor representation as before, $S$ is bilinear of $\psi$ and therefore we can formally integrate out $\psi$. 
\begin{equation}
S[\theta]=const.+\ln\det\nG(\theta)
\end{equation}
 We write out the formal Green's function according to the above action (with respect to the Nambu-like spinor)
\begin{subequations}
\begin{align}
\nG=&G_{0}^{-1}+K_{1}+K_{2},\\
K_{1\, k,k'}=
	&\nth{({\beta{}V)}^{1/2}}(\omega_n-\omega_{n'})\theta(k-k')\sigma_3+
		\nth{{(\beta{}V)}^{1/2}}i\frac{(\vk-\vk')\cdot(\vk+\vk')}{2m}\theta(k-k')\hat{1}\\
K_{2\, k,k'}=
	&\nth{2m}\sum_{q,q'}\nth{{\beta{}V}}(\vq\cdot\vq')\theta(q)\theta(q')\delta(q+q'+k-k')\sigma_3
\end{align}
\end{subequations}
where $G_{0}$ is the same as (Eqs. \ref{eq:pathInt2:nGDelta}, \ref{eq:pathInt2:nG}).  Here $k=(\omega_n,\vk)$ and $k'=(\omega_{n'},\vk')$.  Like in the single channel, $\hat{1}$ is $3\times3$ identity matrix, and 
\begin{equation}
\sigma_3=\mtrx{1&0&0\\0&-1&0\\0&0&-1}
\end{equation}
As the expansion in Eq. \ref{eq:pathInt:expand}, we can look for the expansion of $\nG$ over $K_{1,2}$.  
\begin{equation}\tag{\ref{eq:pathInt:expand}}
\tr\ln \nG=\tr\ln\hat{G_{0}}^{-1}+\tr(\hat{G_{0}}\hat{K})-\nth{2}\tr(\hat{G_{0}}\hat{K}\hat{G_{0}}\hat{K})+\cdots
\end{equation}
For the first order, $\tr(\hat{G_{0}}\hat{K})$, 
\begin{align}
\tr(\hat{G_{0}}\hat{K_1})=&\sum_{k}{G_{0\,k}K_{1\,k,k}}=0\\
\tr(\hat{G_{0}}\hat{K_2})=&\sum_{k}{G_{0\,k}K_{2\,k,k}}\nonumber\\
	=&-\nth{2m}\nth{{\beta{}V}}\sum_{k}\tr(\hat{G}_{0\,k}\sigma_3)\sum_{q}q^2\theta{q}\theta{-q}\nonumber\\
	=&-\frac{n}{2m}\sum_{q}q^2\theta{(q)}\;\theta{(-q)}
\end{align}
Here we use the fact $\nth{{\beta{}V}}\sum_{k}\tr(\hat{G}_{0\,k}\sigma_3)=n$. $\tr(\hat{G_{0}}\hat{K_2})$ is already in the second order of $\theta$, and we only need to keep the expansion of $K_2$ to this order. On the other hand, we have to go to the second order of $K_1$ for the second order of $\theta$. 
\begin{align}		
\tr(\hat{G_{0}}{K_1}\hat{G_{0}}{K_1})=&\sum_{k,q}\tr(\hat{G}_{0,k+q}K_{1\,k+q,k}\hat{G}_{0\,k}K_{1\,k,k+q})\label{eq:pathInt2:GKGK}\\
=&\nth{{\beta{}V}}\sum_{k,q=(\omega_m,\vq)}\theta(q)\theta(-q)\Big[(-\omega_m^2)\tr(\hat{G}_{0,k+q}\sigma_3\hat{G}_{0\,k}\sigma_3)\nonumber\\
&\quad+\nth{m^2}\sum_{i,j=(x,y,z)}q_iq_j(k_i+\frac{q_i}{2})(k_j+\frac{q_j}{2})\tr(\hat{G}_{0,k+q}\hat{G}_{0\,k})\Big]\label{eq:pathInt2:GKGK2}\\
\equiv&\sum_{q}\theta(q)\theta(-q)\big[-\pi^{(0)}(q)\omega_m^2+\sum_{i,j=(x,y,z)}\pi^{(\perp)}_{ij}(q)q_iq_j\big]\label{eq:pathInt2:pi}
\end{align}
Here we introduce $q=(\omega_m,\vq)=k-k'$.  \emph{We can take $q=0$ in $\pi^{(0)}(q)$ and $\pi^{(\perp)}(q)$ for low frequency and momentum ($\omega_m,\nth{2m} |\vq|^2\ll\Delta_{1,2}$).  } Use the previous result of Sec. \ref{sec:diagonalGreen}, we can calculate the Green's function in the lowest and the first order of $\zeta$ in $\pi$'s.  After some long but straightforward algebra (see Appendix. \ref{sec:calculatePi}), we find
\begin{gather}
\pi^{(0)}(0)\approx\sum_{\vk}\frac{\Delta_{1}^{2}}{E_{\vk}^{3}}
	-\sum_{\vk}\frac{\Delta_{1}^{2}\Delta_{2}^{2}\xi_{\vk}}{2E_{\vk}^{5}(\xi_{\vk}+\eta)}
\label{eq:pathInt2:pi0}\\
\pi^{(\perp)}(0)=0
\end{gather}
Combining all these together, we have a new action for the phase fluctuation $\theta$
\begin{equation}
S[\theta]=\int{dx}\sum_{q}\theta(q)\theta(-q)\big[\nth{2}\pi^{(0)}(0)\omega_m^2-\frac{n}{2m}q^2\big]
\end{equation}
The correlation determines the velocity of the Anderson-Bogoliubov collective mode.  The second term in Eq. \eqref{eq:pathInt2:pi0}, $-\sum_{\vk}\frac{\Delta_{1}^{2}\Delta_{2}^{2}\xi_{\vk}}{2E_{\vk}^{5}(\xi_{\vk}+\eta)}\approx-\sum_{\vk}\frac{\Delta_{1}^{3}}{2E_{\vk}^{3}}\frac{\xi_{\vk}}{E_{\vk}}\nth{E_{\vk}}\zeta$, is the only correction in the next order.  Thus, this mode   is qualitatively similar as  the one in the single-channel with some correction in the order of $\zeta\ll1$ (see Appendix \ref{sec:pathApp:consistency}).

\subsection{The out-of-phase phase fluctuation}
Another interesting phase fluctuation of order parameters is the  phase fluctuation in two channels out of sync.  This mode is unique for a two-channel system without any counterpart in the single-channel model.  When  the phase fluctuation of two channels are out of  sync,  the inter-channel coupling strength changes.  Thus, it is  expected to be  a gapped (massive) mode.  We narrow down to the mode that the phases of two atoms  ($\psi_{b}$ and $\psi_{c}$) are opposite and leave all other modes constant.  
\begin{equation*}
\mtrx{\psi_{a}(x)\\\psi_{b}(x)\\\psi_{c}(x)}\rightarrow{}
	\mtrx{\psi_{a}(x)\\\psi_{b}(x)e^{+i\theta(x)}\\\psi_{c}(x)e^{-i\theta(x)}}
\qquad{}
\mtrx{\bar\psi_{a}(x)\\\bar\psi_{b}(x)\\\bar\psi_{c}(x)}\rightarrow{}
	\mtrx{\bar\psi_{a}(x)\\\bar\psi_{b}(x)e^{-i\theta(x)}\\\bar\psi_{c}(x)e^{+i\theta(x)}}
\end{equation*}
The order parameters do not have a simple transformation because they are connected to two channels via $2\times2$ interaction matrix $\tilde{U}$, which mixes two channels (Eq. \ref{eq:pathInt2:DeltaPhi}).  
\begin{equation*}
\begin{pmatrix}\Delta_{1}(x)\\\Delta_{2}(x)\end{pmatrix}\rightarrow{}
	\mtrx{U&Y\\Y^{*}&V}\begin{pmatrix}\psi_{b}\psi_{a}(x)e^{+i\theta(x)}\\\psi_{c}\psi_{a}(x)e^{-i\theta(x)}\end{pmatrix}
%	=\begin{pmatrix}\Delta_{1}^{(0)}(x)\\\Delta_{2}^{(0)}(x)\end{pmatrix}+
%	\mtrx{U&Y\\Y^{*}&V}\begin{pmatrix}\av{\psi_{b}\psi_{a}}(x)(e^{+i\theta(x)}-1)\\\av{\psi_{c}\psi_{a}}(x)(e^{-i\theta(x)}-1)\end{pmatrix}
	%\approx\mtrx{U&Y\\Y^{*}&V}\mtrx{\psi_{b}\psi_{a}(x)(1+i\theta)\\\psi_{b}\psi_{a}(x)(1-i\theta)}
\end{equation*}
This term cannot be easily written in terms of mean-field value $\Delta_i$.   On the other hand, as mentioned before, we freeze all  other modes to their mean-field value except $\theta$.  We therefore use another two-component of atom pairs$({\psi_{b}\psi_{a}},{\psi_{c}\psi_{a}})$, which is the linear recombination of $(\Delta_{1},\Delta_{2})$.  
%On the other hand, we do not have a full two-dimensional functional variables $\Delta_i$. 
%\begin{equation*}
%\begin{pmatrix}\bar\Delta_{1}(x)&\bar\Delta_{2}(x)\end{pmatrix}\rightarrow{}
%	\begin{pmatrix}\bar\Delta_{1}(x)e^{-i\theta(x)}&\bar\Delta_{2}(x)e^{+i\theta(x)}\end{pmatrix}
%\end{equation*}

%The original action can be expended in the similar fashion as Eq. \ref{eq:pathInt2:actionPhase}. Indeed it is almost the same except there is one more terms due to inter-channel coupling term.  
%\begin{align*}
%&\Delta^{(0)}{}^{\dg}\tilde{U}^{-1}\Delta^{(0)}-(\bar\psi\bar\psi)\Delta^{(0)}-{\Delta^{(0)}}{}^{\dg}(\psi\psi)\\
%=&\bar\Delta^{(0)}_{1}{}^{\dg}\tilde{U}^{-1}_{11}\Delta^{(0)}_{1}+\bar\Delta^{(0)}_{2}\tilde{U}^{-1}_{22}\Delta^{(0)}_{2}
%	+\bar\Delta^{(0)}_{1}\tilde{U}^{-1}_{12}\Delta^{(0)}_{2}e^{-i2\theta(x)}
%	+\bar\Delta^{(0)}_{2}\tilde{U}^{-1}_{21}\Delta^{(0)}_{1}e^{+i2\theta(x)}\\
%&-(\bar\psi\bar\psi)\Delta^{(0)}-{\Delta^{(0)}}{}^{\dg}(\psi\psi)\\
%\end{align*}
Comparing to the in-phase phase rotation, the out-of-phase one does modify the action's magnitude. Write the interaction term in the momentum space,
\begin{align*}
&\Delta^{}{}^{\dg}\tilde{U}^{-1}\Delta^{}\\
=&\mtrx{\av{\bar{\psi}_{b}\bar{\psi}_{a}}e^{-i\theta}&\av{\bar{\psi}_{c}\bar{\psi}_{a}}e^{+i\theta}}\tilde{U}^{}\tilde{U}^{-1}\tilde{U}
\mtrx{\av{\psi_{b}\psi_{a}}e^{+i\theta}\\\av{\psi_{c}\psi_{a}}e^{-i\theta}}\\
%=&\mtrx{\av{\psi_{b}\psi_{a}}e^{-i\theta}&\av{\psi_{c}\psi_{a}}e^{+i\theta}}\mtrx{U&Y\\Y^{*}&V}
%\mtrx{\av{\psi_{b}\psi_{a}}e^{+i\theta}\\\av{\psi_{c}\psi_{a}}e^{-i\theta}}
=&\mtrx{\htd_{1}^{*}&\htd_{2}^{*}}\mtrx{U&Ye^{-i2\theta}\\Y^{*}e^{+i2\theta}&V}\mtrx{\htd_{1}\\\htd_{2}}\\
=&\Delta^{(0)}{}^{\dg}\tilde{U}^{-1}\Delta^{(0)}
+\mbr{Y(e^{-i2\theta}-1)\htd_{1}^{*}\htd_{2} +Y^{*}(e^{+i2\theta}-1)\htd_{1}\htd_{2}^{*}}
\end{align*}
Note that here we take the mean-field expectation of pairs of fermions $\tilde{h}_{1}=\av{\psi_{b}\psi_{a}}=\sum_{\vk}h_{1\vk}$, $\tilde{h}_{2}=\av{\psi_{c}\psi_{a}}=\sum_{\vk}h_{2\vk}$, and their complex conjugates\footnote{$\psi\psi$ and $\bar\psi\bar\psi$ are complex conjugate in the mean-field level.}.  Comparing with the action of the in-phase phase fluctuation (Eq. \ref{eq:pathInt2:actionPhase}), we find one more terms in the new action.  
\begin{subequations}\label{eq:pathInt2:actionPhase2}
\begin{align}
S[\theta,\bar\psi_{i},\psi_{i}]=&S_0[\bar\psi_{i},\psi_{i}]+S_1[\theta,\bar\psi_{i},\psi_{i}]+S_2[\theta,\bar\psi_{i},\psi_{i}]
	+S_3[\theta]\\
S_0[\bar\psi_{i},\psi_{i}]=&\int{dx}
\Big\{\sum_{j=(a,b,c)}\bar\psi_{j}(\partial_\tau-\nth{2m}\nabla^{2}-\mu+\eta_{j})\psi_{j}\nonumber\\
&\quad+\Delta^{(0)}{}^{\dg}\tilde{U}^{-1}\Delta^{(0)}-(\bar\psi\bar\psi)\Delta^{(0)}-{\Delta^{(0)}}{}^{\dg}(\psi\psi)\Big\}\\
S_1[\theta,\bar\psi_{i},\psi_{i}]=&\int{dx}\Big\{
   i\,(\partial_{\tau}\theta)(\bar\psi_{b}\psi_{b}-\bar\psi_{c}\psi_{c})\\
   &\quad+\nabla\theta\cdot\nth{2mi}[\bar\psi_{b}\nabla\psi_{b}-\bar\psi_{c}\nabla\psi_{c}
   -(\nabla\bar{\psi}_{b})\psi_{b}   +(\nabla\bar{\psi}_{c})\psi_{c}]\Big\}\\
S_2[\theta,\bar\psi_{i},\psi_{i}]=&\int{dx}\nth{2m}(\nabla\theta)^{2}(\bar\psi_{b}\psi_{b}+\bar\psi_{c}\psi_{c})\\
S_3[\theta]=&\int{dx}\mbr{Y(e^{-i2\theta}-1)\htd_{1}^{*}\htd_{2} +Y^{*}(e^{+i2\theta}-1)\htd_{1}\htd_{2}^{*}}
\end{align}
\end{subequations}
Let us look at $S_{3}$ first. We can expand it around small $\theta$.  
\begin{align}
S_3[\theta]=&\int{dx}\mbr{Y(e^{-i2\theta}-1)\htd_{1}^{*}\htd_{2} +Y^{*}(e^{+i2\theta}-1)\htd_{1}\htd_{2}^{*}}\\
  =&\int{dx}-i2\mbr{Y\htd_{1}^{*}\htd_{2} -Y^{*}\htd_{1}\htd_{2}^{*}}\theta
  	-2\mbr{Y\htd_{1}^{*}\htd_{2} +Y^{*}\htd_{1}\htd_{2}^{*}}\theta^2
\end{align}

From Eq. \ref{eq:pathInt2:h1}
\begin{equation*}
 h_{1\vk}=\Delta_{1}\frac{E_{1\,\vk}+\xi_{\vk}+\eta}{(E_{1\,\vk}+E_{2\,\vk})(E_{1\,\vk}+E_{3\,\vk})}
\end{equation*}
If $\Delta_{1}\to\Delta_{1}e^{i\varphi}$, all $h_{1\vk}$ has argument $\varphi$ (or $\pi+\varphi$). Thus $\htd_{1}$ has argument $\varphi$ (or $\pi+\varphi$) as well.  Also consider the mean-field equation  $\Delta_{1}=U\htd_{1}+Y\htd_{2}$, We can rewrite 
\begin{equation}
Y\htd_{1}^{*}\htd_{2}=\htd_{1}^{*}\Delta_{1}-\htd_{1}^{*}U\htd_{1}
\end{equation}
$U$ is real for a Hermite interaction.  So this quantity is real.  In addition,  $Y^{*}\htd_{1}\htd_{2}^{*}$ is complex conjugate of $Y^{}\htd_{2}\htd_{1}^{*}$. So they are equal because they are both real.  From all these, we conclude that the linear term of $\theta$ vanishes.  This is actually expected for a perturbation around the saddle point.  Now we can rewrite this term
\begin{equation}
S_{3}[\theta]=\int{dx}-4Y\htd_{1}^{*}\htd_{2}\theta^{2}=\sum_{q}\theta({q})\theta({-q})(-4Y\htd_{1}^{*}\htd_{2})
\end{equation}
$Y\htd_{1}^{*}\htd_{2}$ is actually the expectation of coupling for the two channel correlation.  It is not hard to see that this value  is negative in the minimum (saddle point).  

The other two terms in the expansion of the action, $S_{1}$ and $S_{2}$, go in the same way as the in-phase phase fluctuation except an extra  $\nth{2}$ because only hyperfine species ($b$ and $c$) participates.  They actually produce the same result except only a half of the particles $(b,c)$ participates. ($n_{b}+n_{c}=n_{a}=n/2$)   The full action about $\theta$ is 
\begin{equation}\label{eq:pathInt2:outofphase}
S[\theta]=\int{dx}\sum_{q}\theta(q)\theta(-q)\big[\nth{4}\pi^{(0)}(0)(\omega_m^2-\omega_{0}^{2})-\frac{n}{4m}q^2\big]
\end{equation}
where
\begin{equation}
\omega_{0}^{2}=-\frac{16Y\htd_{1}^{*}\htd_{2}}{\pi^{(0)}(0)}
\end{equation}
As expected, the out-of-phase phase mode is a gapped mode, with a spectrum starting from $\omega_{0}$. When two channels' phases fluctuate out-of-phase, the original self-consistent mean-field equation cannot be satisfied and this fluctuation is coupled with the fermionic modes.  Its general spectrum is expected to related to the pair-breaking energy, but Eq. \ref{eq:pathInt2:outofphase} can only be calculated numerically in general.  Particularly, $Y\htd_{1}^{*}\htd_{2}$ is hard to estimate except at the ends of crossover. At the BCS end, $Y\htd_{1}^{*}\htd_{2}$ is close the expectation of the interaction energy calculated according to the effective interaction in the open-channel.  We can estimate it as $Y\htd_{1}^{*}\htd_{2}\sim{}-N(0)\Delta_{1}^{2}$, where $N(0)$ is the density of the state at the Fermi energy.  $\pi^{(0)}(0)\sim{}N(0)$.  So we can estimate that $\omega_{0}\sim\Delta_{1}$.  This mode is inside the fermionic excitation spectrum.  At the BEC end, the interaction energy in the open-channel can be estimated roughly as $N_{\text{open}}\abs{\mu}$; $\pi^{(0)}(0)\sim{}\mathcal{V}_{0}\Delta_{1}^{2}/\abs{\mu}^{3/2}$. Note that here the open-channel atom number $N_{\text{open}}$ might be significantly smaller than the total atom number. Here, $\Delta_{1}\sim{}n_{o}^{1/2}a_{s}^{-1/2}$ and $\abs{\mu}\sim{}a_{s}^{-2}$.  Put all these together, $\omega_{0}\sim{}n^{1/2}\abs{\mu}^{5/4}/\Delta_{1}\sim{}a_{s}^{-2}$. It is in the order of the ionization (pair breaking) energy again and is only related to two-body physics.   
%\begin{subappendices}
%\include{pathAppendix}
%\end{subappendices}

\chapter{Conclusions\label{ch:conclusion}}
% !TeX root =thesis.tex
% 
In this thesis, we study the narrow Feshbach resonance in  the three-species case where two channels share the same species.  

In general, for the narrow resonance without a shared species, the main correction to the single-channel result comes from the  extra counting of the atoms in the open-channel, which leads to the extra shift $2\mu$ in $\tilde{a}_{s}$, and the closed-channel, which leads to the extra number equation.  Two number equations exist, one for the open-channel and one for the closed-channel.  The open-channel number equation resembles the number equation of the single-channel model.

When there is a common species,  however, the  Pauli exclusion between two channels due to the common species in the three-species narrow resonance, calls for careful  consideration.  Our treatment follows the spirit of the so-called ``universality'' idea.  For a dilute system with a short-range potential, such as the dilute ultracold alkali gas, the short-range part of a two-body correlation does not significantly change from two-body  to many-body.  This particular feature justifies using the two-body quantities (e.g. the s-wave scattering length $a_{s}$) as the boundary condition for the many-body correlations.  The most part of the correlations, where no two particles are close in short-range, is essentially free from interaction. This long-range part of the correlations does change from two-body to many-body.  Unique in our problem is the involvement of the Feshbach resonance.  A resonance makes the system extremely sensitive to even small change in the relative weight between two channels. We use the fact that,  in the Feshbach resonance, the closed-channel bound state, $\phi_{0}$, as a short-range object,  is still not very sensitive to the resonance; hence, we can, within the universality idea,  apply a simple boundary condition to the closed-channel correlation. However, the indirect interaction mediated by the closed-channel dominates the direct interaction within the open-channel. Consequently, a small change in the closed-channel bound-state weight \emph{does} affect the short-range wave function in the open-channel, originally expressed through the boundary condition using the s-wave scattering length $a_{s}$. 

When the spatial extension of  the closed-channel bound state is  of the order of the   inter-particle distance, $a_{0}$, or even larger, the Feshbach resonance in the many-body context is a genuine three-species many-body problem and no simple solution is available.  By contrast, when the bound-state's spatial extension is much smaller than the interparticle distance, the ratio ($a_{c}/a_{0}$) serves as the expansion parameter and we can extract  the effect of the inter-channel Pauli exclusion perturbatively.  In essence, we can then  ignore the many-body effects within  closed-channel bound-states, while only taking into consideration  the Pauli exclusion between channels.  A few new parameters need to be introduced and can be calibrated from experiments, such as $\lambda_{1}$ (Eq. \ref{eq:pathInt2:lambda1}), $\lambda_{2}$ (Eq. \ref{eq:pathInt2:lambda2}).  Mean field properties can still be determined through gap equations and number equations similar as in the single-channel case.  The excitation modes are also close to the original single-channel result with correction of the order of $a_{c}/a_{0}$, where $a_{c}$ is the spatial extension of the uncoupled closed-channel  bound state and $a_{0}$ is the average interparticle distance.

We can distinguish three different regimes for the many-body energy scale $E_F$ when compared with the two-body physics.  
\begin{enumerate}
\item The regime where $E_F$ is much smaller than the characteristic energy $\delta_c$ (See Eq. \ref{eq:intro:deltaC} in Sec. \ref{sec:intro:twobody}). Around resonance, the closed-channel weight is negligible in this regime. We can then ignore the closed-channel completely and approximate the effective open-channel interaction by a  pseudo-potential characterized by the s-wave scattering length, $a_s$. 
\item The regime where  $E_F$ is larger than $\delta_c$, but smaller than the closed-channel bound state binding energy $E_b$, or the bare detuning $\eta$.  The closed-channel weight is significant in this regime and cannot be ignored, which means $n_{\text{open}}+n_{\text{close}}=n_{\text{total}}$. Also, we need to take into account the fact that, in the resonance formula of $a_s$, shifting should be counted from the Fermi surface instead of from zero as in two-body physics.    These effects, considered previously\cite{GurarieNarrow}, are  also carefully analyzed in this thesis.  Furthermore, the Pauli exclusion between two channels needs to be taken into account when these two channels share a common species.  Nevertheless, at  low momentum, where the open-channel has most of its weight, the closed-channel bound state only has very small weight.  Consequently, we can still treat the Pauli exclusion between channels perturbatively using the expansion coefficient  $E_F/\eta$. 
\item The regime where  $E_F$ is larger than the binding energy $E_b$ or the bare detuning $\eta$. We have a genuine three-species many-body problem. This regime can be achieved in two ways.  One way requires a very large $E_F$, which indicates a very dense system.  However, it would be hard to imagine that  the original dilute alkali gas model still applies in this case. Various approximations in the model would probably break down beforehand, such as the pseudo-potential approximation with $a_s$. The second way requires a very small $\eta$, which indicates a genuine three-species many-body problem.  This remains an open problem.   
\end{enumerate}

This thesis focuses on the second regime.  Careful analysis leads us to distinguish two related but different concepts, the total weight of the closed-channel  and the weight of the closed-channel in a particular momentum level.  
       Within our assumptions, namely, the spatial extension of a closed-channel bound state is much smaller than the interparticle distance, a very low occupation in low-momentum levels ($k\lesssim{}k_{F}$) of the closed-channel persists despite the large total weight of the closed-channel.     A suitable small parameter $E_F/\eta$ (or $a_0/a_c$), upon which a perturbation theory can be developed over a non-perturbative zeroth order solution, emerges with this discovery.  Pairing in many-body problems is known to be  non-perturbative and has to be handled with proper non-perturbative techniques.  Nonetheless, we can concentrate the non-perturbative part into the zeroth order  (broad resonance) solution, while treating the other corrections perturbatively.   The resulting theory  handles the two channels in steps self-consistently\footnote{Here by ``self-consistent'', we refer to the self-consistency in treating the close-channel according to zeroth order in the open-channel pairing.  The BCS-type treatment of pairing in the open-channel   is known to be not self-consistent. }. 
       
       In  summary, a two-channel model has a two-component order parameter $(\Delta_{1},\Delta_{2})$: one component for each channel.  The order parameter for the closed-channel can be determined by the number equation of the closed-channel (\eef{eq:pathInt2:closeD2})
\begin{equation}\tag{\ref{eq:pathInt2:closeD2}}
N_{close}\approx\sum_{\vk}\frac{\Delta_{2}^2}{(\xi_{\vk}+\eta)(2\xi_{\vk}+\eta)}
\end{equation}
 Where $\xi_{\vk}=\hbar^{2}k^{2}/2m-\mu$ and $\eta$ is the energy difference between two channels. The renormalized gap equation is given in \eef{eq:pathInt2:gapRenorm}
 \begin{equation}\tag{\ref{eq:pathInt2:gapRenorm}}
1=\frac{4\pi{\tilde{a}_{s}(\mu,\lambda_{1})}}{m}\sum(\nth{2E_{\vk}}-\nth{2\epsilon_{\vk}}-\frac{\Delta_{2}^{2}\xi_{\vk}}{4(\xi_{\vk}+\eta){E_{\vk}^{3}}})
	-\frac{\lambda_{2}}{\Delta_{1}}
\end{equation}
where $E_{\vk}=\sqrt{\xi_{\vk}^{2}+\Delta_{1}^{2}}$. Here, $\tilde{a}_{s}(\mu,\lambda_{1})$ is the two-body open-channel effective  s-wave scattering length with  an additional shifting $2\mu+\lambda_{1}$ as Eq. \ref{eq:pathInt2:asKshift}. 
\begin{equation}\tag{\ref{eq:pathInt2:asKshift}}
{a}_{s}=a_{\text{bg}}(1+\frac{\mathcal{K}}{\delta-2\mu-\lambda_{1}})
\end{equation}
  $\lambda_{1}$ and $\lambda_{2}$ are two new parameters describing the overlapping of the two channels, and they can be calibrated from  experiments (see Eqs. \ref{eq:pathInt2:lambda1} \ref{eq:pathInt2:lambda2}, and Sec. \ref{sec:pathInt2:lambda}).  The open-channel number equation is \eef{eq:pathInt2:openNum}
\begin{equation}\tag{\ref{eq:pathInt2:openNum}}
\begin{split}
N_{open}\approx\sum_\vk\mbr{\frac{E_\vk-\xi_\vk}{2E_\vk}(1+\frac{\Delta_{1}}{\eta}\zeta)-\frac{\Delta_{1}^{3}}{4E_\vk^{3}}\zeta
	}	
\end{split}
\end{equation}
Here $\zeta=\Delta_{2}^{2}/\Delta_{1}\eta\ll{}E_{\vk}$ appears in multiple places as the small expansion parameter.  The open-channel number equation and the associated gap equation need to be solved self-consistently to get  the mean field result ($\Delta_{1}$ and $\mu$).  One of the most noteworthy new phenomena is probably that the open-channel gap $\Delta_{1}$ saturates in the BEC side of  a narrow resonance. 

There are three fermionic excitation modes. Their spectrums with the first order correction due to the inter-channel Pauli exclusion are given by Eqs. \ref{eq:pathInt2:xiExpand}-\ref{eq:pathInt2:xiExpand3}
\begin{align}
E_{1\vk}&\approx{}E_{\vk}+u_{\vk}^{2}\Delta_{1}\zeta\tag{\ref{eq:pathInt2:xiExpand}}\\
E_{2\vk}&\approx{}E_{\vk}-v_{\vk}^{2}\Delta_{1}\zeta\tag{\ref{eq:pathInt2:xiExpand2}}\\
E_{3\vk}&\approx{}\epsilon_{\vk}+\eta+\frac{\zeta}{2}\Delta_{1}
\tag{\ref{eq:pathInt2:xiExpand3}}
\end{align}
With a two-component order parameter, the bosonic collective fluctuation modes are rich.  We explored two  modes about  phase fluctuations.  The two-component in-phase fluctuation is massless and the low-energy one.  It is similar to the Anderson-Bogoliubov modes in the single-channel problem with a small correction in the order of $\zeta$.  The new out-of-phase fluctuation is gapped and the minimum excitation energy is in the order of the pair-breaking energy ($\Delta_{1}$ in the BCS-like states, $\mu>0$ and $\sqrt{\mu^{2}+\Delta_{1}^{2}}$ in the BEC-like states, $\mu<0$).  

       In our approach, we take the broad resonance result (or the single-channel crossover) as our zeroth order solution, upon which the expansion is performed.  It is however known that the simple BCS-type pairing treatment is not adequate  to quantitatively describe the whole BEC-BCS crossover region.  Therefore the zeroth order solution used in this thesis (simple BCS type ansatz or saddle point) can be improved through further theoretical development.  Nevertheless, we expect the perturbative approach used here to build the narrow resonance from the single-channel crossover result to be still valid then.  Once the zeroth order solution (for a broad resonance or a single channel BEC-BCS crossover model) is patched over with whatever advancement, the correction of the narrow resonance in such a parameter regime, can still be obtained with a procedure similar as the one discussed in this thesis.

\begin{unsure}
The theory we have developed here is for zero-temperature only. This limit simplifies the calculations considerably.  However, an extension to finite temperature along the same spirit should be possible.  The two-component order parameters should persist at low-temperature.  At higher temperature, these components  are likely to decay at different temperatures.  The open-channel order parameter, associated with a much lower energy scale (of the order of  the Fermi energy or even lower), should turn to zero first.  Then the system becomes a normal gas with a Feshbach resonance.  From the above discussion, many-body corrections due to the narrow resonance (both the intra- and inter-channel Pauli exclusion) seem to be agnostic to whether the system is in superfluid state or not.  Thus, we expect that these many-body corrections can be carried out in a similar fashion.   There should still be corrections over the detuning due to the chemical potential and the  inter-channel Pauli exclusion ($\lambda_{1}$) although a system is more likely to be a Fermi liquid in the BCS side and a gas of normal fermion-dimer-molecules on the BEC side in such a temperature.   
\end{unsure}

\appendix

%\include{Appendix.tex}
% !TeX root =thesis.tex
\chapter{The variational approach  using  the BCS-type ansatz\label{ch:mean}}
We can also formulate the problem  using the  BCS-type ansatz.  It is not difficult to calculate  the expectation of the free energy as well as  the wave function that optimizes the free energy within this ansatz.  The optimization process gives us gap equations, which determine the  wave function with the constraint from number equations.   %with those common parameters as chemical potential, gap.  
We will see that this method yields  the mean-field solution consistent with the one of the path-integral approach.  However, this method is mean-field by nature and  difficult  to be extended for studying collective excitations.  

We rewrite the Hamiltonian Eq. \ref{eq:pathInt2:ham2} in momentum space, and restore the momentum dependence of the interaction coefficients.  
\begin{equation}\label{eq:uvw:hamiltonian}
\begin{split}
 H=&\sum_\vk\epsilon^a_\vk{}a^+_\vk{}a^{}_\vk+\sum_\vk\epsilon^b_\vk{}b^+_\vk{}b^{}_\vk+\sum_\vk\epsilon^c_\vk{}c^+_\vk{}c^{}_\vk\\
  &-\sum_{\vk\vk'}U_{\vk\vk'}a^+_\vk{}b^+_{-\vk}{}b^{}_{-\vk'}a^{}_{\vk'}
	-\sum_{\vk\vk'}V_{\vk\vk'}a^+_\vk{}c^+_{-\vk}{}c^{}_{-\vk'}a^{}_{\vk'}\\
 &-\sum_{\vk\vk'}Y_{\vk\vk'}a^+_\vk{}b^+_{-\vk}{}c^{}_{-\vk'}a^{}_{\vk'}
	-\sum_{\vk\vk'}Y^*_{\vk\vk'}a^+_{\vk'}{}c^+_{-\vk'}{}b^{}_{-\vk}a^{}_{\vk}
\end{split} 
\end{equation}
Here we only keep the zero center-of-mass momentum paring terms as in the original BCS work\cite{BCS}. 
We take the free atom at zero magnetic field as the  zero energy.  
\begin{equation*}
\epsilon_{\vk}^i=k^2/(2m)+\eta^i, \qquad (i=a,b,c)
\end{equation*}
 $\eta^i$ is the Zeeman energy of the $i^{\text{th}}$ atom at magnetic field $B$. The hermiticity  of the Hamiltonian imposes
 \begin{equation}
 U_{\vk'\vk}=U^*_{\vk\vk'},\qquad{} V_{\vk'\vk}=V^*_{\vk\vk'}
\end{equation}
  We then introduce the BCS-type ansatz 
\begin{equation}\label{eq:ansatz}
 \ket{\Psi}=\prod_\vk\br{u_\vk+v_\vk{}a^\dg_\vk{}b^\dg_{-\vk}+w_\vk{}a^\dg_\vk{}c^\dg_{-\vk}}\ket{0}
\end{equation}
$\ket{0}$ is the particle vacuum state.  We require $\abs{u_\vk}^2+\abs{v_\vk}^2+\abs{w_\vk}^2=1$ for normalization.  This ansatz is constructed like the original BCS ansatz for superconductivity.  An alternative ansatz would be $\prod_\vk(u_\vk+v_\vk{}a^\dg_\vk{}b^\dg_{-\vk})(u^{'}_\vk+w_\vk{}a^\dg_\vk{}c^\dg_{-\vk})\ket{0}$, which is actually the same as Eq. \ref{eq:ansatz}, because  the cross term vanishes due to the Pauli exclusion on the common species.   As the original BCS ansatz, this ansatz does not have the fixed particle number.  We  will just require the expected value of number operators match the total particle number. For all  interaction terms, we get  two types of contributions,
for example, 
\begin{equation*}
\av{U_{\vk\vk'}a^\dg_\vk{}b^\dg_{-\vk}{}b^{}_{-\vk'}a^{}_{\vk'}}
=\sum_{\vk}U_{\vk\vk}\abs{v_\vk}^2+\sum_{\vk\neq\vk'}U_{\vk\vk'}v^{}_{\vk'}u^*_{\vk'}u^{}_\vk{}v^*_\vk
\end{equation*}
The first term is the Hatree term and the second term is the more interesting pairing term.

This gives the full free energy as 
\begin{equation}\label{eq:uvw:F}
 \begin{split}
  &F\equiv\av{H-\mu{}N}\\
    =&\sum(\xi^a_\vk+\xi^b_\vk)\abs{v_\vk}^2+\sum(\xi^a_\vk+\xi^c_\vk)\abs{w_\vk}^2\\
    &-\sum_{\vk}U_{\vk\vk}\abs{v_\vk}^2-\sum_{\vk\neq\vk'}U_{\vk\vk'}v^{}_{\vk'}u^*_{\vk'}u^{}_\vk{}v^*_\vk\\
    &-\sum_{\vk}V_{\vk\vk}\abs{w_\vk}^2
      -\sum_{\vk\neq\vk'}V_{\vk\vk'}w^{}_{\vk'}u^*_{\vk'}u^{}_\vk{}w^*_\vk\\
    &-\sum_{\vk}Y_{\vk\vk}w^{}_{\vk}v^*_\vk{}
      -\sum_{\vk\neq\vk'}Y_{\vk\vk'}w^{}_{\vk'}{u^{*}_{\vk'}}v^*_\vk{}u^{}_\vk\\
    &-\sum_{\vk}Y^*_{\vk\vk}w^*_{\vk}v^{}_{\vk}{}
      -\sum_{\vk\neq\vk'}Y^*_{\vk\vk'}w^*_{\vk}{u^{}_{\vk}}v^{}_{\vk'}{}u^{*}_{\vk'}
 \end{split}
\end{equation}
where we have set
\begin{equation*}
 \xi^a_\vk=\epsilon^a_\vk-\mu^a,\qquad\xi^b_\vk=\epsilon^b_\vk-\mu^b,\qquad\xi^c_\vk=\epsilon^c_\vk-\mu^b
\end{equation*}
 Two chemical potentials are added to make sure the $n_a=n_b+n_c=\nth{2}n$.  In principle, $\mu^{a}$ does not need to be equal to $\mu^{b}$  as there is no exchange or conversion between $(a)$ and $(b,c)$.  In fact, the structure of the ansatz guarantees that $n_a=n_b+n_c$. Therefore, we set $\mu^{a}=\mu^{b}$ for simplicity and drop the superscript on chemical potential hereinafter. 
We will also drop the Hartree terms because they are  only related to the density and thus can be absorbed into the chemical potentials.   In the second summation, we will ignore the fact that the summation only goes through $\vk\neq\vk'$ because the corrections due to this restriction lead to only the higher order terms. 
 
 \section{Exact gap equations and number equations}
We introduce two new parameters $h_{1\vk}$ and $h_{2\vk}$, which corresponds to the mean field values (equal time average) of the anomalous Green's functions,  (in the same way as  $h_{1,2}$  in Sec \ref{sec:pathInt2:meanfield}), 
\begin{gather}
u_{\vk}^2+v^{2}_{\vk}+w^{2}_{\vk}=1\\
u_{\vk}v_{\vk}=h_{1\vk}\\
u_{\vk}w_{\vk}=h_{2\vk}
\end{gather}
We can solve $u_{\vk}$, $v_{\vk}$, $w_{\vk}$ in terms of  $h_{1\vk}$ and $h_{2\vk}$ (all parameters are taken as real)\footnote{It is not hard to prove that the parameters that optimize the free energy are real, within an overall phase factor, when the interactions are real. (See also footnote \ref{foot:pathInt2:real} in Chapter \ref{ch:path2}, page \pageref{foot:pathInt2:real}.)}.  One complication in solving above equations is that $u_\vk$, $v_\vk$ and  $w_\vk$ are all  monotonic functions of $\vk$, while $h_{1\vk}$ or $h_{2\vk}$ is not in the BCS end.  So, we need to be careful when taking the square root.  We introduce $\sgn_k$ for such purpose.  
\begin{equation}\label{eq:uvw:uvwh12}
\begin{split}
u_{\vk}^2=&\frac{1}{2} \left(1+\sgn_{k}\sqrt{1-4 h_{1\vk}^2-4 h_{2\vk}^2}\right)\\
v^{2}_{\vk}=&\frac{2 h_{1\vk}^2}{1+\sgn_{k}\sqrt{1-4 h_{1\vk}^2-4 h_{2\vk}^2}}
=\frac{ h_{1\vk}^2}{2( h_{1\vk}^2+ h_{2\vk}^2)} \left(1-\sgn_{k}\sqrt{1-4 h_{1\vk}^2-4 h_{2\vk}^2}\right)\\\
w^{2}_{\vk}=&\frac{2 h_{2\vk}^2}{1+\sgn_{k}\sqrt{1-4 h_{1\vk}^2-4 h_{2\vk}^2}}
=\frac{h_{2\vk}^2}{2( h_{1\vk}^2+ h_{2\vk}^2)} \left(1-\sgn_{k}\sqrt{1-4 h_{1\vk}^2-4 h_{2\vk}^2}\right)
\end{split}
\end{equation}
  where $\sgn_k=1$  for all $k$ in BEC cases, and  for large $k$ in BCS cases. In BCS cases,  $\sgn_{k}=-1$ when $k$ is small.  In the single-channel crossover, $\sgn_{k}=\sgn(\epsilon_{k}-\mu)$, and the turning point is at $u_\vk^2=1/2$ which corresponds to zero chemical potential, $\mu=0$.  The two-channel crossover is more delicate to treat.  The turning point is still at $u_\vk^2=1/2$, but it is no longer exactly at $\mu=0$.  The $\sgn_{k}$ function  is very important in number equations Eqs. \ref{eq:20100909:number}.  They however can be  absorbed later on, after we convert the formulas back into notations of  $(u_{\vk},\,v_{\vk}, \,w_{\vk})$ or $\xi_{k}$.

%Below, we adopt $\sgn_k=1$. (This works for BEC and most part of BCS; and in the final equations, the sign function is taken care of automatically by using $E_{\vk}$.  It is easy to verify that  we arrive at the same final result in the other cases where $\sgn_{k}=-1$. ) 
Eq. \ref{eq:uvw:uvwh12}'s  derivatives over $h_{1\vk}$ are\footnote{Here we follow the same convention by taking $h_{1\vk}$ and $h_{2\vk}$ as real}
\begin{equation}
\begin{split}
\pdiff{u_{\vk}^2}{h_{1\vk}}=&-\frac{2 h_{1\vk}}{\sqrt{1-4 h_{1\vk}^2-4 h_{2\vk}^2}}\sgn_{k}\\
\pdiff{v_{\vk}^2}{h_{1\vk}}=&\frac{2 h_{1\vk}\sgn_k}{\sqrt{1-4 h_{1\vk}^2-4 h_{2\vk}^2}}-\frac{8 h_{1\vk} h_{2\vk}^2\sgn_{k}}{\sqrt{1-4 h_{1\vk}^2-4 h_{2\vk}^2} \left(1+\sgn_{k}\sqrt{1-4 h_{1\vk}^2-4 h_{2\vk}^2}\right)^2}\\
\pdiff{w_{\vk}^2}{h_{1\vk}}=&\frac{8 h_{1\vk} h_{2\vk}^2\sgn_{k}}{\sqrt{1-4 h_{1\vk}^2-4 h_{2\vk}^2} \left(1+\sgn_{k}\sqrt{1-4 h_{1\vk}^2-4 h_{2\vk}^2}\right)^2}\\
=&\sgn_{k}\frac{h_{1\vk} h_{2\vk}^2 \left(1-\sgn_{k}\sqrt{1-4 h_{1\vk}^2-4 h_{2\vk}^2}\right)^2}{2\sqrt{1-4 h_{1\vk}^2-4 h_{2\vk}^2} (h_{1\vk}^2 + h_{2\vk}^2)^2}
\end{split}
\end{equation}
The second equation is not very obvious, but can be obtained by noting that $\pdiff{v_{\vk}^2}{h_{1\vk}}=-\pdiff{u_{\vk}^2}{h_{1\vk}}-\pdiff{w_{\vk}^2}{h_{1\vk}}$.   Similarly, derivatives over $h_{2\vk}$ are
\begin{equation}
\begin{split}
\pdiff{u_{\vk}^2}{h_{2\vk}}=&-\frac{2 h_{2\vk}}{\sqrt{1-4 h_{1\vk}^2-4 h_{2\vk}^2}}\sgn_{k}\\
\pdiff{v_{\vk}^2}{h_{2\vk}}=&\frac{8 h_{1\vk} ^{2}h_{2\vk}\sgn_{k}}{\sqrt{1-4 h_{1\vk}^2-4 h_{2\vk}^2} \left(1+\sgn_{k}\sqrt{1-4 h_{1\vk}^2-4 h_{2\vk}^2}\right)^2}\\
=&\sgn_{k}\frac{h_{1\vk}^{2}h_{2\vk} \left(1-\sgn_{k}\sqrt{1-4 h_{1\vk}^2-4 h_{2\vk}^2}\right)^2}{2\sqrt{1-4 h_{1\vk}^2-4 h_{2\vk}^2} (h_{1\vk}^2 + h_{2\vk}^2)^2}\\
\pdiff{w_{\vk}^2}{h_{2\vk}}=&\frac{2 h_{2\vk}\sgn_k}{\sqrt{1-4 h_{1\vk}^2-4 h_{2\vk}^2}}-\frac{8 h_{1\vk} ^{2}h_{2\vk}\sgn_{k}}{\sqrt{1-4 h_{1\vk}^2-4 h_{2\vk}^2} \left(1+\sgn_{k}\sqrt{1-4 h_{1\vk}^2-4 h_{2\vk}^2}\right)^2}
\end{split}
\end{equation}
We can obtain the gap equations by differentiating the free energy with respect to $h_{1\vk}$ and $h_{2\vk}$.
\begin{subequations}\label{eq:20100909:fullgap}
\begin{multline}
\frac{h_{1\vk}\sgn_k}{\sqrt{1-4 h_{1\vk}^2-4 h_{2\vk}^2}} \xi^{ab}_{\vk}+\frac{4 h_{1\vk} h_{2\vk}^2\sgn_{k}}{\sqrt{1-4 h_{1\vk}^2-4 h_{2\vk}^2} \left(1+\sgn_{k}\sqrt{1-4 h_{1\vk}^2-4 h_{2\vk}^2}\right)^2}\eta\\
-\sum_{\vk'}U_{\vk\vk'}h_{1\vk'}-\sum_{\vk'}Y_{\vk\vk'}h_{2\vk'}=0
\label{eq:20100909:fullgapa}
\end{multline}
\begin{multline}
\frac{h_{2\vk}\sgn_k}{\sqrt{1-4 h_{1\vk}^2-4 h_{2\vk}^2}} \xi^{ac}_{\vk}-\frac{4 h_{1\vk}^{2} h_{2\vk}\sgn_{k}}{\sqrt{1-4 h_{1\vk}^2-4 h_{2\vk}^2} \left(1+\sgn_{k}\sqrt{1-4 h_{1\vk}^2-4 h_{2\vk}^2}\right)^2}\eta\\
-\sum_{\vk'}V_{\vk\vk'}h_{2\vk'}-\sum_{\vk'}Y_{\vk\vk'}h_{1\vk'}=0
\label{eq:20100909:fullgapb}
\end{multline}
\end{subequations}
where $\eta=\epsilon^{ac}_{\vk}-\epsilon^{ab}_{\vk}=\eta^{c}-\eta^{b}$ is the bare Zeeman energy difference and is larger than most other energy scales, such as $E_{F}$.  It is close to  the binding energy of the closed-channel bound state when it is not too far away from the resonance point.   We see one  disadvantages of the ansatz method: unlike the path integral approach, where $h_{1\vk}$ and $h_{2\vk}$ have a very specific forms, Eqs. (\ref{eq:pathInt2:h1}, \ref{eq:pathInt2:h2}), and ready for further approximation and analysis,  we do not have much guide for these two quantities here and need to make some guess.  

We can define order parameters
\begin{gather}
\Delta_{1\vk}=\sum_{\vk'}U_{\vk\vk'}h_{1\vk'}+\sum_{\vk'}Y_{\vk\vk'}h_{2\vk'}\\
\Delta_{2\vk}=\sum_{\vk'}V_{\vk\vk'}h_{2\vk'}+\sum_{\vk'}Y_{\vk\vk'}h_{1\vk'}
\end{gather}
Both of them should have little $\vk$-dependence at low momentum for short-range interactions.  We will drop the $\vk$ subscripts in both $\Delta$ in the following. 
Two number equations can be obtained.  One for each channel.
\begin{subequations}\label{eq:20100909:number}
\begin{gather}
N_{\text{open}}=2\sum_{\vk}v_{\vk}^{2}
	=\sum_{\vk}\frac{ h_{1\vk}^2}{( h_{1\vk}^2+ h_{2\vk}^2)} \left(1-\sgn_{k}\sqrt{1-4 h_{1\vk}^2-4 h_{2\vk}^2}\right)
	\label{eq:meanfield:opennum}\\
N_{\text{close}}=2\sum_{\vk}w_{\vk}^{2}
	=\sum_{\vk}\frac{ h_{2\vk}^2}{( h_{1\vk}^2+ h_{2\vk}^2)} \left(1-\sgn_{k}\sqrt{1-4 h_{1\vk}^2-4 h_{2\vk}^2}\right)
\end{gather} 
\end{subequations}

\section{Approximate solution of the mean field equations}
Here instead of solving these equations, we just demonstrate that they are consistent to the saddle point solution obtained in the path-integral method up to the first order of correction $\zeta$.

Let us look at the closed-channel gap equation (Eq. \ref{eq:20100909:fullgapb}) first.  It can be rewritten as 
\begin{multline*}
\frac{h_{2\vk} \sgn_{k}}{\sqrt{1-4 h_{1\vk}^2-4 h_{2\vk}^2}}\br{2 \xi_{\vk}+\eta-\frac{4 h_{1\vk}^{2}}{\left(1+\sgn_{k}\sqrt{1-4 h_{1\vk}^2-4 h_{2\vk}^2}\right)^2}\eta}
-\sum_{\vk'}V_{\vk\vk'}h_{2\vk'}-\sum_{\vk'}Y_{\vk\vk'}h_{1\vk'}=0
\end{multline*}
We can put some terms back into the notation of $(u_{\vk},\,v_{\vk},\, w_{\vk})$.
\begin{gather}
\sgn_{k}\sqrt{1-4 h_{1\vk}^2-4 h_{2\vk}^2}=(u^{2}_{\vk}-v_{\vk}^{2}-w_{\vk}^{2})\\
1+\sgn_{k}\sqrt{1-4 h_{1\vk}^2-4 h_{2\vk}^2}=2u_{\vk}^{2}\label{eq:meanfield:1sqrt}\\
\frac{4 h_{1\vk}^{2} }{\left(1+\sgn_{k}\sqrt{1-4 h_{1\vk}^2-4 h_{2\vk}^2}\right)^2}=\frac{v_{\vk}^{2}}{u_{\vk}^{2}}
\end{gather}
We rewrite the closed-channel gap equation using these relations
\begin{equation*}
\frac{h_{2\vk}}{u^{2}_{\vk}-v_{\vk}^{2}-w_{\vk}^{2}}\br{2 \xi_{\vk}+\eta-\frac{v_{\vk}^{2}}{u_{\vk}^{2}}\eta}-\sum_{\vk'}V_{\vk\vk'}h_{2\vk'}-\sum_{\vk'}Y_{\vk\vk'}h_{1\vk'}=0
\end{equation*}
At high momentum ($\gg{}k_{F}$),  the third term in the parenthesis is negligible and the the denominator $u^{2}_{\vk}-v_{\vk}^{2}-w_{\vk}^{2}\approx1$.  This equation is then very similar to the two-body \sch equation (Eq. \ref{eq:intro:close}). For the closed-channel, the interaction, or the closed-channel bound state, $\phi_{0}$, has non-zero value for a range  in the momentum space much larger than $k_{F}$.  Thus we expect the closed-channel correlation $h_{2\vk}\propto\phi_{0}f(\vk)$ where $f(\vk)$ approach one at high momentum.  This is in fact the same argument as Eq. \ref{eq:pathInt2:hphif}  in the beginning of Chapter \ref{ch:path2}.  The next thing to notice is that the third term in the parenthesis might dominate the other two terms in  certain situation (low momentum, BCS) because the denominator  $u_{\vk}^{2}$ can be very small in such regions.  A nature way to compensate this big term is to set\footnote{$u_{\vk}^{2}$ in Chapter \ref{ch:path2} is defined to be $(\xi_{\vk}+E_{\vk})/2E_{\vk}$ and does not carry an obvious physical meaning as $u_{\vk}^{2}$ in this appendix. They are the same in the lowest order of our expansion over $\zeta$ for the narrow resonance. This does not affect our analysis because the closed-channel correlation is actually a first order quantity.} $f(\vk)=u_{\vk}^{2}$.  Now we arrive the same conclusion for the closed-channel correlation as in the path-integral approach (Eq. \ref{eq:pathInt2:hphi})
\begin{equation}\label{eq:meanfield:h2phi}
h_{2\vk}=\tilde\alpha{}\phi_{0\vk}u_{\vk}^{2}
\end{equation}
Note that $\tilde\alpha$ is not fixed yet and should not be identified as $\alpha$ in the path-integral approach (Eq. \ref{eq:pathInt2:hphi}) \textit{a prior}. The equation becomes,
\begin{equation*}
\frac{\tilde\alpha{}\phi_{0\vk}u_{\vk}^{2} }{u^{2}_{\vk}-v_{\vk}^{2}-w_{\vk}^{2}}\br{2 \xi_{\vk}+\eta-\frac{v_{\vk}^{2}}{u_{\vk}^{2}}\eta}-\sum_{\vk'}V_{\vk\vk'}\tilde\alpha{}\phi_{0\vk}u_{\vk}^{2}-\sum_{\vk'}Y_{\vk\vk'}h_{1\vk'}=0
\end{equation*}
Using the similar approach as in the path-integral method, we multiply the equation with $\phi_{0\vk}^{*}$ and integrate the equation in order to find $\tilde{\alpha}$. In the denominator of the first term $u^{2}_{\vk}\gg{}v_{\vk}^{2}$, $u^{2}_{\vk}\gg{}w_{\vk}^{2}$ in the most of the integral domains (up to order of $\kappa$($\eta$) in the momentum space).  The equation is approximately 
\begin{equation*}
\tilde\alpha{}\sum_{\vk}\frac{\phi_{0\vk}\phi_{0\vk}^{*}u_{\vk}^{2}}{u^{2}_{\vk}}\br{2 \xi_{\vk}+\eta}-\tilde\alpha\sum_{\vk\vk'}V_{\vk\vk'}{}\phi_{0\vk}\phi_{0\vk}^{*}u_{\vk}^{2}-\sum_{\vk\vk'}\phi_{0\vk}^{*}Y_{\vk\vk'}h_{1\vk'}=0
\end{equation*}
And it is not hard to find 
\begin{equation}
\tilde\alpha\approx\frac{\sum_{\vk\vk'}{\phi_{\vk}^{*}}{Y_{\vk\vk'}}{h_{1\vk'}}}{{-E_{b}+\eta-2\mu-\tilde\lambda_{1}}}
\end{equation}
 And $\tilde\lambda_{1}$ is simpler than Eq. \ref{eq:pathInt2:lambda1}
\begin{equation*}
\tilde\lambda_{1}(\eta)\equiv-\sum_{\vk\vk'}{\phi_{\vk}^{*}}{v_{\vk'}^{2}V_{\vk\vk'}}\phi_{\vk'}
\end{equation*}
But it should be equal to $\lambda_{1}$ from the path-integral approach at the first order of $\zeta$ (Eq. \ref{eq:pathInt2:lambda1}) because this term is the dominated one (see Sec. \ref{sec:pathInt2:lambda}, Eq. \ref{eq:pathInt2:lambda1es}). We will just use $\lambda_{1}$ hereinafter. $\tilde\alpha$ is the same as $\alpha$ in the path-integral approach if $h_{1\vk}$ here is the same as the one in the path-integral approach,  which we will show in the following. Particularly for low momentum, $\phi_{0\vk}\sim\nth{\eta+\epsilon_{\vk}}$ (see Appendix \ref{sec:pathInt2:short-range}) and it is not hard to find at low momentum
\begin{equation}\label{eq:meanfield:aphi}
\alpha\phi_{0\vk}\overset{{k\lesssim{k_F}}}{=}\frac{\Delta_2}{\eta}
\end{equation}

   Now let us check the open-channel gap equation.  We can rewrite it as 
   \begin{equation*}
   \frac{h_{1\vk} \sgn_{k}}{\sqrt{1-4 h_{1\vk}^2-4 h_{2\vk}^2}}\br{ \xi^{ab}_{\vk}+\frac{4 h_{2\vk}^2}{\left(1+\sgn_{k}\sqrt{1-4 h_{1\vk}^2-4 h_{2\vk}^2}\right)^2}\eta}=\Delta_{1}
   \end{equation*}
Again, using Eq. \ref{eq:meanfield:1sqrt}, we replace the denominator with $2u_\vk^2$.  Also we replace $h_{2\vk}$ with $\tilde\alpha{}\phi_{0\vk}u_{\vk}^{2}$ using Eq. \ref{eq:meanfield:h2phi}.  
 \begin{equation*}
   \frac{h_{1\vk}\sgn_{k}}{\sqrt{1-4 h_{1\vk}^2-4 h_{2\vk}^2}}\br{ 2\xi_{\vk}+({\tilde\alpha}\phi_{0\vk})^2\eta}=\Delta_{1}
   \end{equation*}
 At low momentum, the second term in the parenthesis can be simplified as $\Delta_2^2/\eta=\zeta\Delta_{1}$  using Eq. \ref{eq:meanfield:aphi}
   \begin{equation*}
   \frac{h_{1\vk}\sgn_{k}}{\sqrt{1-4 h_{1\vk}^2-4 h_{2\vk}^2}}\br{ 2\xi_{\vk}+\zeta\Delta_{1}}=\Delta_{1}
   \end{equation*}
   We can solve $h_{1\vk}$ in terms of $\Delta_1$.  
   \begin{equation*}
   h_{1\vk}^2=\frac{\Delta_1^2(1-4h_{2\vk}^2)}{(2\xi_{\vk}+\zeta\Delta_{1})^2+4\Delta_1^2}
   \approx\frac{\Delta_1^2(1-4h_{2\vk}^2)}{4E_\vk^2+4\xi_\vk\Delta_{1}\zeta}
   \approx\frac{\Delta_1^2}{4E_\vk^2}(1-\frac{\xi_\vk\Delta_{1}}{E_\vk^2}\zeta)(1-4\frac{\Delta_{1}}{\eta}u_\vk^4\zeta)
   \end{equation*}
   Here we take the low momentum value of $h_{2\vk}^{2}\approx\frac{\Delta_{1}}{\eta}u_\vk^4\zeta$ in the last step. This term is much smaller  than $\frac{\xi_\vk\zeta}{E_\vk^2}$ because of the factor $\frac{\Delta_{1}}{\eta}$ at the low momentum.  We can neglect this term. Now we have 
    \begin{equation}\label{eq:meanfield:h1approx}
   h_{1\vk}^2\approx\frac{\Delta_1^2}{4E_\vk^2}(1-\frac{\xi_\vk\Delta_{1}}{E_\vk^2}\zeta)
   \end{equation}
   This is equivalent in the first order of $\zeta$ to the $h_{1\vk}$ in the path-integral approach (Eq. \ref{eq:pathInt2:h1}).

It is easy to see that $h_{2\vk}^2$ is the higher order as $\zeta$ and $w_\vk\ll1$ all the time.  So we always have 
$h_{2\vk}\approx{}w_\vk$ except in low momentum.  However, the summation in the low momentum is very small, one order higher in $\zeta$.  The closed-channel number equation is then simply $N_{\text{closed}}=\sum{}w_\vk^2\approx{}\sum_{\vk}h_{2\vk}^2$, which is the same result in the path-integral approach as we discussed in Eq. \ref{eq:pathInt2:closeD2}.

We move to the open-channel number equation Eq. \ref{eq:meanfield:opennum}.  At the momentum smaller than the characteristic momentum of the closed-channel bound state $\kappa$  ($\eta\sim{}E_{b}=\kappa^{2}/2m$), factor $\frac{ h_{1\vk}^2}{( h_{1\vk}^2+ h_{2\vk}^2)} $ is approximately $1$,
\begin{equation*}
\frac{ h_{1\vk}^2}{( h_{1\vk}^2+ h_{2\vk}^2)} =1-\frac{h_{2\vk}^2}{( h_{1\vk}^2+ h_{2\vk}^2)}
\overset{k\ll{}\kappa}\approx1-\frac{h_{2\vk}^2}{h_{1\vk}^2}
\approx1-\frac{4E_{\vk}^{2}}{\eta\Delta_{1}}u_{\vk^{4}}\zeta
\sim1+O(\zeta^{2})
\end{equation*}
while on the high momentum 
\begin{equation*}
\frac{ h_{1\vk}^2}{( h_{1\vk}^2+ h_{2\vk}^2)}
\overset{k\gtrsim\kappa}\approx\frac{\frac{\Delta_{1}^{2}}{4E_{\vk}^{2}}}{\frac{\Delta_{2}^{2}}{\eta+\epsilon_{\vk}}}
\sim\frac{\Delta_{1}^{2}}{\epsilon_{\vk}\Delta_{2}}\ll1
\end{equation*}
Therefore the factor $\frac{ h_{1\vk}^2}{( h_{1\vk}^2+ h_{2\vk}^2)} $ limit the summation into  low momentum.  On the other hand, in the low momentum we have, $\frac{h_{2\vk}^{2}}{h_{1\vk}^{2}}\sim\frac{\Delta_{1}}{\eta}\zeta$, thus we can neglect $h_{2\vk}^{2}$ in the square root comparing to $h_{1\vk}^{2}$. And using Eq. \ref{eq:meanfield:h1approx}
\begin{equation*}
\sqrt{1-4 h_{1\vk}^2-4 h_{2\vk}^2}\approx\sqrt{1-4h_{1\vk}^{2}+O({\zeta}^{2})}
	\approx\frac{\xi_{\vk}}{E_{\vk}}+\frac{\Delta_{1}^{2}}{4E_{\vk}^{3}}\zeta+O({\zeta}^{2})
\end{equation*}
So we have the open-channel number equation 
\begin{equation}
N_{\text{open}}\approx\sum_{\vk}1-\frac{\xi_{\vk}}{E_{\vk}}-\frac{\Delta_{1}^{3}}{4E_{\vk}^{3}}\zeta
\end{equation}
This is consistent to the open-channel number equation derived from the path-integral approach  (Eq. \ref{eq:pathInt2:openNum}, where the second term is negligible comparing to the third term).

Thus we have shown that the mean-field solution of the path-integral approach can solve the gap equations and number equations of the ansatz method up to the first order of $\zeta$. 

\chapter{Materials for Chapter \ref{ch:path2}}
% !TeX root =thesis.tex

\section{Diagonalization of  the matrix Eq. (\ref{eq:pathInt2:G2})\label{sec:diagonalize}}
We need to find a unitary transformation $L$ to diagonalize the  matrix 
\begin{equation}\tag{\ref{eq:pathInt2:G2}}
T_k^{\dg}G_{\omega_{n},\vk}^{-1}T_k=i\omega_nI+\mtrx{-E_k&0&u_k\Delta_2\\0&+E_k&v_k\Delta_2\\u_k\Delta_2&v_k\Delta_2&+\xi_k+\eta}
\end{equation}
We drops all the $k$ subscripts in this section because matrices in this section are decoupled in momentum and we only deal with one particular momentum $\vk$ a time. We notice that the first term is proportional to an identity matrix and does not change by unitary transformation, we only need to concentrate on the second term.  We rescale all elements with $E_{\vk}$ for simplicity in the following of this section. 
\begin{equation*}
y=\frac{\Delta_2}{E_{\vk}},\qquad
 t=\frac{\xi_k+\eta}{E_{\vk}},\qquad
\end{equation*}
 
\begin{equation*}
R=
\begin{pmatrix}
-1&0&uy\\
0&1&vy\\
uy&vy&t
\end{pmatrix}
\end{equation*}
The secular equation of $R$ is ($\abs{x\,I-R}=0$)
\begin{equation}\label{eq:pahtApp:secular}
(x^{2}-1)(x-t)-y^{2}x+(u^{2}-v^{2})y^{2}=0
\end{equation}
We use $u^{2}+v^{2}=1$ here.  We  assume at the zeroth order, the three eigenvalues are $-1$, $1$ and $t$.  ($t$ has weak dependency on energy as $(\xi_{k}+\eta)/E_{k}$, however, at the low energy region of interest, we ignore $\xi_{k}$.) Both $y$ and t are larger than 1, however, we will verify that given condition $y^{2}\ll{t}$, the correction is indeed small and the expansion is reasonable (See Sec.\ref{sec:pathApp:consistency}).  \emph{Indeed,  close-channel component can still be smaller than the open-channel component at low-k (in the order of $k_{F}$)  due to the close-channel bound state is much smaller than the interparticle distance even when the total close-channel atom number  is more than that of open-channel.  And here all the quantities are about low-k unless specifically noticed.} 
We expand the system to the first order of the dimensionless parameter $\tilde\zeta=y^{2}/{t}$ (\eef{eq:pathInt2:zetaDef})%\footnote{Note that now $\zeta$ has no momentum dependency.}
, and find
\begin{equation}
\begin{array}{ccc}
x^{(0)}&\quad{}x^{(1)}&\quad{}Eigenvector\nonumber\\
-1&-u^{2}\tilde\zeta&\mtrx{1&\frac{uvy^{2}}{2t}&-\frac{uy}{t}}\\
1&-v^{2}\tilde\zeta&\mtrx{-\frac{uvy^{2}}{2t}&1&-\frac{vy}{t}}\\
t&\nth2\tilde\zeta&\mtrx{\frac{uy}{t}&\frac{vy}{t}&1}
\end{array}
\end{equation}
Now it is easy to write down the corresponding diagonal matrix and the unitary transformation
\begin{equation}
B=i\omega_{n}I+E\mtrx{-1-u^{2}\tilde\zeta&0&0\\0&1-v^{2}\tilde\zeta&0\\0&0&t\nth2\tilde\zeta}
\end{equation}
\begin{equation}
L=\mtrx{1&-\frac{uvy^{2}}{2t}&\frac{uy}{t}\\\frac{uvy^{2}}{2t}&1&\frac{vy}{t}\\-\frac{uy}{t}&-\frac{vy}{t}&1}
\end{equation}
Here $L$ is not exactly unitary transformation, it is only unitary in the first order of  $\tilde\zeta$. We have 
\[
B=i\omega_{n}I+E\,L^{\dg}RL+o(\tilde\zeta)
\]
Alternatively, we can write $L$ as 
\begin{equation}
L=I+
\mtrx{0&-\frac{\Delta_{1}\Delta_{2}}{4E^{2}}&u\\
\frac{\Delta_{1}\Delta_{2}}{4E^{2}}&0&v\\
-u&v&0
}\frac{\Delta_{2}}{\eta}
\end{equation}
Here we use $uv=\Delta_{1}/2E$.

In the above treatment, the small parameter $\tilde\zeta$ is momentum dependent.  If we restore the subscript $\vk$ and scale it  back with $E_{\vk}$
\begin{equation}
\tilde\zeta=\frac{\Delta_{2}^{2}}{E_{\vk}(\xi_{\vk}+\eta)}
\end{equation}
A momentum-dependent small parameter is not very convenient to work with, so we take its maximum value in low momentum ($\lesssim{}E_{F}$).  In the BCS-like states ($\mu>0$), $\min{E_{k}}=\Delta_{1}$, $\min{\xi_{\vk}}=0$; in the BEC-like states ($\mu<0$), $\min{E_{k}}=\sqrt{\Delta_{1}^{2}+\mu^{2}}$ and $\min{\xi_{\vk}}=\abs{\mu}$. We take the smaller values and have our expanding small parameter $\zeta$(\eef{eq:pathInt2:zetaDef})
\begin{equation}
\zeta=\max\tilde{\zeta}=\frac{\Delta_{2}^{2}}{\Delta_{1}\eta}
\end{equation}

\section{Derivation of  the mean-field equations \eqref{eq:pathInt2:mf}\label{sec:pathInt2:deriveMF}}
We have fermion correlation as a $3\times3$ matrix (Eq. \ref{eq:pathInt2:nG}), 
\begin{equation}\tag{\ref{eq:pathInt2:nG}}
\mathcal{G}^{-1}=\begin{pmatrix}
i\omega_{n}-\xi_{k}&\Delta_{1}&\Delta_{2}\\
\bar{\Delta}_{1}&i\omega_{n}+\xi_{k}&0\\
\bar{\Delta}_{2}&0&i\omega_{n}+\xi_{k}+\eta
\end{pmatrix}
\end{equation}
Here we work in the momentum space, in which the system is nicely decoupled at least to the mean-field order.  And we therefore drop all the $k$ subscript in the rest of section. A general $3\times3$ matrix inverted as such, 
  \begin{equation}
  \mtrx{A_{11}&A_{12}&A_{13}\\A_{12}^{*}&A_{22}&0\\A_{13}^{*}&0&A_{33}}^{-1}=
  \nth{|A|}
  \mtrx{A_{22}A_{33}&-A_{12}A_{33}&-A_{13}A_{22}\\
  	-A_{12}^{*}A_{33}&A_{11}A_{33}-A_{13}A_{13}^{*}&A_{12}^{*}A_{13}\\
	-A_{13}^{*}A_{22}&A_{12}A_{13}^{*}&A_{11}A_{22}-A_{12}A_{12}^{*}}
  \end{equation}
where $|A|$ is the determent of $A$.   At the mean field level, all $\Delta_{i}$'s are real constants.  We denote the mean field value of $\mathcal{G}$ as $G_{0}$.  The determent of ${G}_{0}^{-1}$ can be expressed as 
\begin{equation}
|G_{0}^{-1}|=(i\omega_{n}-E_{1})(i\omega_{n}+E_{2})(i\omega_{n}+E_{3})
\end{equation}
where $E_{i}$'s are defined in Eqs. (\ref{eq:pathInt2:xiExpand}-\ref{eq:pathInt2:xiExpand3}).
And $G_{0}$ can be obtained according to the above rule. Now we can find the last term in \ref{eq:pathInt2:mf01}, 
\begin{equation}
\begin{split}
\tr\mbr{{G_{0}}\cdot\cmtrx{0&1&0\\0&0&0\\0&0&0}}&=\sum_{\vk\omega_{n}}G_{0\,(21)}\\
&=\sum_{\vk}\sum_{\omega_{n}}\frac{-\Delta_{1}^{*}(i\omega_{n}+\xi+\eta)}{(i\omega_{n}-E_{1})(i\omega_{n}+E_{2})(i\omega_{n}+E_{3})}\\
&=\sum_{\vk}\Delta_{1}^{*}\frac{E_{1}+\xi+\eta}{(E_{1}+E_{2})(E_{1}+E_{3})}\equiv\sum_{\vk}h_{1\,\vk}
\end{split}
\end{equation}
Here we perform the zero-temperature Matsubara summation  in the third equal sign with the normal trick (see sec. 4.2.1 in \cite{Altland}, sec. 25 in \cite{Fetter}, also refer to Footnote \ref{foot:intro:sum} at Page. \pageref{foot:intro:sum}). Because of zero-temperature,  within three roots, $E_{1}$, $-E_{2}$ and $-E_{3}$, we only need to take into account two negative roots $-E_{2}$ and $-E_{3}$, assuming the correction is small. 
Similarly
\begin{equation}
\begin{split}
\tr\mbr{{G_{0}}\cdot\cmtrx{0&0&1\\0&0&0\\0&0&0}}&=\sum_{\vk\omega_{n}}G_{0\,(31)}
=\sum_{\vk}\Delta_{2}\frac{E_{1}+\xi}{(E_{1}+E_{2})(E_{1}+E_{3})}\equiv\sum_{\vk}h_{2\,\vk}
\end{split}
\end{equation}
And we have 
 \begin{align*}
(\tilde{U}^{-1})_{11}\bar{\Delta}_{1}+(\tilde{U}^{-1})_{21}\bar{\Delta}_{2}-\sum_{\vk}h_{1\,\vk}=0\\
(\tilde{U}^{-1})_{12}\bar{\Delta}_{1}+(\tilde{U}^{-1})_{22}\bar{\Delta}_{2}-\sum_{\vk}h_{2\,\vk}=0
 \end{align*}
Invert the interaction matrix $\tilde{U}$ and we have Eq.  \ref{eq:pathInt2:mf}.

\section{The wave function for a short-range potential}\label{sec:pathInt2:short-range}
Here we discuss some possible generalization on the wave function for a short-range potential.  This topic has been studied by Zhang \cite{shizhongUniv}. We will use some similar ideas.  Outside the range $r_{c}$ of a short-range potential,  an atom is free and  the  \sch equation is very simple.
\begin{equation}
-\frac{\hbar^{2}}{2m}\nabla^{2}\psi=E\psi
\end{equation}
The equation has a simple solution for s-wave, $\psi=A'{e^{-\kappa{r}}}/{r}$ ($\kappa$ is imaginary for a scattering state). For a bound state. normalization $A'$ is determined  by connecting it with the short-range part of the wave function, $\varphi_0$, and then requiring the full wave function normalized to $1$. 

Let us discuss the bound-state first, where $\kappa>0$.  In the momentum space, there is also a universal behavior at low momentum, where $kr_{c}\ll1$.   
\begin{equation*}
\psi_{k}=\nth{(2\pi)^{3/2}}\int{d\vr}(\varphi_{0}+A'\frac{e^{-\kappa{r}}}{r})e^{-i\vk\cdot\vr}
\end{equation*}
The first part for $\varphi_{0}$ corresponds to the short-range part of the wave-function. 
\begin{equation*}
\psi_{k}=\varphi_{0\,k}+\nth{(2\pi)^{3/2}}\int{d\vr}(A'\frac{e^{-\kappa{r}}}{r})e^{-i\vk\cdot\vr}=\varphi_{0\,k}-A\nth{k^{2}+\kappa^{2}}
\end{equation*}
The first term has  very little $k$ dependence for low memontum $k\ll1/r_{c}$ and the second terms is more important in this range. 
Furthermore, if the bound-state is  close to threshold, the most weight is outside $r_{c}$, we can neglect the first term and we have universal behavior at low-momentum while the normalization factor $A$ can be easily determined.
\begin{equation}
A=\sqrt{\frac{8\pi\kappa}{\mathcal{V}_{0}}}
\end{equation}
Where $\mathcal{V}_{0}$ is the total volume of the system. And the wave-function is
\begin{equation}\label{eq:pathInt2:phi2body}
\psi_{\vk}\approx\sqrt{\frac{8\pi\kappa}{\mathcal{V}_{0}}}\frac{1}{k^{2}+\kappa^{2}}\approx\sqrt{\frac{8\pi\kappa}{\mathcal{V}_{0}}}\frac{1}{\kappa^{2}}
\end{equation}
The second approximation is when the momentum is low ($\lesssim{}k_{F}$).

   Besides  the bound-state, if the interaction is weak and short-range, the low energy scattering state is well described by the s-wave scattering state $\psi\propto1/r-1/a$ (Eq. \ref{eq:intro:Bethe}), and its Fourier transformation in the momentum space has the similar form $1/k^{2}$.  When considering many-body physics, in the low momentum below or around the Fermi momentum,  wave function  is modified by the many-body effect; but in the medium momentum, (still much smaller than $1/r_{c}$), this $1/k^2$ universal behavior is preserved.  The distribution of particle in such momentum, $k_{F}\ll{k}\ll{1/r_{c}}$, is $1/k^{4}$. This is actually the ``high-momentum'' (medium here) behavior ($C/k^{4}$) described in Tan's work about universality\cite{Tan2008-1,Tan2008-2}. 

On the other hand, at very higher momentum ($k\gg1/r_{c}$), the second term in the above is very small.  This is because the smooth tail part of the wave function ($\phi_0$) cancels out and contributes little in high frequency momentum oscillation.  The high-frequency Fourier component in momentum space is solely determined by the wave function within the potential range ($r_c$).   This can be extend beyond the two-body wave function to the two-body correlation as long as the long-wave-length part is smooth.  In all cases, two-body, or many-body, very high-frequency of two-body correlation follows the two-body wave function.  

Incidentally, the two-body \sch equation in the momentum space reads as 
\begin{equation}
\frac{\hbar^{2}k^{2}}{2m}\psi_{\vk}+\sum_{\vk'}U_{\vk\vk'}\psi_{\vk'}=E\psi_{\vk}
\end{equation}
At very high momentum (determined by the interaction strength and  potential range), the first term dominates, and we have the asymptotic form of the wave-function similar as Eq. \ref{eq:pathInt2:phi2body},
\begin{equation}
\lim_{k\rightarrow\infty}\psi_{\vk}=\tilde{A}\nth{k^{2}+\kappa^{2}}
\end{equation}
where $-\frac{\hbar^{2}\kappa^{2}}{2m}=E$.  Note that this behavior is for a different reason and $\tilde{A}$ is not necessarily equal $A$ at low momentum discussed before.

\section{Evaluation of $\pi^{(0)}(0)$ and $\pi^{\perp}(0)$\label{sec:calculatePi}}
Here we  calculate $\pi^{(0)}$ and $\pi^{\perp}$ (Eqs. \ref{eq:pathInt2:GKGK2}, \ref{eq:pathInt2:pi}) to the first order of $\zeta$ (Eq. \ref{eq:pathInt2:zetaDef}) using the expansion of the Green's function (Eqs. \ref{eq:pathInt2:Gexpand}) described in Sec. \ref{sec:diagonalGreen}.
\begin{equation}\label{eq:pathInt2:pi0long}
\begin{split}
\pi^{(0)}(0)=&\sum_k\tr(\hat{G}_{0\,k}\sigma_3\hat{G}_{0\,k}\sigma_3)\\
	\approx&\sum_k\tr\big(T_{\vk}B_{k}^{-1}T_{\vk}^{\dg}\sigma_3T_{\vk}B_{k}^{-1}T_{\vk}^{\dg}\sigma_3\big)\\
	&\quad+\tr\Big(T_{\vk}\delta_{\vk}B_{k}^{-1}T_{\vk}^{\dg}\sigma_3T_{\vk}B_{k}^{-1}T_{\vk}^{\dg}\sigma_3
	-T_{\vk}B_{k}^{-1}\delta_{\vk}T_{\vk}^{\dg}\sigma_3T_{\vk}B_{k}^{-1}T_{\vk}^{\dg}\sigma_3\\
	&\qquad+T_{\vk}B_{k}^{-1}T_{\vk}^{\dg}\sigma_3T_{\vk}\delta_{\vk}B_{k}^{-1}T_{\vk}^{\dg}\sigma_3
	-T_{\vk}B_{k}^{-1}T_{\vk}^{\dg}\sigma_3T_{\vk}B_{k}^{-1}\delta_{\vk}T_{\vk}^{\dg}\sigma_3\Big)
	\end{split}
\end{equation}
Note that $k$ stands for both the momentum and  the Matsubara frequency, $(\omega_{n},\vk)$. $\delta_{\vk}$ is defined in Eq. \ref{eq:pathInt2:L1}. 
\begin{equation*}
\delta_{c}\equiv\mtrx{0&-\frac{\Delta_{1}{}\Delta_{2}{}}{4E^{2}_{\vk}}&u_{\vk}\\
\frac{\Delta_{1}{}\Delta_{2}{}}{4E^{2}_{\vk}}&0&v_{\vk}\\
-u_{\vk}&-v_{\vk}&0
}\frac{\Delta_{2}{}}{\eta}
\end{equation*}
Introduce two matricies $M_{k}$ and $\widetilde{M}_{k}$, 
\begin{gather}
M_{k}\equiv{}T_{\vk}^{\dg}\sigma_3T_{\vk}B_{k}^{-1}T_{\vk}^{\dg}\sigma_3T_{\vk}B_{k}^{-1}\\
\widetilde{M}_{k}\equiv{}B_{k}^{-1}T_{\vk}^{\dg}\sigma_3T^{}_{\vk}B_{k}^{-1}T_{\vk}^{\dg}\sigma_3T^{}_{\vk}
\end{gather}
And we can rewrite $\pi^{(0)}(0)$ as
\begin{equation}
\pi^{(0)}(0)=\sum_k\tr\big({M}_{k}\big)+2\tr\Big(\delta_{\vk}\widetilde{M}_{k}-\delta_{\vk}M_{k}\Big)
\end{equation}
Here we use the cyclical  property of the trace $\tr(AB)=\tr(BA)$.  
It is straightforward to calculate
\begin{equation*}
T_{\vk}^{\dg}\sigma_3T_{\vk}B_{k}^{-1}=
\begin{pmatrix}
{\frac{\xi_{\vk}}{E_{\vk}(i\omega_{k}-E_{1\,\vk})}}&\frac{\Delta_{1}}{E_{\vk}(i\omega_{k}+E_{2\,\vk})}&0\\
{\frac{\Delta_{1}}{E_{\vk}(i\omega_{k}-E_{1\,\vk})}}&-\frac{\xi_{\vk}}{E_{\vk}(i\omega_{k}+E_{2\,\vk})}&0\\
0&0&-\nth{i\omega_{k}+E_{3\,\vk}}\\
\end{pmatrix}
\end{equation*}
Now it is easy to calculate the first term
\begin{equation}
\begin{split}
\sum_k\tr\big(M_{k}\big)=&
\sum_{k}\mbr{
\frac{2\Delta_{1}^{2}}{E_{\vk}^{2}(i\omega_{k}-E_{1\,\vk})(i\omega_{k}+E_{2\,\vk})}+
\br{\frac{\xi_{\vk}^{2}}{E_{\vk}^{2}(i\omega_{k}-E_{1\,\vk})^{2}}+\frac{\xi_{\vk}^{2}}{E_{\vk}^{2}(i\omega_{k}+E_{1\,\vk})^{2}}
-\nth{(i\omega_{k}+E_{3\,\vk})^{2}}}
}
\end{split}
\end{equation}
Only root $-E_{2\,\vk}$ in the first term contributes in the Matsubara frequency summation at zero temperature.
\begin{equation}\label{eq:pathInt2:pi0-1}
\sum_k\tr\big(M_{k}\big)=\sum_{\vk}\frac{2\Delta_{1}^{2}}{E_{\vk}^{2}(E_{1\,\vk}+E_{2\,\vk})}
\approx\sum_{\vk}\frac{\Delta_{1}^{2}}{E_{\vk}^{3}}-\sum_{\vk}\frac{\Delta_{1}^{2}\Delta_{2}^{2}\xi_{\vk}}{2E_{\vk}^{5}(\xi_{\vk}+\eta)}
\end{equation}

For the lowest order of the second term in Eq. \eqref{eq:pathInt2:pi0long}, we only need to take the lowest order of $B_{k}$
\begin{equation}
B_{k}=\mtrx{i\omega_{k}-E_{\vk}&0&0\\0&i\omega_{k}+E_{\vk}&0\\0&0&i\omega_{k}+\xi_{\vk}+\eta}
\end{equation}
It is easy to verify at this approximation
\begin{equation}\label{eq:pathInt2:pi0-2}
\tr\Big(\delta_{\vk}\widetilde{M}_{k}-\delta_{\vk}M_{k}\Big)=0
\end{equation}
Combine Eq. \eqref{eq:pathInt2:pi0-1} and Eq. \eqref{eq:pathInt2:pi0-2}, we have 
\begin{equation}
\pi^{(0)}(0)\approx\sum_{\vk}\frac{\Delta_{1}^{2}}{E_{\vk}^{3}}-\sum_{\vk}\frac{\Delta_{1}^{2}\Delta_{2}^{2}\xi_{\vk}}{2E_{\vk}^{5}(\xi_{\vk}+\eta)}\end{equation}

$\pi^{\perp}(0)$ can actually be calculated   exactly
\begin{equation}
\pi_{ij}^{\perp}(0)=\sum_k(k_i)(k_j)\tr(\hat{G}_{0\,k}\hat{G}_{0\,k})
\end{equation}
\begin{equation}
\begin{split}
	\tr(\hat{G}_{0\,k}\hat{G}_{0\,k})=&\sum_k\tr\big(T_{\vk}L_{\vk}B_{k}^{-1}L_{\vk}^{\dg}T_{\vk}^{\dg}T_{\vk}L_{\vk}B_{k}^{-1}L_{\vk}^{\dg}T_{\vk}^{\dg}\big)\\
	=&\sum_k\tr\big(B_{k}^{-1}B_{k}^{-1}\big)\\
	=&\sum_{\vk,i}(\sum_{\omega_{n}}(i\omega_{k}-\xi_{i})^{-2})\\
	=&0
\end{split}
\end{equation}

\section{Consistency of the expansion over $\zeta$\label{sec:pathApp:consistency}}
In our treatment here, one crucial assumption in expansion is the smallness of $\Delta_{2}/\eta$ comparing to $1$.  Here we check it.  We have the closed-channel gap equation (\eef{eq:pathInt2:mfclose})
\begin{equation}
\Delta_{2}=\sum{}Yh_{1\vk}+\sum{}Vh_{2\vk}\tag{\ref{eq:pathInt2:mfclose}}
\end{equation}
The first term on the right is relatively small comparing to the second term.  We just keep the second term for estimation.  Furthermore,  we assume $h_{2\,\vk}=\sqrt{N_{c}}\phi_{0\,\vk}$, where $\phi_{0\,\vk}$ is the normalized wave function of the  isolated closed-channel potential satisfying \sch equation (\eef{eq:pathInt2:phi})
\begin{equation}\tag{\ref{eq:pathInt2:phi}}
-E_{b}^{(0)}\phi_{0\,\vp}=2\epsilon_{\vp}\phi_{0\,\vp}-\sum_{\vk}V \phi_{0\,\vk}
\end{equation}
Rearranging it, we have (especially at low momentum)
\begin{equation*}
\sum_{\vk}V \phi_{0\,\vk}=(2\epsilon_{\vp}+E_{b})\phi_{0\,\vp}\approx{\eta}\phi_{0\,\vp}
\end{equation*}
Here $E_{b}$ is the binding energy of the closed-channel bound state and $\eta$ is the Zeeman energy difference. The second approximation is correct at low momentum (smaller or in the same order of the Fermi momentum) as $\epsilon_{\vp}\ll{}E_{b}\approx\eta$ not too far away from the resonance.  Put all these together, we have
\begin{equation*}
\Delta_{2}\approx\alpha{}E_{b}\phi_{k=0}
\end{equation*}
If we assume a simple exponentially decayed wave function as in Eq. \ref{eq:pathInt2:phi2body} from Sec. \ref{sec:pathInt2:short-range} 
\begin{equation}\tag{\ref{eq:pathInt2:phi2body}}
\phi_{\vk}=\sqrt{\frac{8\pi\kappa}{\mathcal{V}_{0}}}\frac{1}{k^{2}+\kappa^{2}}\approx\sqrt{\frac{8\pi\kappa}{\mathcal{V}_{0}}}\frac{1}{\kappa^{2}}
\end{equation}
Here  $\mathcal{V}_{0}$ is the total volume and $\kappa$ is the characteristic momentum of the closed-channel bound state, $\eta\approx{}E_{b}=\hbar^{2}\kappa^{2}/2m$.  The second approximation above is only for  low momentum.  Collect all these together, we have
\begin{equation}
\Delta_{2}\approx\sqrt{N_{c}}\eta\sqrt{\frac{8\pi\kappa}{\mathcal{V}_{0}}}\frac{1}{\kappa^{2}}
\sim\eta\sqrt\frac{n_{c}}{\kappa^{3}}
\sim\eta\br{\frac{k_{Fc}}{\kappa}}^{\frac{3}{2}}
\sim\eta\br{\frac{E_{Fc}}{\eta}}^{\frac{3}{4}}
\end{equation}
%:
$k_{Fc}$ is the Fermi momentum corresponding to density of atoms in the close-channel, which is much smaller than the characteristic momentum for the bound-state, $\kappa$.   Therefore we have $\Delta_{2}\ll\eta$, even when $n_{c}$ is close to the total density $n$. 

Now we check whether the corrections in Fermionic spectrum (Eqs. \ref{eq:pathInt2:xiExpand}-\ref{eq:pathInt2:xiExpand3}), 
are indeed small comparing to the zeroth order terms.  
\begin{align}
E_{1\vk}\approx{}E_{\vk}+u_{\vk}^{2}\Delta_{1}\zeta\tag{\ref{eq:pathInt2:xiExpand}}\\
E_{2\vk}\approx{}E_{\vk}-v_{\vk}^{2}\Delta_{1}\zeta\tag{\ref{eq:pathInt2:xiExpand2}}\\
E_{3\vk}\approx{}\epsilon_{\vk}+\eta+\frac{\zeta}{2}\Delta_{1}
\tag{\ref{eq:pathInt2:xiExpand3}}
\end{align}
Here, we mostly only concern of case of low momentum ($k\sim{}k_{F}$).  In Eq. \ref{eq:pathInt2:xiExpand3}, 
\begin{equation*}
\frac{\zeta\Delta_{1}}{E_{3\vk}}\sim{}\frac{\Delta_{2}^{2}}{\eta^{2}}\sim\br{\frac{k_{Fc}}{\kappa}}^{3}\ll1
\end{equation*}
Eq. \ref{eq:pathInt2:xiExpand} and Eq. \ref{eq:pathInt2:xiExpand2} are slightly more complicated.  Both of them involve $\frac{\Delta_{2}^{2}}{E_{\vk}\eta}$,  at the  BCS limit, the closed-channel density is small, $k_{F\,c}$ is small and that makes this ratio small; when close to the (narrow) resonance, where $n_{c}$ is comparable to to the total density, at low energy, $\Delta_{1}$ is in the order of the Fermi energy, so does $E_{\vk}$.   We have (we no longer distinguish $k_{F\,c}$ with $k_{F}$)
 \begin{equation}\label{eq:pathApp:zetaEs}
 \zeta=\frac{\Delta_{2}^{2}}{\Delta_{1}\eta}\sim\frac{\eta^{2}\frac{k_{Fc}^{3}}{\kappa^{3}}}{k_{F}^{2}\eta}\sim\frac{k_{F}}{\kappa}\ll1
\end{equation}

More deeply, in the secular equation that leads to spectrum, Eq. \ref{eq:pahtApp:secular}.  We rewrite it without scaling to $E_{\vk}$,  (We drop subscript $\vk$ in the following equations for simplicity)
\begin{equation*}
(x^{2}-E^{2})(x-\xi-\eta)-\Delta_{2}^{2}x+\Delta_{2}^{2}E(u^{2}-v^{2})=0
\end{equation*}
It is not hard to use definition of $u$ and $v$ to find $u^{2}-v^{2}=\xi/E$, and express $E^{2}=\xi^{2}+\Delta_{1}^{2}$. Therefore we have
\begin{equation*}
(x-\xi)(x+\xi)(x-\xi-\eta)-\Delta_{1}^{2}(x-\xi-\eta)-\Delta_{2}^{2}(x-\xi)=0
\end{equation*}
Here the first term is for free particles, and let us estimate the relative size of the last two terms.  For low-momentum solution, we simply use $\Delta_{1}\sim{}E_{F}$, we find
\begin{equation*}
\frac{\Delta_{1}^{2}(x-\xi-\eta)}{\Delta_{2}^{2}(x-\xi)}\sim\frac{E_{F}^{2}\eta}{\Delta_{2}^{2}E_{F}}\sim\frac{\kappa}{k_{F}}\gg1
\end{equation*}
This justifies our choice to neglect the last term when finding the lowest-order solution and then use the last term for correction.

 In another word, the above estimation is just saying that the total occupation number of the closed-channel at a low momentum level is much smaller than 1 in all regions of resonance (narrow or broad) because the closed-channel bound state is much smaller than the interparticle distance.  This factor gives us a small factor, $\zeta\sim\frac{r_{c}}{a_{0}}\sim\frac{k_{F}}{\kappa}\sim\sqrt\frac{E_{F}}{\eta}$, upon which we can do the expansion.

\backmatter

\bibliography{../citation}
\bibliographystyle{apalike}

\end{document}